\pdfoutput=0
\documentclass{ws-ijmpe}
\usepackage{amsmath,amssymb,mathrsfs}
\usepackage{graphicx}
\usepackage{hyperref}
\usepackage{ifpdf}
\def\gsim{\mbox{~{\protect\raisebox{0.4ex}{$>$}}\hspace{-1.1em}
	{\protect\raisebox{-0.6ex}{$\sim$}}~}}
\def\lsim{\mbox{~{\protect\raisebox{0.4ex}{$<$}}\hspace{-1.1em}
	{\protect\raisebox{-0.6ex}{$\sim$}}~}}
\def\Eq#1{Eq.~(\ref{#1})}

\def\Fig#1{Fig.~\ref{#1}}
\def\Sect#1{Section~\ref{#1}}
\def\Ref#1{Ref.\cite{#1}}
\def\Refs#1{Refs.\cite{#1}}
\def\Table#1{Table~\ref{#1}}

\def\dd{{\rm d}}
\def\st{\begin{equation}}
\def\stp{\end{equation}}
\def\bg{\begin{eqnarray}}
\def\nd{\end{eqnarray}}
\def\llangle{\left\langle}
\def\rrangle{\right\rangle}
\def\N{{\mathcal N}}
\def\x{{\bf x}}
\def\p{{\bf p}}
\def\pr{{\mathcal P}}
\def\b{{\bf b}}
\def\nc{{\, ,}}
\def\np{{\, .}}
\newcommand{\nn}{\nonumber \\ }
\def\NN{{\scriptscriptstyle \rm NN}}

\def\pseudo{{\scriptscriptstyle \rm pseudo}}

\begin{document}

\markboth{Derek A. Teaney}{Viscous Hydrodynamics and the Quark Gluon Plasma}

\catchline{}{}{}{}{}

\title{Viscous Hydrodynamics and the Quark Gluon Plasma\\
}

\author{\footnotesize Derek A. Teaney
}

\address{
Department of Physics and Astronomy, Stony Brook University,  \\
Stony Brook, New York 11794-3800, United States \\
derek.teaney@stonybrook.edu}

\maketitle

\begin{history}
\received{\today}

\end{history}

\begin{abstract}
One of the most striking results from the Relativistic Heavy Ion Collider
is the strong elliptic flow.  This review summarizes what is observed
and how these results are combined  with reasonable theoretical assumptions 
to estimate the
shear viscosity of QCD near the phase transition. 
A data comparison with viscous hydrodynamics and kinetic theory 
calculations indicates that the shear viscosity to entropy 
ratio is surprisingly small,  $\eta/s < 0.4$.   The preferred range 
is $\eta/s \simeq (1\leftrightarrow 3) \times 1/4\pi$.
\end{abstract}

\section{Introduction}
\label{intro}
One of the most striking observations from the Relativistic Heavy Ion Collider
(RHIC) is the very large elliptic flow\cite{Adams:2005dq,Adcox:2004mh}.  
The primary goal of this report is to explain as succinctly as possible precisely what is 
observed and how the shear viscosity can be estimated from these observations.
The resulting estimates \cite{Molnar:2001ux,Teaney:2003kp,Romatschke:2007mq,Xu:2007jv,Xu:2008av,Drescher:2007cd,Song:2007ux,Dusling:2007gi,Molnar:2008xj,Ferini:2008he}  indicate that the shear viscosity to entropy ratio
$\eta/s$ is close to the limits suggested by the uncertainty principle\cite{Danielewicz:1984ww}, 
and the result of $\N=4$ Super Yang Mills (SYM) theory at strong coupling\cite{Policastro:2001yc,Kovtun:2004de}
\[
   \frac{\eta}{s}=\frac{1}{4\pi} \np
\]
These estimates imply that the heavy ion experiments are probing quantum kinetic processes in this theoretically interesting, but poorly understood regime. 
Clearly a complete understanding of nucleus-nucleus collisions at high energies is extraordinarily difficult.
We will attempt to explain 
the theoretical basis  for these recent claims and the uncertainties in the estimated values of  $\eta/s$.  Additionally, since the result has raised considerable interest outside
of the heavy ion community, this review will try to make the analysis
accessible to a fairly broad theoretical audience.

\subsection{Experimental Overview}

\begin{figure}
\begin{minipage}[c]{6.5in}
\begin{minipage}[c]{2.5in}
\includegraphics[width=2.3in]{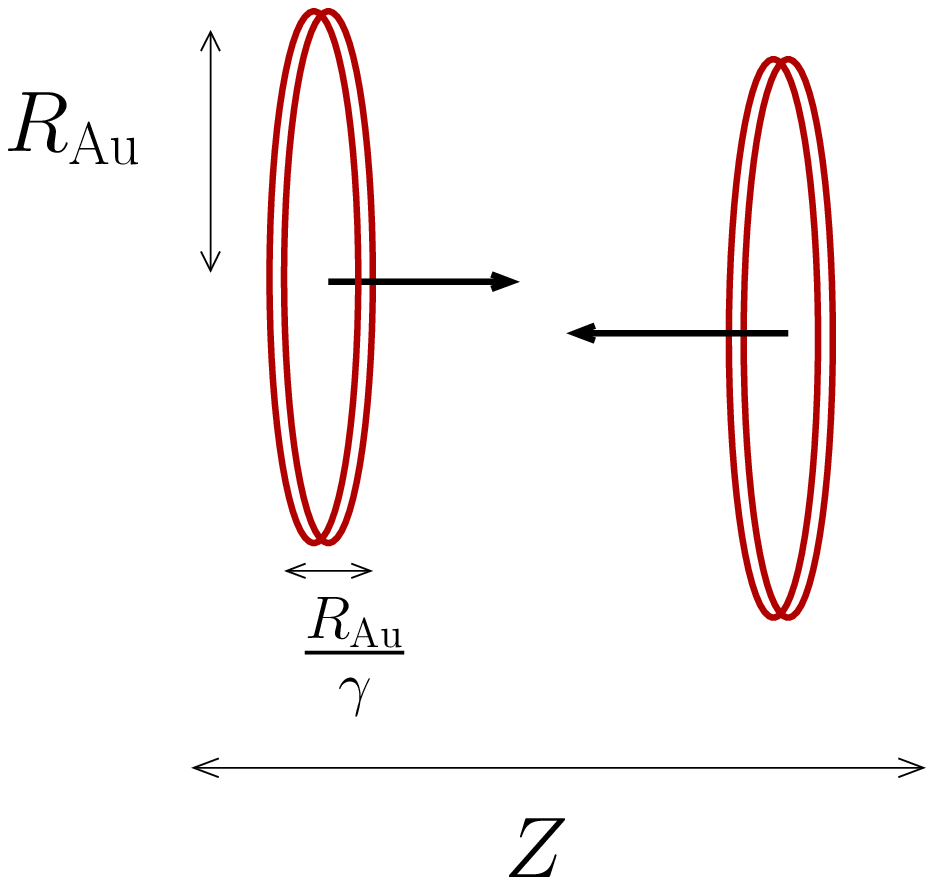}
\end{minipage}
\begin{minipage}[c]{2.5in}
\includegraphics[width=2.3in]{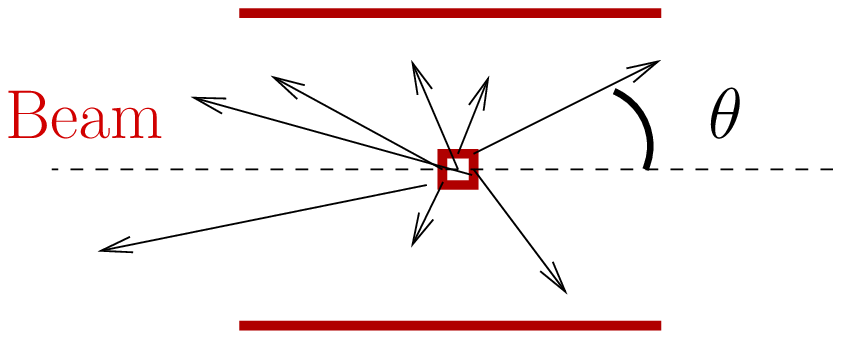}
\end{minipage}
\end{minipage}
\caption{Overview of a heavy ion event.  In the 
left figure the two nuclei collide along 
the beam axis usually labeled as $Z$. 
At RHIC the  nuclei are length contracted by 
a factor of $\gamma\simeq 100$.
The right figure shows the collision vertex of a typical event as viewed
in a schematic particle detector 
and shows a few of the thousands of  
charged particle tracks recorded per event. The angle $\theta$ is usually 
reported in pseudo-rapidity variables as discussed in the text.
\label{geometrylong}}
\end{figure}

In high energy nucleus-nucleus collisions at 
RHIC approximately $\sim 7000$ particles are produced in a single gold-gold
event with collision energy,
$\sqrt{s}=200\, {\rm GeV/nucleon}$. Each nucleus has 197 nucleons and 
the two nuclei are initially length contracted by a factor of a hundred.
The transverse size of the nucleus is $R_{\rm Au} \sim 5
\, {\rm fm}$  and the duration of the event is roughly of order $\sim R_{\rm Au}/c$. 
\Fig{geometrylong} shows the pre-collision geometry. Also shown is
a schematic of the collision vertex and a schematic particle detector.

\begin{figure}
\begin{center}
\includegraphics[height=2.0in]{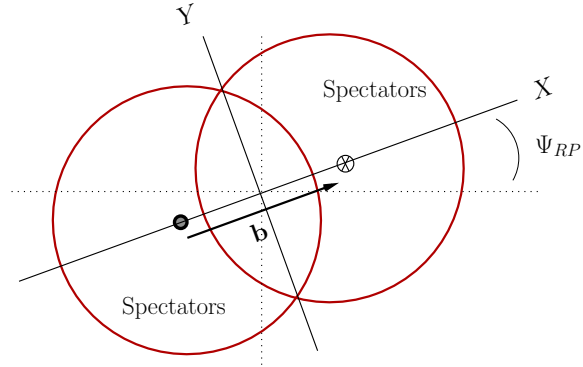}
\end{center}
\caption{A schematic of the transverse plane in a heavy ion event. Both
the magnitude and direction of the  impact parameter $\b$ can be determined
on an event by event basis. $X$ and $Y$ label the reaction plane 
axes and the dotted lines indicate the lab axis.  $\Psi_{RP}$ is 
known as the reaction plane angle. 
\label{nucleus_rot}}
\end{figure}

Usually the two nuclei collide off-center at impact parameter
${\bf b}$ and oriented at an angle
$\Psi_{RP}$ with respect to the lab  axes
as  shown in Fig.~\ref{nucleus_rot}. During the collision the spectator nucleons (see \Fig{nucleus_rot}) 
continue down the beam pipe, leaving behind an excited almond shaped
region.
 The impact parameter ${\bf b}$ is a transverse vector
${\bf b}=(b_x, b_y)$ pointing from the center of one nucleus to the
center of the other.  As discussed in \Sect{elliptic} both the magnitude and direction of
${\bf b}$ can be determined on an event by event basis.   
We will generally work
with reaction plane coordinates $X$ and $Y$ rather than  
lab coordinates.

The elliptic flow is defined as the anisotropy of
particle production with respect to the reaction plane 
(see \Fig{nucleus_rot} and \Fig{nucleus})
\st
    v_{2} \equiv \llangle \frac{p_X^2 - p_Y^2}{p_X^2 + p_Y^2}  \rrangle \nc
\stp
or the second Fourier coefficient of the azimuthal distribution, 
$\llangle \cos(2(\phi- \Psi_{RP}) ) \rrangle$. 
Elliptic flow can also be measured as a function of transverse momentum $p_T =\sqrt{p_X^2 + p_Y^2}$
by expanding the differential yield of particles in a Fourier series
\st
 \frac{1}{p_T} \frac{dN}{dy dp_{T} d\phi} =  \frac{1}{2\pi p_T} \frac{dN}{dy dp_T } \left(1 + 2 v_2(p_T) \cos2(\phi-\Psi_{RP}) + \ldots \right)   \np
\stp
Here ellipses denote still higher harmonics, $v_4$, $v_6$ and 
so on. In addition the flow can be measured as a function
of impact parameter, particle type, and rapidity. 
For a mid-peripheral collision ($b\simeq 7\,{\rm fm}$) 
the average elliptic flow  $\llangle v_2\rrangle $ is approximately $7\%$. 
This is surprising large. For instance, the ratio of particles in the $X$ direction to the $Y$ is  $1+2v_2:1-2v_2 \simeq 1.3:1$.  At 
higher transverse momentum the elliptic flow grows and 
at $p_T \sim 1.5\,{\rm GeV}$  elliptic flow can be as large as 15\%. 

\subsection{An Interpretation of Elliptic Flow}

\begin{figure}
\begin{center}
\includegraphics[height=2.0in]{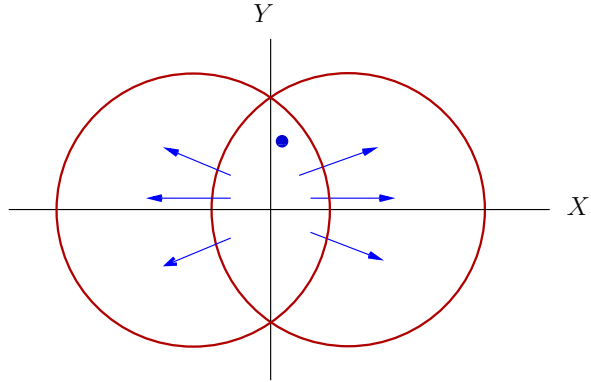}
\end{center}
\caption{ 
The conventional explanation for the observed elliptic flow. The spectators
continue down the beam  pipe
leaving behind an excited oval shape  which expands
preferentially along the short axis of the ellipse.
The  finally momentum asymmetry in the particle distribution $v_2$ reflects the 
response of the excited medium to this geometry. The dot with transverse 
coordinate $\x = (x,y)$ is illustrated to explain a technical point
in \Sect{elliptic}.
\label{nucleus}
}
\end{figure}

The generally accepted explanation for the observed flow
 is illustrated in \Fig{nucleus}.
Since the pressure gradient in the $X$ direction is 
larger than in the  $Y$ direction, the nuclear medium
expands preferentially along the short axis of the ellipse. 
Elliptic flow is such a useful observable because it is a
rather direct probe of the response of the  QCD medium to the high energy
density created during the event. If the mean free path is large compared to
the size of the interaction region, then the produced particles will not
respond to the initial geometry.  On the other hand, if the transverse size of
the nucleus is  large compared to the interaction length scales involved,  
hydrodynamics is the appropriate theoretical framework to calculate 
the response of the medium to the geometry. In a pioneering paper  by
Ollitrualt,
the elliptic flow observable was proposed and analyzed based partly
on the conviction that ideal hydrodynamic models would vastly over-predict
the flow\cite{OllitraultSph,OllitraultPrivate}.  

However, calculations based on ideal hydrodynamics do a fair to reasonable job
job in reproducing the observed elliptic flow\cite{Hirano:2004en,Teaney:2001av,Kolb:2000fh,Huovinen:2001cy,Nonaka:2006yn}. This has been reviewed elsewhere \cite{Huovinen:2006jp,Kolb:2003dz}. 
Nevertheless, the hydrodynamic interpretation
requires that the relevant mean free paths and relaxation times be small
compared to the nuclear sizes and expansion rates. This review 
will assess the consistency  of the hydrodynamic interpretation  
by categorizing viscous corrections. 
 The principle tool is viscous hydrodynamics which needs to be extended into the
relativistic domain in order to address the problems associated with nuclear 
collisions. This problem has received considerable attention 
recently and progress has been achieved both at a conceptual 
\cite{BRSSS,York:2008rr,Betz:2008me,Bhattacharyya:2008jc,Huovinen:2008te} and
practical level \cite{Romatschke:2007mq,Song:2007ux,Song:2008si,Dusling:2007gi,Luzum:2008cw,Molnar:2008xj}.  Generally macroscopic approaches, such as viscous hydrodynamics, and 
microscopic approaches, such as kinetic theory, are converging  on the implications of the measured elliptic flow \cite{Huovinen:2008te,Greco:2008fs,Ferini:2008he,Xu:2004mz,Xu:2007jv,Gombeaud:2007ub}. There has never been an even remotely successful model of 
the flow with $\eta/s > 0.4$.  Since $\eta/s$ 
is a measure of the relaxation time relative to $\hbar/k_B T$  (see \Sect{transport}), this 
estimate of $\eta/s$ places the kinetic processes measured at RHIC in 
an interesting and fully quantum regime.

\section{Elliptic Flow -- Measurements and Definitions}
\label{elliptic}
The goal of this section is to review the progress that has been achieved in
measuring the elliptic flow.  This progress  has produced an 
increasingly self-consistent  hydrodynamic interpretation of the observed elliptic flow results. This section will also  collect the various 
definitions which are needed to categorize the response of the excited
medium to the initial geometry.

\subsection{Measurements and Definitions } 

As discussed in the introduction (see \Fig{nucleus_rot}) both the magnitude
and direction of the impact parameter can be determined on an event by event
basis.
The magnitude of the impact
parameter can be determined by selecting events with definite multiplicity for
example.  For instance, on average the top 10\% of events with the highest
multiplicity correspond to the 10\% of events with the smallest impact parameter.
Since the cross section is almost purely geometrical  in this energy range this
top 10\% of events may be found by a purely geometrical argument.  This line of
reasoning gives that the top 10\% of events are 
produced by collisions with an impact parameter in the range 
\st
 0 < b < b_{*}\nc
\qquad  \mbox{where}  \qquad  10\% = \frac{\pi {b_{*}}^2}{\sigma_{\rm tot} } \nc 
\stp
and  $\sigma_{\rm tot} \simeq \pi (2R_{A})^2$ is the total inelastic cross
section. After categorizing the top 10\% of events
we can categorize the top 10-20\% of events and so on.  The general  relation
is
\st
  \left(\frac{b}{2R_A}\right)^2 \simeq   \mbox{\% Centrality} \np
\stp
Here we have neglected 
fluctuations  and many other effects. For instance there is a
very small probability  that an event with impact parameter 
$b=4\,{\rm fm}$ will
produce the same multiplicity as an event with $b=0\,{\rm fm}$.   A full
discussion of these and many other issues is given in \Ref{Miller:2007ri}.
The end result is that the magnitude of the impact parameter ${\bf b}$ 
can be determined
to within half a femptometer or so\cite{SteinbergPrivate}. 

\begin{figure}
\begin{center}
\includegraphics[height=3.0in]{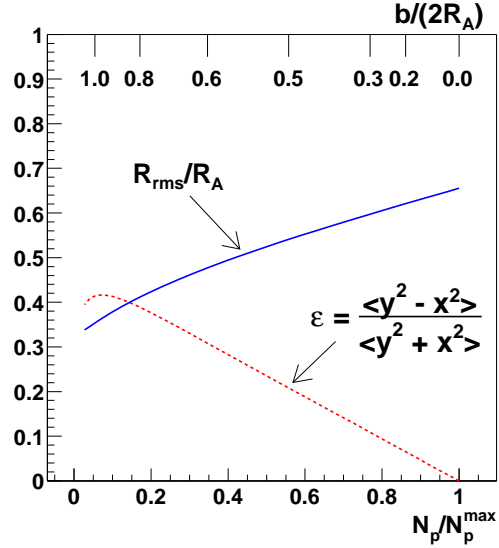} 
\end{center}
\caption{ The standard Glauber eccentricity $\epsilon_{\rm s,part}$ as a function of the number of   
participants. $N_p^{\rm max}\simeq 340$ 
is the maximum  number of participants in a central AuAu event and $R_A\simeq6.3\,{\rm fm}$ is the gold radius. 
The top axis shows the translation between impact parameter and participants.  
The root mean square radius $R_{\rm rms}$ and the standard Glauber eccentricity  
are given in \Eq{epsilonpart} and \Eq{Rrms}.
\label{eccentricity}
}
\end{figure}

Now that the impact parameter is quantified, a useful definition 
is the number of participating nucleons (also called ``wounded" nucleons). 
The number of nucleons per unit volume 
in the  rest frame  of the nucleus is $\rho_{A}(\x-\x_o,z)$,
were 
$\x-\x_o$ is the transverse displacement from  
a nucleus centered at $\x_o$, and $z$ is the longitudinal
direction.  These distributions are known experimentally  
and are reasonably modeled by a Woods-Saxon form\cite{Miller:2007ri}. 
The number of nucleons per unit transverse area is 
\st
   T_{A}(\x - \x_o) =  \int_{-\infty}^{\infty} \dd z \, \rho_{A} (\x - \x_o, z) \np
\stp
Then, after reexamining \Fig{nucleus}, 
we find that the  probability that a nucleon at $\x=(x,y)$ 
will suffer an inelastic interaction
passing through the right nucleus  
centered $\b/2 = (+{\rm b}/2, 0)$  is
\[
     1 -  \exp\left(-\sigma_{\NN} T_{A}(\x -\b/2)  \right)  \nc
\]
where  
$\sigma_{\NN}\simeq 40\,{\rm mb}$ 
is the inelastic nucleon-nucleon cross section. 
The number of nucleons which suffer an inelastic collision per unit area is
then
\bg
 \frac{\dd N_{p}}{\dd x \dd y} &= &
           T_A({\bf x}_\perp+{\bf b}/2) \left[1 - \exp\left(-\sigma_{\NN} 
T_A({\bf x}_\perp - {\bf b}/2) \right)\right] \nonumber  \\
 & & +
         \,  T_A({\bf x}_\perp-{\bf b}/2) \left[1  - \exp\left(-\sigma_{\NN}
         \,  T_A({\bf x}_\perp+{\bf b}/2) \right) \right] \, \np
\nd
Finally,  the total number of participants ($i.e.$ the the number of 
nucleons which collide) is
\st
 N_{p} = \int \dd x\,\dd y \, \frac{\dd N}{\dd x \dd y}  \np
\stp
For a central collision of two gold nuclei the  number
of participants $N_p\simeq 340$  nearly equals the total number nucleons 
in the two nuclei, $N=394$, leaving about fifty spectators.  By 
comparing the top axis in  Fig. ~\ref{eccentricity} to 
the bottom axis, the relationship between  participants, impact 
parameter $b$, and centrality can be determined.

The reaction plane angle,  $\Psi_{RP}$, is also determined experimentally.
Here we will describe the Event Plane method which is conceptually the 
simplest.
Assume first that the reaction plane angle is known. Then the particle 
distribution can be expanded in harmonics about the reaction plane 
\st
 \frac{dN}{d\phi} \propto 1 + 2 v_2 \cos(2(\phi - \Psi_{RP}))  + \ldots 
\stp 
If the number of particles is very large one could simply make 
a histogram of the angular distribution of particles in an event  
with respect to the lab axis.
Then the reaction plane  angle could be determined by  finding
where the histogram is maximum. This is the basis of the event plane
method\cite{Voloshin:2008dg}. For all the particles in the event we form the vector
\bg
   \vec{Q} &=& \left(Q_x \,  , Q_y\right) =  \left(\sum_i \cos 2\phi_i\,  , \, \sum_{i} \sin 2\phi_i \right) \np
\nd
Using the continuum approximation, $Q_x \simeq \int \dd\phi \, dN/d\phi \cos(2\phi)$, 
we can estimate the reaction plane angle $\Psi_{RP}$, from the $\vec{Q}$-vector
\st
   \frac{\vec{Q}}{|\vec{Q}|} \equiv \left(\cos(2\Psi_2), \sin(2\Psi_2) \right)  \simeq  \left(\cos(2\Psi_{RP}) \, , \, \sin(2\Psi_{RP}) \right) \np
\stp 
Then we can estimate the elliptic flow as $v_2^{\rm obs} \simeq \llangle \cos(2(\phi_i - \Psi_2) ) \rrangle$.   The  estimated angle $\Psi_2$
differs from $\Psi_{RP}$  due to statistical fluctuations.  Consequently
$v_2^{\rm obs}$ will be systematically smaller than  $v_2$ since $\Psi_2$ 
is not $\Psi_{RP}$. This leads to a correction to the estimate given 
above which is known as the reaction plane resolution. The final result, after 
considering the dispersion of  $\Psi_2$ relative to the true  reaction
plane angle $\Psi_{RP}$ is 
\st
  v_2 = \frac{v_2^{\rm obs}}{\mathscr R}   \qquad\mbox{where} \qquad {\mathscr R} = \llangle \cos2(\Psi_2 - \Psi_{RP}) \rrangle  \np
\stp
In practice the resolution parameter ${\mathscr R}$
is estimated by dividing a given event into sub-events and
looking at the dispersion in $\Psi_2$ between different sub-events.

There is {\it a lot} more to the determination of the event 
plane in practice. Fortunately the various methods have been reviewed 
recently\cite{Voloshin:2008dg}.   
An important criterion for the validity of these
methods  is that the magnitude of elliptic flow be large compared 
to statistical fluctuations
\st
    v_2^2  \gg  \frac{1}{N} \np
\stp
For $v_2 \simeq 7\%$ and  $N\simeq 500$ we have $Nv_2^2 \simeq 2.5$.  
Since this number is not particularly large the simple method described
above is not completely adequate in practice.
The resolution parameter  is $\mathscr R \simeq 0.7$ in the STAR experiment.
At the LHC, estimates suggest that the resolution parameter $\mathscr R$ 
could be as large as\cite{RaimondPrivate} $\mathscr R\simeq 0.95$.
Current methods  use  two particle,
four particle, and higher cummulants to remove the effects of correlations  and fluctuations.  These advances  are discussed more completely in \Sect{eccentricity_sect}  and
have played an important role in the current estimates  of the shear viscosity. 
The current measurements provide a unique 
theoretical opportunity to study systematically  how  hydrodynamics begins 
to develop in mesoscopic systems. 

We would like to measure the response of nuclei to the geometry. 
To this end, we categorize the overlap region with  an asymmetry parameter 
$\epsilon_{s,\rm part}$
\st
 \epsilon_{\rm s,part}  = \frac{\llangle y^2 - x^2 \rrangle }{\llangle y^2 + x^2 \rrangle} \np
\label{epsilonpart}
\stp
Traditionally the average $\llangle \ldots \rrangle$ is taken with respect 
to the number of participants  in the transverse plane, for example
\st
 \llangle y^2-x^2\rrangle = \frac{1}{N_p} \int \dd x \dd y\,   (y^2-x^2) \frac{\dd N_p}{\dd x \, \dd y} \np
\stp
We will explain the  ``s,part" label shortly; for the moment 
we return to \Fig{eccentricity},  which plots
the asymmetry parameter  versus centrality  and also shows the 
the root mean square radius
\st
   R_{\rm rms} = \sqrt{ \llangle x^2 +y^2 \rrangle } \nc
\label{Rrms}
\stp
which is important for categorizing the size of viscous corrections.
  
\subsection{Interpretation}
\label{interpretation}

\begin{figure}
\begin{center}
\includegraphics[height=4.5in]{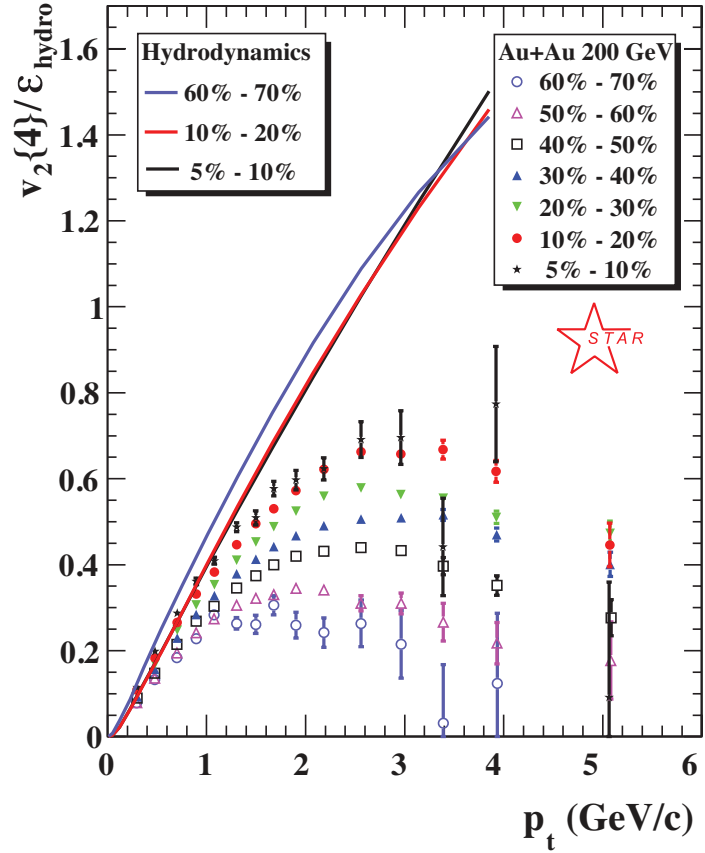}
\end{center}
\caption{
Elliptic flow $v_2(p_T)$ as measured by the STAR collaboration\protect \cite{STARV2,STAR:2008ed} for different centralities.  
The measured elliptic flow has been divided by the eccentricity -- $\epsilon_{\rm hydro}\equiv \epsilon_{s,\rm part}$ in this work. The curves are ideal hydrodynamic calculations based
on \protect\Refs{Huovinen:2006jp,Huovinen:2001cy} rather than the viscous hydrodynamics discussed in much of this review. 
\label{raimond_v2pt} 
}
\end{figure}

We have collected the essential definitions of $\epsilon$,  centrality,
and $v_2$, and  are now in a position to return to the physics. 
The scaled elliptic flow $v_2/\epsilon$
measures the response of the medium to the initial geometry.
\Fig{raimond_v2pt} shows $v_2(p_T)/\epsilon$ as a
function of centrality, 0-5\% being the most central and 60-70\% being the most
peripheral. Examining this figure we see a gradual transition  from a weak to a
strong dynamic response with growing system size.  The
interpretation adopted in this review is that this  change is a consequence of
a system transitioning from a kinetic to a hydrodynamic regime. 

There are several theoretical curves based upon calculations
of ideal hydrodynamics\cite{Huovinen:2001cy,Kolb:2000fh} 
which for $p_T <  1\,{\rm GeV}$ approximately
reproduce the observed elliptic flow in the most central collisions.
Since ideal hydrodynamics 
is scale invariant (for a scale invariant equation of state) the expectation
is that the response $v_2/\epsilon$ of this theory should 
be independent of system size or centrality. This reasoning is borne out 
by the more elaborate hydrodynamic calculations shown in the figure.
On the other hand, the data show a gradual transition as a function of
increasing centrality, rising towards the ideal hydrodynamic calculations  in
a systematic way.
These trends are captured by models with a finite mean free path\cite{Drescher:2007uh}. 

The data show other trends as a function of centrality. In more 
central collisions the linearly rising trend, which resembles
the ideal hydrodynamic calculations,
extends to larger and larger transverse momentum.
We will see in \Sect{kinetics}
that viscous corrections to ideal hydrodynamics grow as 
\st
   \left(\frac{p_T}{T}\right)^2 \frac{\ell_{\rm \scriptscriptstyle mfp}}{L} \nc
\stp
where $L$ is a characteristic length scale. Thus 
these viscous corrections restrict the applicable momentum range in hydrodynamics\cite{Teaney:2003kp}.
In more central collisions, where $\ell_{\rm \scriptscriptstyle mfp}/L$ 
is smaller,  the transverse momentum range described by 
hydrodynamics extends to increasingly large $p_T$.  These qualitative
trends are reproduced by the more involved viscous calculations  
discussed in \Sect{viscous_model}. 

\begin{figure}
\begin{center}
\includegraphics[height=3.0in]{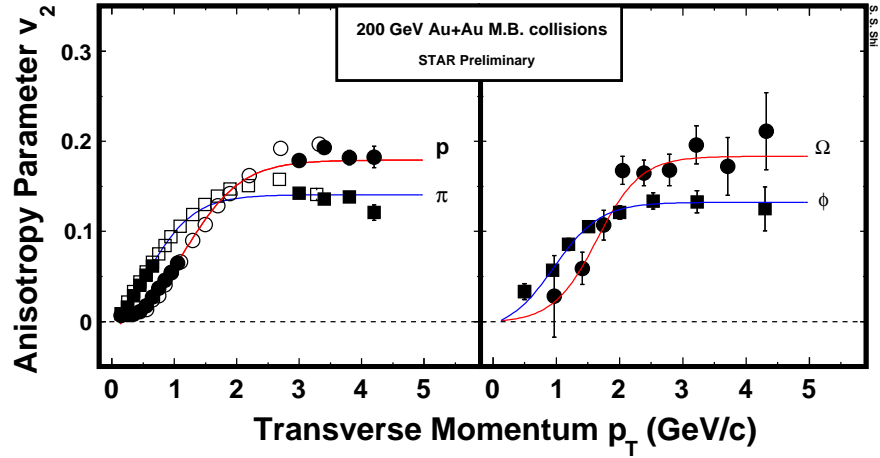}
\end{center}
\caption{A comparison of the elliptic flow of pions and protons 
to the elliptic flow  of the multi-strange $\phi$ and $\Omega^{-}$ 
hadrons\protect \cite{omega_ref}.
\label{omega} }
\end{figure}

To conclude this section, we turn to \Fig{omega} which compares the elliptic 
protons and pions to the 
flow of the multi-strange hadrons $\Omega^{-}$ and $\phi$. 
(These hadrons  
have valence quark content $sss$ and $s\bar s$ respectively.)
The important point
is that the $\Omega^{-}$ is nearly twice as heavy as the proton
and more importantly, does not have a strong resonant interaction
analogous to the $\Delta$. For these reasons the hadronic relaxation time
of the $\Omega^{-}$ is expected to be much longer than the duration 
of the heavy ion event\cite{omega_long}. Nevertheless the $\Omega$ shows nearly
the same elliptic flow as the protons.  This provides fairly convincing
evidence that the majority of the elliptic flow develops during 
a deconfined phase which 
hadronizes to produce a flowing $\Omega^{-}$ baryon. 

\subsection{The Eccentricity and Fluctuations}
\label{eccentricity_sect}

\begin{figure}
\begin{center}
\includegraphics[height=2.5in]{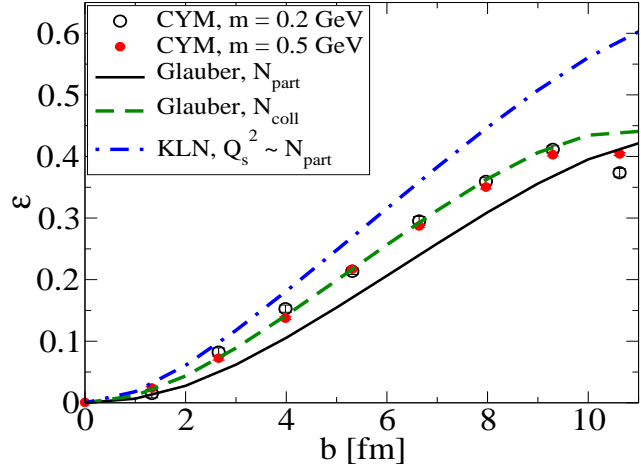} 
\end{center}
\caption{Figure from \protect \Ref{Lappi:2006xc} showing various estimates for the initial eccentricity in  heavy ion collisions. The physics of the KLN
eccentricity is described in the text. In the KLN model the 
eccentricity is expected
to increase by about 20\% when going from RHIC to the LHC\protect \cite{Drescher:2007uh}.
\label{raju_eccentricity} }
\end{figure}

Clearly much of the interpretation of elliptic flow 
relies on a solid understanding 
of the eccentricity. There are several issues here. First 
there is the theoretical uncertainty in this  average quantity.  For 
example, so far we have defined the ``standard Glauber participant eccentricity" 
in \Eq{epsilonpart}. An equally good definition is 
provided by collision scaling. For instance, one measure  used
in heavy ion collisions is the number of binary nucleon-nucleon collisions per transverse area
\st
   \frac{\dd^2N_{\rm coll} }{\dd x \dd y} = \sigma_{\NN} T_A(\x + \b/2) T_A(\x - \b/2 ) \nc
\stp 
Then the eccentricity is defined with this $N_{\rm coll}$ weight in analogy with \Eq{epsilonpart}.
\Fig{raju_eccentricity} shows the  ``standard Glauber $N_{\rm coll}$ 
eccentricity". 
Another more sophisticated model is provided by the KLN model 
which is based on the ideas of gluon saturation and the Color Glass Condensate (CGC) \cite{Kharzeev:2007zt,Kharzeev:2002ei} as implemented
in Refs.~\cite{Hirano:2005xf,Drescher:2006pi}.  
This model is a safe upper bound on what can be expected for the 
eccentricity from saturation physics and  is also shown in
\Fig{raju_eccentricity}.    
We can not describe the details of this model and its implementation 
here. However, the physical reason why this model has a sharper eccentricity 
is the readily understood: the center of one nucleus (nucleus $A$) passes through the edge 
of the other nucleus (nucleus $B$). 
Since the density of gluons per unit area in the 
initial wave function is larger in the center of a nucleus 
relative to the edge,  
the typical momentum scale of nucleus $A$ ($\sim Q_{s,A}$)
is larger nucleus $B$ ($\sim Q_{s,B}$).
It is then difficult for the long wavelength (low momentum) 
gluons in  $B$ to liberate
the short wavelength gluons in $A$. The result is that the 
production of gluons  falls off more quickly  near the $x$ edge 
relative to the $y$ edge making the eccentricity   larger. Clearly
this physics is largely correct although the magnitude of the effect
is uncertain. 
Another CGC estimate of $\epsilon$ is  based on  classical
simulations of Yang-Mills fields. The simulations
include similar  saturation physics  but model the 
production and non-perturbative sectors differently. 
The eccentricity from these simulations is also shown
in \Fig{raju_eccentricity} and is similar to the $N_{\rm coll}$  
eccentricity\cite{Lappi:2006xc}. Thus the predictions of the KLN model seem to be a safe upper bound
for the  eccentricity in heavy ion collisions. 
Note that an important phenomenological consequence of the 
the KLN model is that the eccentricity grows with beam  energy  and 
is expected to increase about 20\% from the RHIC to the LHC\cite{Drescher:2007uh}.

\begin{figure}
\begin{center}
\includegraphics[height=2.0in]{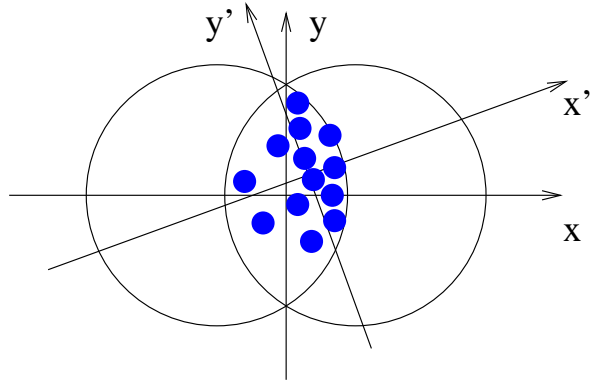}
\end{center}
\caption{
A figure from \protect \Ref{Bhalerao:2006tp} illustrating the participant plane 
eccentricity $\epsilon_{PP}$ in a single event.
\label{rotated_ellipse}
}
\end{figure}

Another important aspect in heavy ion collisions  when interpreting
the elliptic flow data is fluctuations in the initial eccentricity. 
These fluctuations  are not accounted for in \Fig{raju_eccentricity}.
The history is complicated and is reviewed in \Refs{Ollitrault:2009ie,Voloshin:2008dg}. 
There are fluctuations in the initial eccentricity of the 
participants especially in peripheral AuAu and CuCu collisions. 
Thus rather than using the continuum approximation given in 
\Eq{epsilonpart} it is better to 
implement a Monte-Carlo Glauber calculation 
and estimate the  eccentricity using the ``participant plane eccentricity". 
\Fig{rotated_ellipse} illustrates the issue: In a given event the  
ellipse is 
tilted and the eccentricity depends on the distribution of participants.
This event by event eccentricity is denoted $\epsilon_{PP}$ in the literature. 
Clearly the experimental goal is to extract the {\it response} 
coefficient $C$ relating
the elliptic flow to the eccentricity on an event by event basis
\st
    v_2  = C \epsilon_{PP} \np
\stp

If the flow methods measured $\llangle v_2 \rrangle$,  then we 
could  simply divide the measured flow to determine the response coefficient, $C=\llangle v_2 \rrangle / \llangle \epsilon_{PP} \rrangle$.
The PHOBOS collaboration deciphered
the  confusing CuCu data by recognizing the need for $\epsilon_{PP}$ and
following this procedure\cite{Alver:2006wh}.
However, it was generally realized (see in particular. \Ref{Miller:2003kd}) that the elliptic flow methods
 do not measure precisely $\llangle v_2\rrangle$. Some methods (such as two particle correlations $v_2\left\{2\right\}$)
are sensitive  to $\sqrt{\llangle v^2_2 \rrangle}$, while other methods (such 
as the event   plane method $v_2\left\{EP\right\}$) measure  
something closer to $\llangle v_2 \rrangle$. What 
precisely the event plane method measures depends on 
the reaction plane resolution in a known way\cite{Ollitrault:2009ie}.
So just dividing the measured flow by the average participant eccentricity is not entirely 
correct.
The appropriate quantity to divide by depends on the method 
\cite{Miller:2003kd,Bhalerao:2006tp,Voloshin:2007pc}.
In a Gaussian approximation for the eccentricity fluctuations 
this can  be worked out analytically. For instance, the two particle correlation
elliptic flow  $v_{2}\left\{2\right\}$ (which measures
$\sqrt{\llangle v_2^2 \rrangle}$),  should be divided by 
$
   \sqrt{ \llangle \epsilon_{PP}^2 \rrangle } \np
$
An 
important corollary of this analysis is that $v_2\left\{4 \right\}$  ($v_2$ 
measured from four particle correlations) 
can  be divided by  $\epsilon_s$ of \Eq{epsilonpart} 
to yield a good estimate of the coefficient
$C$. This is the policy adopted in \Fig{raimond_v2pt}. 
Unfortunately, in the most peripheral  AuAu bins and in CuCu the Gaussian 
approximation is poor  due to strong correlations amongst the 
participants\cite{Alver:2008zza}. These correlations 
arise because participants come in pairs and every participant 
is associated with another participant in the other nucleus.
Presumably the last centrality bin in \Fig{raimond_v2pt} could be  
moved up or down somewhat  due to non-Gaussian corrections of this sort.
With a complete understanding of what each method measures, \Ref{Ollitrault:2009ie} 
was able to make a simple model for the 
fluctuations and non-flow and show that $\llangle v_2 \rrangle$ measured by
the different methods are compatible to an extremely good precision.
This work should be extended to the CuCu system where non-Gaussian
fluctuations are stronger and ultimately corroborate the 
PHOBOS  analysis\cite{Alver:2006wh,Alver:2008zza}. This is a worthwhile goal 
because it will clarify the transition into the hydrodynamic regime \cite{Drescher:2007cd}.

\subsection{Summary}

In this section we have gone into considerable experimental detail 
--  perhaps more than necessary to explain the basic ideas. 
The reason for this lengthy summary is because the trends
seen in \Fig{raimond_v2pt} were not always so transparent. The relatively 
coherent  hydrodynamic and kinetic interpretation of the observed
elliptic flow (which was previewed in \Sect{interpretation} and which is discussed more 
completely below) is the result of  careful experimental analysis. 

\section{The Shear Viscosity in QCD}
\label{transport}
In this section we will discuss thermal QCD  in equilibrium with the primary
goal of collecting various theoretical estimates for the shear viscosity in
QCD.

\begin{figure}
\begin{center}
\includegraphics[height=2.5in]{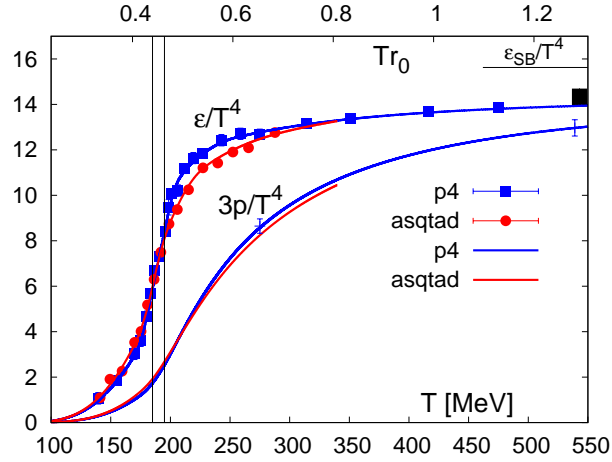}
\end{center}
\caption{ 
Figure from \protect \Ref{Bazavov:2009zn} illustrating 
the energy density and pressure by $T^4$ of QCD computed with $N_{\tau}=8$ lattice data.  (In this figure 
$\epsilon$ is energy density $e(T)$ and the pressure $p$ is denoted with $\pr$ throughout this review.) 
$\epsilon_{SB}/T^4 \equiv e_{SB}/T^4$  is the energy density of a free three flavor
massless QGP (see text).
\label{eos}
}
\end{figure}

The prominent feature 
of QCD at finite temperature is the presence of an approximate 
phase transition from hadrons to 
quarks and gluons.  
The Equation of State (EoS) from 
lattice QCD calculations is  shown in \Fig{eos}, and the 
energy density $e(T)$
shows a rapid change for the temperature range, $T\simeq 170-220\,{\rm MeV}$.
As estimated in \Sect{hydro_basic}, the transition region is 
directly probed during high energy heavy ion collisions.

Well below the phase transition, 
the gas of hadrons  
is very dilute and the thermodynamics is dominated by  
the measured particle spectrum. 
For instance the number of  pions in this low temperatures regime is
\st
   n_{\pi} = d_\pi \int \frac{\dd^3\p }{(2\pi)^3 }  \frac{1}{e^{E_\p/T } - 1}  \nc
\stp 
where $E_\p = \sqrt{p^2 + m_\pi^2}$ and 
$d_{\pi}=3$ counts the three fold isospin degeneracy, 
$\pi^{+},\pi^{-},\pi^{0}$, in the spectrum. If all known particles are included
up to a mass $m_{\rm res} < 2.5\,{\rm GeV}$,  the resulting  Hadron Resonance Gas (HRG) equation of
state does a reasonable job  of reproducing the
thermodynamics up to about $T\simeq 180\,{\rm MeV}$. However, the validity
of this quasi-particle description is unclear above a temperature 
of\cite{Itakura:2007mx}, $T\simeq 140\,{\rm MeV}$.
As the temperature increases, the hadron wave functions overlap until
the medium reorganizes into quark and gluon degrees of freedom.  
Well above the transition the QCD medium evolves  to 
a phase of massless quarks and gluons. The energy density is approximately 
described by
the  Stefan-Boltzmann  equation of state
\st
    e_{\rm glue} = d_{\rm glue} \int 
\frac{\dd^3\p}{(2\pi)^3} 
\frac{E_\p}{e^{E_\p/T} -1 }  \nc
\qquad   
e_{\rm quark} 
= d_{\rm quark} 
\int \frac{\dd^3\p}{(2\pi)^3} 
\frac{E_\p}
{ e^{E_\p/T} +1 } \nc 
\stp
where $d_{\rm glue}={\scriptstyle 2\times8}$ counts spin and color, and $d_{\rm quark}={\scriptstyle 2\times2\times3\times3}$ counts spin, anti-quarks, flavor, and color. 
Performing these integrals we find, $e_{SB} = e_{\rm glue} + e_{\rm quark}\simeq 15.6\,T^4$ as illustrated by the line in the top-right corner of the figure.

We have described the particle content well above and well below 
the transition. Near the approximate phase transition the validity of such 
a simple quasi-particle description is not clear. 
 The transition is a rapid cross-over where  
hadron degrees of freedom evolve into quark and gluon degrees of freedom rather 
than a true phase transition. All correlators
change smoothly, but rapidly, in a temperature range of $T\simeq 170-210\,{\rm MeV}$. 
From a phenomenological perspective the smoothness of the transition
suggests that the change from quarks to hadrons should be thought
of as a soft process  rather than an abrupt change.

Lattice QCD simulations have determined the equation of state rather well\cite{}. However, in addition to the equation of state,  we need to estimate the
transport coefficients to assess whether the heavy ion reactions  produce enough material, over
a large enough space-time volume to be described  in thermodynamic terms. 
 The shear and bulk viscosities govern  the transport of energy
and momentum and are clearly the most important.

Later in \Sect{hydro_basic} and \Sect{kinetics} we will 
describe the role of 
shear viscosity in the reaction dynamics. 
In this section we summarize the shear viscosities found in various theoretical computations which will place these dynamical conclusions in context.
A good way to implement this theoretical summary
 is to form shear viscosity 
to entropy ratio\cite{Kovtun:2004de},  $\eta/s$. 
To motivate this ratio we remark that it seems
difficult to transport energy faster than a quantum time scale set by 
the inverse temperature\footnote{In this paragraph we will restore $\hbar$ 
and the Boltzmann constant, $k_B$.}, 
\[
\tau_{\rm quant}  \sim \frac{\hbar}{k_B T} \np
\]
A sound wave propagating with speed $c_s$ will diffuse (or spread out) due to the 
shear viscosity. Linearized hydrodynamics shows that this process is controlled by the momentum diffusion coefficient, $D_{\eta}\equiv \eta/(e + \pr)$,
 where   $e + \pr$ is the enthalpy (see for example \Ref{Teaney_corr}).
Noting that the diffusion coefficient has units of $({\rm distance})^2/{\rm time}$, 
a kinetic theory estimate 
for the diffusion process  yields 
\st
   D_{\eta} \equiv \frac{\eta}{e + \pr} \sim  v_{\rm th}^2 \tau_R \nc
\stp
where $\tau_R$ is the particle relaxation time and   $v_{\rm th}^2 \sim c_s^2$ 
is the particle velocity. Dividing by   $v_{\rm th}^2$ and 
using the thermodynamic  estimates
\st
   s T \sim  e v_{\rm th}^2 \sim \pr \sim n\,  k_B T  \nc
\stp
we see that 
\st
   \frac{\eta}{s} \, \sim \tau_R T \sim \frac{\hbar}{k_B} \frac{\tau_R}{\tau_{\rm quant} }  \np
\stp
Therefore, $\eta/s$ is the ratio between the medium 
relaxation time and the quantum time scale $\tau_{\rm quant}$ 
in units of $\hbar/k_B$, {\it i.e.}
a measure of the transport time in ``natural units".  

In the dilute 
regime the 
ratio between the medium relaxation time and the quantum time scale is long and
kinetic theory can be used to calculate the shear viscosity to entropy ratio. 
First we consider a simple classical massless gas with particle 
density $n$ and a constant hard sphere cross section $\sigma_o$. 
The equation of state of  this gas is $e= 3\pr = 3n T$  and the 
shear viscosity  is computed using kinetic theory\cite{constantsig} 
\st
\label{shearvisc0}
    \eta \simeq 1.2 \frac{T}{\sigma_o } \nc
\stp
The entropy is $s = (e+\pr)/T$ and the resulting shear to entropy ratio is 
\st
   \frac{\eta}{s} \simeq 0.3 \, \frac{T}{n\sigma_o} \np
\stp
In what follows, this calculation will provide a qualitative understanding
of more sophisticated kinetic calculations. 

\begin{figure}
\begin{center}
\includegraphics[height=3.6in]{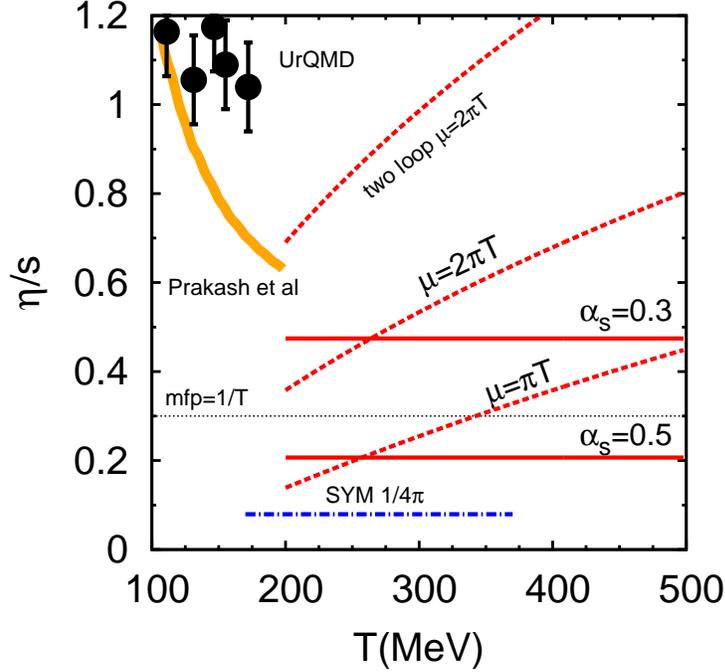}
\end{center}
\caption{(Color Online) A compilation of values of $\eta/s$. The
results from Prakash {\it et al} are from \protect \Ref{Prakash:1993bt}  and 
describe  a meson gas of pions and kaons (and indirectly $K^{*}$ and $\rho$) 
computed with measured cross sections. The black points are based 
on a Kubo analysis of the UrQMD code which includes many higher resonances
\protect \cite{Demir:2008tr}. The red lines are different implementations of the AMY (Arnold, Moore, Yaffe) calculation
of shear viscosity \protect \cite{Arnold:2003zc}. In each curve the Debye scale is fixed $m_{D} =2 T$. 
In the dashed red curves the (one loop three flavor) 
running coupling is taken at the scale $\mu$. In 
the solid red curves $\alpha_s$ is kept fixed.  The two loop 
running coupling is shown with $\mu=2\pi T$ for comparison and
the two loop $\mu=\pi T$ (not shown) is similar to the one loop $\mu=2\pi T$ result.
In the AMY curves, changing the Debye mass by $\pm 0.5 T$ 
changes $\eta/s$ by $\sim \pm 30\%$.
Finally the thin dashed line 
indicates a simple model discussed in the text with $\ell_{\rm mfp}=1/T$.
\label{etabys} }
\end{figure}

In the dilute hadronic regime,
$\eta/s$ was calculated in \Ref{Prakash:1993bt} using 
measured elastic cross sections for a gas of pions and kaons.
 In the $\pi\pi$ phase shifts there is a prominent
$\rho$ resonance, while in the $\pi K$ channel there is a prominent 
$K^{*}$ resonance. Thus the equation of state of this gas is well modeled 
by an ideal gas of $\pi, K, \rho$ and $K^{*}$ \cite{BethUhlenbeck,RajuA}.  The viscosity 
of this mixture was computed in \Ref{Prakash:1993bt} and the current author digitized 
this viscosity, computed the entropy, and determined the $\eta/s$ ratio. 
This is shown in \Fig{etabys}.  Slightly larger values   were obtained in \Ref{Itakura:2007mx} which also estimated the range of
validity for hadronic  kinetic theory,   $T \lsim 140\,{\rm MeV}$.  
Finally a more involved Kubo analysis of the UrQMD  hadronic transport model \cite{Demir:2008tr} 
 (which includes many resonances) is also displayed in \Fig{etabys}.

At asymptotically high temperatures the coupling constant $\alpha_s$ is 
weak and the shear viscosity can be computed using perturbation theory.
Initially, only $2\rightarrow 2$ elastic scattering 
was considered, and the shear viscosity was computed in a leading 
log plasma with self consistent screening\cite{Baymetal}.   
Later it was recognized\cite{Baier:2000sb,Aurenche:1998nw}  that collinear Bremsstrahlung processes 
are important for the calculation of shear viscosity and this realization
ultimately resulted 
in a complete leading order calculation \cite{Arnold:2003zc}.  We  can
estimate $\eta/s$ in the perturbative plasma using 
\Eq{shearvisc0} with $s\propto T^3$ and $\sigma \propto  \alpha_s^2/T^2$,
\st
      \frac{\eta}{s} \sim \frac{1}{\alpha_s^2} \np
\stp
The final result from a complete calculation 
has the  form
\st
\label{AMYcalc}
      \frac{\eta}{s} = \frac{1}{\alpha_s^2} F(m_{D}/T)  \nc
\stp
where $F(m_D/T)$ is a function of the Debye mass which was 
computed for $m_D/T$ small and then extrapolated to more realistic values\cite{Arnold:2003zc}.
There are many scales in 
the problem and it is difficult to know what precisely to take 
for the Debye mass  and the coupling constant. At lowest
order in the coupling, the Debye mass is\cite{LeBellac} 
\st
  m_{D}^2 =   \left(\frac{N_c}{3}  + \frac{N_f}{6} \right) g^2 T^2 \nc 
\stp
but this is too large to be considered reliable.
For definiteness we have evaluated the leading coupling constant in \Eq{AMYcalc} at a 
scale of $\pi T$ and set the Debye mass to  $m_{D}=2T$.   The 
resulting value of $\eta/s$  is shown in \Fig{etabys}.  Various other 
alternatives are explored in the figure and underscore the ambiguity 
in these numbers.

Clearly all of the calculations presented  have a  great deal
of uncertainty around the phase transition region. On the 
hadronic side there are 
a large number of inelastic reactions which become important.
On the quark gluon plasma side, the strong dependence on 
the Debye scale and the coupling constant is disconcerting.  
It is very useful to have a strongly coupled theory 
where the shear viscosity to entropy ratio can be computed exactly.
In strongly coupled $\N=4$ SYM theory with a large
number of colors, $\eta/s$ can be computed using 
gauge gravity duality and yields 
the  result\cite{Policastro:2001yc,Kovtun:2004de}
\st
\label{etabys4pi}
     \frac{\eta}{s} = \frac{1}{4\pi} \np
\stp
From the perspective of heavy ion physics  this 
result was important because it showed that there exist 
field theories where $\eta/s$ can be this low. 
Although $\N=4$ has no particle interpretation, 
we note that extrapolating \Eq{shearvisc0} by setting $\ell_{\rm mfp} =1/n\sigma_o=1/\pi T$ 
yields a value for $\eta/s$ which is approximately equal to the SYM result.
In \Fig{etabys} we have displayed this numerology with $\ell_{\rm mfp} = 1/T$
for clarity.

There are many aspects of transport coefficients 
which have not been reviewed here. 
For instance,  there is an ongoing effort to determine the transport coefficients
of QCD from the lattice\cite{Meyer:2008sn,Aarts:2007wj}. While a  precise determination of the 
transport coefficients is very difficult\cite{Aarts:2002cc,PTcorr,Teaney_corr}, the lattice may 
be able to determine enough about the spectral densities to 
distinguish the orthogonal pictures represented by $\N=4$ SYM theory
and  kinetic theory\cite{Meyer:2008sn}.  This is clearly an important goal 
and we refer to \Ref{Schaefer:2009dj} for  theoretical background.
Also throughout this review we have emphasized
the shear viscosity and neglected the
bulk viscosity. This is because on the hadronic side of the
phase transition  the bulk viscosity is a thousand times
smaller than the shear viscosity in the regime where
it can be reliably calculated \cite{Prakash:1993bt}.  
Similarly on the high temperature
QGP side of the phase transition the bulk viscosity is also
a thousand times smaller than shear\cite{Arnold:2006fz}. 
However, near a second order phase transition the bulk viscosity
can become very large\cite{Kharzeev:2007wb,Moore:2008ws,Huebner:2008as}.
Nevertheless the rapid cross-over seen in \Fig{eos} is not 
particularly close to a second order phase transition
and universality arguments can be questioned 
 (see \Ref{Bazavov:2009zn} for a discussion in the context of the chiral susceptibility.) 
Given the ambiguity at this moment it seems prudent to leave the bulk viscosity to future review.

\section{Hydrodynamic Description of Heavy Ion Collisions}
\label{hydro_basic}
In the previous sections we analyzed the phase diagram 
of QCD and estimated the transport coefficients in different
phases.  In this section
we will study the hydrodynamic modeling of heavy ion collisions.  

In \Sect{ideal_bj} we will consider ideal hydrodynamics  and assume that 
the mean free paths are small enough to support this interpretation. 
Subsequently we will study viscous hydrodynamics in \Sect{viscous_bj}.
\Sect{applicable} will analyze the ratio of the viscous terms to the ideal 
terms and use the estimates of the transport coefficients given 
above to assess the validity of the hydrodynamic interpretation. 
\Sect{second} will discuss the recent advances in interpreting 
the hydrodynamic equations beyond  the Navier Stokes limit. 
This work will lay the foundation for the more detailed 
hydrodynamic models presented  in \Sect{viscous_model}. 

\subsection{Ideal Hydrodynamics} 
\label{ideal_intro}

The stress tensor of an ideal fluid   and 
its equation of motion are simply 
\st
\label{Tideal}
  T^{\mu\nu}  = e u^{\mu} u^{\nu} + \pr \Delta^{\mu\nu} \nc \qquad  \partial_{\mu} T^{\mu\nu} = 0  \nc
\stp
where $e$ is the energy density, $\pr(e)$ is the pressure, and $u^{\mu} =(\gamma, \gamma{\bf v} )$ is the four velocity.
Here we will use the metric $(-, +, + ,+)$  and define  the projection tensor,
$\Delta^{\mu\nu} = g^{\mu\nu} + u^{\mu} u^{\nu}$, with $u^{\mu}u_{\mu} = -1$ and $\Delta^{\mu\nu} u_{\mu} = 0$. 
This decomposition of the stress tensor is simply a reflection of the fact 
that in  the local rest frame of a thermalized medium the stress tensor must 
have the form,  ${\rm diag}(e, \pr, \pr, \pr)$. In developing viscous hydrodynamics
we will define two  derivatives which are the time derivative $D$,  
and the spatial derivatives $\nabla^{\mu}$ in the local rest frame
\st
D\equiv u^{\mu}\partial_\mu \nc \qquad \nabla^{\mu}  \equiv \Delta^{\mu\nu} \partial_{\mu} \np
\stp
Using $\partial_\mu = -u_\mu D + \nabla_{\mu}$ and  $u_{\mu} Du^{\mu} =0$,  the ideal equations of motion can be written  
\begin{align}
\label{idealeom}
    De  &= - (e + \pr) \nabla_\mu u^{\mu} \nc \\
    Du^{\mu} &= - \frac{ \nabla^\mu \pr }{e + \pr }   \np
\end{align}
 The first equation says that the change in energy density is due to
the $\pr dV$ work or equivalently  that entropy is conserved.  To see this
we associate $\nabla_\mu u^\mu$ with the fractional change in volume per unit time in the  co-moving frame,  $dV/V = dt \times \nabla_{\mu} u^{\mu}$,  and use the thermodynamic identity, $d(eV)  = Td(sV) - \pr dV$.
The second equation says that the acceleration is due to
the gradients of pressure. The enthalpy  plays the role of the mass density in
a relativistic theory. 

\subsection{Ideal Bjorken Evolutions and Three Dimensional Estimates}
\label{ideal_bj}

In this section we will follow an analysis due to Bjorken\cite{Bjorken:1982qr}
and apply  ideal hydrodynamics  
to heavy ion collisions. 
Bjorken's analysis  was  subsequently
extended in important ways\cite{Baym:1984np,Danielewicz:1984ww,Gyulassy:1997ib}. In a high energy heavy ion
collisions the two nuclei pass through each other and the partons are scarcely
stopped. This statement underlies much of the interpretation of high energy
events and an enormous amount of data is consistent with this assumption.  For
a time which is short compared to the transverse size of the nucleus, the
transverse expansion can be ignored.  

\begin{figure}
\begin{center}
\includegraphics[height=2.5in]{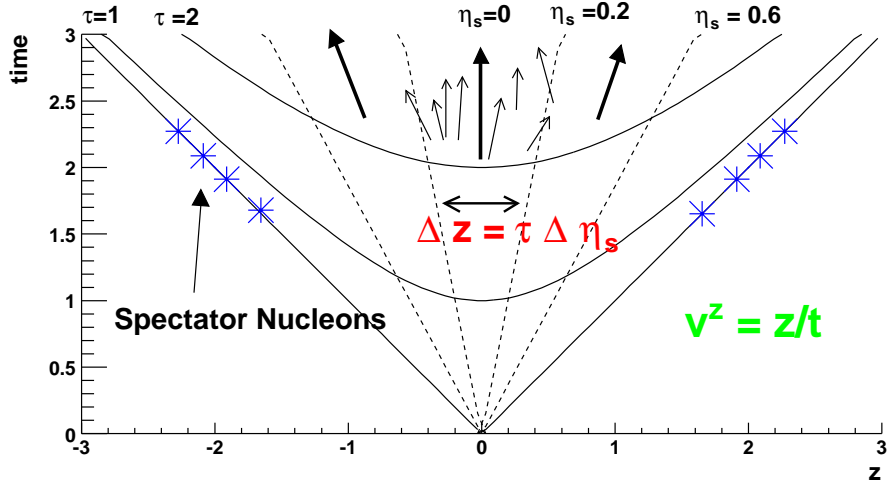}
\caption{A figure motivating for the Bjorken model. 
The space between
the dashed lines of constant $\eta_s$ 
are referred to as a space-time rapidity slice in the
text. Lines of constant proper time $\tau$ are given by the solid 
hyperbolas. The collection of particles in the $\eta_s=0$ 
rapidity slice is indicated by the small arrows for the central ($\eta_s=0$) rapidity 
slice only. The solid arrows indicates the 
average four velocity $u^{\mu}$ in each slice. 
The spectators are those nucleons which do not participate in the collision 
and lie along the light cone.
\label{bj}}  
\end{center}
\end{figure}

Given that the nuclear constituents pass through each other, 
the longitudinal momentum is much much  larger than
the transverse momentum. 
Because of this scale separation
there is a strong identification between the space-time coordinates 
and the typical $z$ momentum. 
For example a particle 
with typical momentum $p_z$ and energy $E$ will be found in 
a definite region of space time
\st
    v^z = \frac{p^{z}}{E}  \simeq \frac{z}{ t} \np
\stp
This kinematics is best analyzed with 
proper time and space-time rapidity variables\footnote{
Here $\eta_s$ denotes the space time rapidity, $\eta_\pseudo$ denotes the 
pseudo-rapidity (see below), $\eta$ denotes the shear viscosity. In raised 
space time indices in $\tau,\eta_s$ coordinates 
we will omit the ``s" when confusion can not arise, {\it e.g.}  
$\pi^{\eta\eta} = \pi^{\eta_s\eta_s}$.
 }, $\tau$ and  $\eta_s$
\[
 \tau \equiv \sqrt{t^2 - z^2 } \nc  \qquad \eta_s \equiv \frac{1}{2} \log\left(\frac{t + z}{t -z} \right) \np
\]
At a proper time $\tau$ particles with rapidity
$y$ are predominantly located at space time 
rapidity $\eta_s$
\st
\label{bjform}
 y \equiv \frac{1}{2} \log \frac{p_z + E}{E - p_z} \simeq   \frac{1}{2} \log \frac{ t + z}{t- z} \equiv \eta_s  \np
\stp
\Fig{bj} illustrates these coordinates and shows schematically 
the identification between  $\eta_s$ and $y$. 
At an initial proper time $\tau_o$,
there is a collection 
of particles predominantly moving  with four velocity  $u^{\mu}$ 
 in each space-time rapidity slice 
\st
\label{bjfluid}
       \frac{1}{2} \log \left( \frac{u^0 + u^z}{u^0 - u^z} \right) \simeq \eta_s \np
\stp

\begin{figure}
\begin{center}
\includegraphics[height=2.5in]{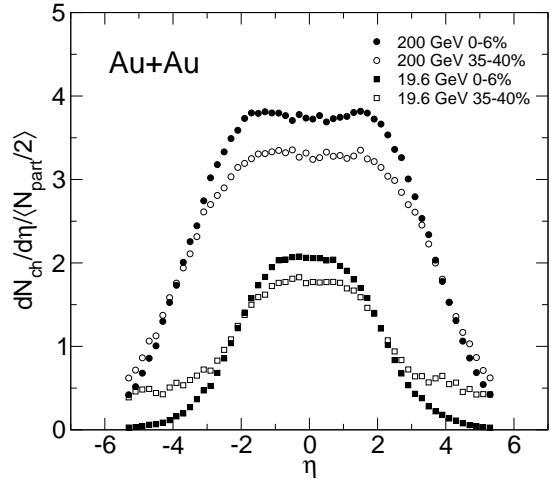}
\caption{The measured charged particle pseudo-rapidity distribution $dN_{\rm ch}/d\eta_{\pseudo}$ for different beam
energies divided by the number of participant pairs, $N_p/2$. 
$N_p/2 \simeq 170$ for a central (0-6\%) AuAu collision. 
This review focuses on $\sqrt{s} =200 \,{\rm GeV}/{\rm nucleon}$.  
\label{pseudo_rapid}}
\end{center}
\end{figure}

The beam rapidity at RHIC is $y_{\rm beam} \simeq 5.3$  and 
therefore roughly speaking the particles are produced in the 
space-time rapidity range $-5.3 < \eta_s < 5.3$. 
It is important to realize  that (up to about a unit or so) each  
space-time rapidity slice is associated with a definite
angle in the detector.   For ultra-relativistic particles $E\simeq p$ 
we have 
\st
   \eta_s \simeq y \simeq \frac{1}{2} \log \left( \frac{p + p_z }{p-p_z} \right)  = \frac{1}{2} \log\left( \frac{1 + \cos\theta}{1-\cos\theta } \right) \equiv \eta_{\pseudo} \nc
\stp 
where a particular $\theta$ is shown in \Fig{geometrylong}. 
The measured pseudo-rapidity distribution of charged particles is shown 
in \Fig{pseudo_rapid}.
We can estimate the energy in a unit of pseudo-rapidity  by taking $\llangle E \rrangle \simeq 0.5\,{\rm GeV}$ as the energy per particle.  Then
the  energy in a pseudo-rapidity unit is 
\[
\frac{\dd E}{\dd \eta_\pseudo} \simeq \llangle E \rrangle \frac{\dd N_{\rm ch}}{\dd \eta_\pseudo} \times 1.5   \simeq  3.0 \, {\rm GeV}\times  (N_p/2) \nc
\]
where $(N_p/2) \simeq 170$ is the number of participant pairs
in a central event.
The factor of $1.5$ has been inserted to account for 
the fact that there are approximately equal numbers of $\pi^{+},\pi^{-}$ 
and $\pi^{0}$ (the most abundant particle) 
but only $\pi^{+}$ and $\pi^{-}$ are counted  in
$dN_{\rm ch}/d\eta_\pseudo$. 
This  estimate  agrees reasonably with the measured 
$dE_T/d\eta_\pseudo \simeq 3.2\, {\rm GeV}\times N_p/2$ from \Ref{Milov:2004sv}.

Bjorken used these kinematic ideas 
to estimate the initial energy density  in the 
$\eta_s=0$ rapidity slice at an initial time, $\tau_o \simeq 1\,{\rm fm}$.
The estimate is  based
on the fairly well supported assumption that the energy which 
finally flows into the  detector $\dd E_T/d\eta_\pseudo$ 
largely reflects the initial energy in a given space-time rapidity
slice
\bg
  \epsilon_{Bj} &\simeq& \frac{1}{A} \frac{\Delta E}{\Delta z} \simeq  
 \frac{1}{A\tau_o} \frac{\Delta E}{\Delta \eta_s}  \simeq 
\frac{1}{A \tau_o} \frac{\dd E_T}{\dd\eta_\pseudo} \nc  \\
               &\simeq& 5.5 \frac{\rm GeV}{{\rm fm}^3}  \np
\nd
In the last line we have estimated the area of a gold nucleus as
$A\simeq 100\,{\rm fm^2}$,  taken $\tau_o \simeq 1\,{\rm fm}$, 
and used the measured  $dE_T/d\eta_\pseudo$.
This estimate is generally considered
a lower limit since during the expansion there is ${\mathcal P} dV$  work
as the particles in one rapidity slice push against the particles
in another rapidity slice\cite{Danielewicz:1984ww,Baym:1984np,Gyulassy:1997ib} 
(See \Fig{bj}). 
Using the equation of state in \Fig{eos} 
we  estimate an  initial temperature, 
 $T(\tau_o) \simeq 250\,{\rm MeV}$.  As mentioned
above this estimate 
is somewhat low for hydrodynamic calculations 
and a more typical temperature is  $T\simeq 310\,{\rm MeV}$, which has
roughly twice the Bjorken density\cite{Kolb:2003dz}.

As seen in \Fig{pseudo_rapid}, the distribution of the energy  density $e(\tau_o, \eta_s)$ 
in space-time rapidity is not uniform.
In the Color Glass Condensate (CGC) picture for instance,  
the final distribution of multiplicity is related to the $x$ distribution of partons inside the nucleus\cite{Kharzeev:2007zt}.  
Bjorken made the additional simplifying assumption  that the 
energy density is uniform in space-time rapidity, {\it i.e.} $e(\tau_o,\eta_s) \simeq e(\tau_o)$. 
With this simplification, the
identification between the fluid  and space time rapidities remains fixed 
as the fluid flows into the forward light cone.  

We have discussed the motivation for the Bjorken model. 
Formally the model consists of the following   
ansatz for the hydrodynamic variables 
\st
\label{bjansatz}
 e(t,\x) = e(\tau)\nc   \qquad u^{\mu}(t,\x) = (u^0,u^{x},u^{y},u^{z} ) = 
(\cosh(\eta_s), 0, 0, \sinh(\eta_s)) \np
\stp
The model is invariant under boosts in the $z$  direction.
Thus given a physical quantity
at mid-rapidity ($\eta_s=0$), one can determine this quantity
at all other rapidities by a longitudinal boost.
We will use curvilinear coordinates where\cite{KSH} 
\st
x^{\mu}=(\tau,\x_\perp,\eta_s) \nc \qquad  g_{\mu\nu} = {\rm diag}(-1,1,1, \tau^2) \nc  \qquad  (u^\tau,u^x,u^y, u^\eta) = (1, 0,0,0) \np
\stp
In this coordinate system boost invariance implies  that everything
is independent of  $\eta_s$. To interpret a tensorial component
in these coordinates, we multiply by $\sqrt{g_{\eta\eta}}=\tau$ for every raised 
$\eta_s$ index, and 
subsequently associate the product with the 
corresponding cartesian component at mid-rapidity.
For example,  $\tau^2 T^{\eta\eta} = \left.  T^{zz} \right|_{\eta_s=0} = \pr $.
Similarly, $\tau u^{\eta} = \left. u^{z} \right|_{\eta_s=0} =0$ 
for boost invariant flow.

Substituting the boost invariant ansatz (\Eq{bjansatz}) 
into the conservation laws yields the 
following equation for the energy density\footnote{A quick way to 
derive this is to work in a neighborhood of $z=\eta_s=0$ where 
$u^{z} \simeq z/t$. Substituting this approximate form into $\partial_\mu T^{\mu\nu} = 0$ in cartesian coordinates, quickly yields \Eq{idealbjeom} with 
the replacement $t\rightarrow \tau$. }
\st
\label{idealbjeom}
  \frac{\dd e}{\dd\tau} = - \frac{e + \pr }{\tau} \np
\stp
Multiplying this equation by the volume of a space time rapidity slice $V=\tau \Delta \eta_s A$ (see \Fig{bj}) we find 
\st
\label{pdv}
   \dd (e \,\tau \Delta \eta_s A) = - \pr\, \dd (\tau \Delta \eta_s A)  \nc
\stp
and we can interpret this result\cite{Danielewicz:1984ww}
as saying that the energy
energy per unit space-time rapidity ($e \, \tau\Delta\eta_s A$) decreases due to the $\pr dV$ work. It is for this reason that the Bjorken estimate $e_{Bj}$ (which assumes that the r.h.s. of \Eq{pdv} equals zero) should 
be considered a lower bound. In general, assuming Bjorken scaling (\Eq{bjansatz}) and the 
conservation laws,  but not assuming local thermal equilibrium, one finds 
\st
    \frac{\dd e}{\dd\tau} = - \frac{e + T^{zz}}{\tau} \nc
\stp
where $T^{zz}\equiv \tau^2 T^{\eta\eta}$  is the effective longitudinal pressure. Viscous corrections will modify $T^{zz}$ from its equilibrium value of $\pr$.

Returning to the equilibrium case,  \Eq{idealbjeom} can be solved for a massless ideal gas equation of state  ($e=3\pr\propto T^4$)
and the time dependence of the temperature is 
\st
   T(\tau) = T_o \left(\frac{\tau_o}{\tau} \right)^{1/3} \nc
\stp
where $T_o$ is the initial temperature.  The temperature decreases rather 
slowly as a function of proper time during the initial one dimensional expansion. This will turn out to be important when discussing equilibration. 
For a massless ideal gas, the entropy is  $s=(e+\pr)/T\propto T^3$ and decreases as
\st
   s(\tau) = s_o\frac{\tau_o}{\tau} \np
\stp

Now we discuss what happens when the initial energy density distribution is not uniform in rapidity.  Due to  pressure gradients in the longitudinal direction, there is some longitudinal acceleration.
This changes the strict identification between the space time
rapidity and the  fluid rapidity given in \Eq{bjfluid}.  It also changes the
temperature dependence given above. One way to quantify this effect is to look at the
results of 3D ideal hydrodynamic calculations and study the differences between
the initial energy distribution in space-time rapidity $\int \dd^2\x_\perp \, e(\tau_o,\x_\perp, \eta_s)$
and the final energy distribution, $\int \dd^2\x_\perp \, e(\tau_f,
\x_\perp, \eta_s)$. Generally, the final distribution in space-time rapidity
is similar to the initial distribution in space-time rapidity \cite{Hirano:2004en,HiranoPrivate}.  Therefore, the effect of
longitudinal acceleration is unimportant  until late times.

 The nuclei have a finite 
transverse size,  $R_{\rm Au} \sim 6\,{\rm fm}$. 
After a time of order
\[
  \tau \sim \frac{R_{\rm Au}}{c}  \nc
\]
the expansion becomes three dimensional. 
To estimate how the temperature evolves during the course of the
resulting 
3D expansion,  
consider a sphere of radius $R$ which expands in all
three directions. The radius and volume increase as
\[
 R \propto \tau\nc  \qquad V \propto \tau^3. 
\]
Since for an ideal expansion the  total entropy in the
sphere is constant, the entropy density decreases as $1/\tau^3$ and the
temperature decreases as 
\st
\label{t3d}
s\propto \frac{1}{\tau^3}\nc \qquad   T  \propto  \frac{1}{\tau} \np
\stp

Here we have estimated how the entropy decreases during a one and three 
dimensional expansion of an ideal massless gas. Now if during 
the course of the collision there are non-equilibrium processes which
generate  entropy that ultimately equilibrates,
the temperature of this final equilibrated gas will be larger than if 
the expansion was isentropic. Effectively the temperature will decrease
more slowly. To estimate this effect in a one dimensional expansion, 
we imagine a free streaming gas where the longitudinal pressure is zero. Then
from \Eq{idealbjeom}  we have
\st
  \frac{de}{d\tau} \sim \frac{e}{\tau} \np
\stp
In the sense discussed above, this equation 
may be integrated to estimate that the temperature and entropy
decrease as 
\st
   T \propto \frac{1}{\tau^{1/4} } \nc \qquad
   s \propto \frac{1}{\tau^{3/4} } \np
\stp
Similarly in  a three dimensional expansion we can estimate how entropy 
production will change the powers  given in \Eq{t3d}.  
Again consider  a sphere of radius $R$ which expands in all
three directions, such that $R \propto \tau$ and $V
\propto \tau^3$. For a free expansion without pressure  the  total 
energy in the
sphere is constant, and the energy density decreases as $1/\tau^3$. 
Similarly, we estimate
that the temperature and entropy density decrease as
\st
   T \propto \frac{1}{\tau^{3/4} } \nc
   \qquad s \propto \frac{1}{\tau^{9/4} } \np
\stp
\begin{figure}[t]
\begin{center}
\includegraphics[height=2.5in]{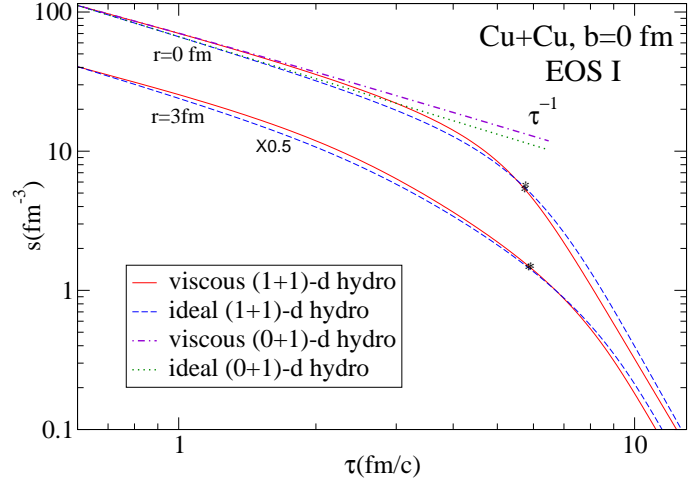}
\end{center}
\caption{Figure from \protect \Ref{Song:2007ux}  showing the entropy density ($s$)  in CuCu simulations as a function of proper time $\tau$ 
using ideal and viscous hydrodynamics.  The top set of lines 
shows the entropy in the center  of the nucleus-nucleus collision, $(r=0\,{\rm fm})$,
and the bottom set of lines shows the analogous curves closer to the edge $(r=3\,{\rm fm})$. 
During an initial one dimensional expansion the entropy density decreases as  $s \propto 1/\tau$.  
Subsequently the entropy  decreases as $s\propto 1/\tau^3$ when the expansion becomes 
three dimensional at a time, $\tau \sim 5\,{\rm fm}$.  The lines
labeled by $(0+1)$ ideal and $(0+1)$ viscous are representative of 
the ideal and viscous Bjorken results \Eq{idealbjeom} and \Eq{visbj} respectively. 
\label{song_1d3d} }
\end{figure}

In summary  we have estimated how the temperature and entropy density depend on the proper time $\tau$ during the course of an ideal and non-ideal $1D$ and $3D$ expansion. This information is recorded in \Table{temptab}.  
These estimates are also nicely realized in actual hydrodynamic simulations.
\Fig{song_1d3d} shows the dependence of entropy density as a function 
of proper  time $\tau$. The figure indicates that the entropy decreases as 
$1/\tau$  during an initial  one dimensional expansion and  
subsequently decreases as $1/\tau^3$  when the expansion becomes
three dimensional at a time of order $\sim 5\,{\rm fm}$. 
These basic
rules will be useful when estimating the relative size of viscous terms in 
what follows. 
\begin{table}
\begin{center}
\begin{tabular}{c | c | c}
Quantity &  1D Expansion & 3D Expansion \\ \hline
  $T $     &  $\left(\frac{1}{\tau}\right)^{1/3 \div 1/4}$ & $\left(\frac{1}{\tau}\right)^{1 \div 3/4} $ \\[5pt] \hline
  $s\,\propto\, T^3 $     &  $\left(\frac{1}{\tau} \right)^{1 \div 3/4}$ & $\left(\frac{1}{\tau} \right)^{3 \div 9/4} $ \\[5pt] 
\end{tabular}
\end{center}
\caption{ Dependence of temperature and entropy as a function of time in a 1D and 3D expansion.  The indicated range,  for instance $1/3\div1/4$,  is an estimate of 
how extreme non-equilibrium effects could modify the ideal power from $1/3$ 
to $1/4$.
\label{temptab} }
\end{table}

\subsection{Viscous Bjorken Evolution and Three Dimensional Estimates} 
\label{viscous_bj}

This section will analyze viscosity in the context of the Bjorken 
model with the primary goal of assessing the validity of hydrodynamics 
in heavy ion collisions. 
In viscous hydrodynamics the stress tensor
is expanded in  all possible gradients.
Using lower order equations of motion any time derivatives of conserved quantities can be rewritten as spatial derivatives.
First the  stress tensor is
decomposed into ideal and viscous pieces
\st
   T^{\mu\nu} = T^{\mu\nu}_{\rm ideal} + \pi^{\mu\nu} + \Pi\Delta^{\mu\nu} \nc
\stp
where $T^{\mu\nu}_{\rm id}$ is the ideal stress tensor (\Eq{Tideal}) and $\Pi$ is the bulk stress.
$\pi^{\mu\nu}$ is the symmetric traceless shear tensor and satisfies the 
orthogonality constraint, $\pi^{\mu\nu} u_\nu = 0$. The equations of motion 
are the conservation laws $\partial_\mu T^{\mu\nu}=0$  together 
with a constituent relation.
The constituent relation  expands $\pi^{\mu\nu}$ and $\Pi$ in 
terms gradients of the conserved charges $T^{00}$ and $T^{0i}$ 
or their thermodynamic conjugates, temperature $T$ and four velocity $u^{\mu}$ .
To first order in this expansion, the equations of motion 
are
\st
\label{stress}
\partial_\mu T^{\mu\nu} = 0 \nc \qquad  \pi^{\mu\nu} = - \eta  \sigma^{\mu\nu}\nc \qquad \Pi = - \zeta  \nabla_\mu u^\mu   \nc
\stp
where $\eta$ and $\zeta$ are the shear and bulk viscosities respectively, and
we have defined the symmetric traceless combination
\st
  \sigma^{\mu\nu} = \nabla^\mu u^\nu + \nabla^\nu u^\mu - \frac{2}{3} \Delta^{\mu\nu} \nabla_\lambda u^\lambda   \np
\stp
For later use we also define the bracket $\llangle\ldots \rrangle$ operation
\st
 \llangle A^{\mu\nu} \rrangle \equiv 
\frac{1}{2} \Delta^{\mu\alpha} \Delta^{\nu\beta} \left(A_{\alpha\beta} + A_{\beta \alpha} \right) - \frac{1}{3} \Delta^{\mu\nu} \Delta^{\alpha\beta}A_{\alpha\beta}  \nc
\stp
which takes a tensor and renders it symmetric, traceless and orthogonal to $u^{\mu}$.
Note that $\sigma^{\mu\nu} = 2\llangle\partial^{\mu} u^{\nu}\rrangle$.

We  now extend the Bjorken model to the viscous case following \Ref{Danielewicz:1984ww}.
The bulk viscosity is neglected in the following analysis 
and we refer to \Sect{transport} for a more complete discussion. 
Substituting the Bjorken ansatz (\Eq{bjansatz})
into the conservation laws  and the associated constituent relation (\Eq{stress}) yields the time evolution of the energy density
\st
\label{visbj}
  \frac{de}{d\tau} = - \frac{e + \pr -  \frac{4}{3}\eta/\tau}{\tau} \np
\stp
The system is expanding in the $z$ direction  and consequently
the pressure in the $z$ direction is reduced from 
its ideal value.  Formally this arises due to the gradient
$\partial_z u^z = 1/\tau$ and the  constituent relation  \Eq{stress} 
\st
  T^{zz} = \pr - \frac{4}{3} \frac{\eta}{\tau}  \np
\stp
Thus during a viscous Bjorken expansion the system will do less longitudinal 
work than in the ideal case. 

\subsection{The Applicability of Hydrodynamics and $\eta/s$}
\label{applicable}

Comparing the viscous equation of motion \Eq{visbj} to the 
ideal equation of motion \Eq{idealbjeom},
we see that the hydrodynamic expansion is controlled by
\st
 \frac{\eta}{e + \pr} \frac{1}{\tau} \ll 1  \np
\stp 
This is a very general result and is a function of time and temperature.  Using the thermodynamic relation $e + \pr = s T$, we  divide this  condition into 
a constraint on a medium parameter $\eta/s$ and a constraint on an experimental parameter $1/\tau T$
\st
\underbrace{\frac{\eta}{s}}_{\mbox{medium parameter} } \times \underbrace{\frac{1}{\tau T}}_{\mbox{experimental parameter}}  \ll 1 \np
\stp
If the experimental conditions are favorable enough, 
it is appropriate to apply hydrodynamics regardless of the value of $\eta/s$. 
This is the case for sound waves in air where although $\eta/s$ is  significantly larger
than the quantum  bound, 
hydrodynamics  remains a good effective theory.
However, for the application to heavy ion collisions, the experimental conditions are so unfavorable that only if $\eta/s$ is close to the quantum bound  will 
hydrodynamics be an appropriate description.  

For instance,
we estimated the experimental condition  in \Sect{ideal_intro}
\st
     \frac{1}{\tau_o T_o} =  0.66  \left(\frac{1\, {\rm fm} }{\tau_o} \right) \left( \frac{300\,\mbox{MeV}}{T_o} \right) \np
\stp 
Here we have evaluated this experimental parameter at a specific initial time $\tau_o$ and will return to the time evolution of these
estimates in the next section. 
In \Sect{transport} we estimated the medium parameter $\eta/s$  and 
can now place these results in context
\st
    0.2 \left( \frac{\eta/s}{0.3} \right) 
\left(\frac{1\, {\rm fm} }{\tau_o} \right) \left( \frac{300\,\mbox{MeV}}{T_o} \right)  \ll 1 \np
\stp
From this condition we see that hydrodynamics will begin to be a good approximation  for $\eta/s \lsim 0.3$ or so. This estimate is borne out by the more detailed calculations presented 
in \Sect{viscous_model}. Reexamining \Fig{etabys}, 
we see that the value of $\eta/s \simeq 0.3$ is at the low 
end of the perturbative QGP estimates given in the figure and it is
difficult to reconcile the observation of strong collective 
flow with a quasi-particle picture of quarks and gluons.
Thus the estimates of $\eta/s$ coming 
from the RHIC experiments, which  are based on the hydrodynamic interpretation
of the observed flow, should be accepted only with considerable care.

\subsection{Time Evolution}
\label{time}

In the previous section we have estimated the applicability  of hydrodynamics at a
time $\tau_o \approx 1\,{\rm fm}$. In this section we will estimate 
how the size of the viscous terms depends on time. For this
purpose we will keep in mind a kinetic theory estimate for the shear viscosity
\st
 \eta \sim \frac{T}{\sigma} \nc
\stp
and estimate how the gradient expansion parameter in \Eq{visbj} depends on
time. We will contrast a conformal gas 
with $\sigma\propto 1/T^2$ ({\it e.g.} perturbation theory or $\N=4$ SYM)
to a gas with fixed
cross section, $\sigma=\sigma_o$.
 There are clearly important scales in the quark gluon plasma as the medium
approaches the transition point.  For instance spectral densities of
current-current correlators near the transition point show a very discernible
correlation where the $\rho$ meson will form in the hadron phase
\cite{Karsch:2001uw,Asakawa:2000tr,Aarts:2007wj}.  Thus the intent of studying
this extreme limit with a constant cross section is to show 
some of the possible effects of these scales. Further the constant cross
section kinetic theory has been used to analyze the centrality dependence of elliptic flow\cite{Drescher:2007cd}.

First consider a theory where the temperature $T$ is the only scale  and 
also
consider a 1D Bjorken expansion.
The shear viscosity is proportional to $T^3$ and the enthalpy  scales as
$T^4$, so the hydrodynamic expansion parameter scales as
\st
\frac{\eta}{(e + \pr)}\frac{1}{\tau} \sim  \frac{1}{\tau T} \sim \frac{1}{\tau^{2/3} } \np
\stp
In the last step we have used the fact that for a scale invariant gas undergoing an ideal Bjorken expansion the temperature decreases as $1/\tau^{1/3}$.  
In general if we have some non-equilibrium processes which produce entropy 
during the course of the expansion, the temperature will decrease more 
slowly than estimated in the ideal gas case  --
see \Table{temptab}.
 The result is that we do not expect the temperature
to decrease more slowly than  $1/\tau^{1/4} $, and  we can estimate that  the hydrodynamic expansion parameter evolves as 
\st
\frac{\eta}{(e + \pr)}\frac{1}{\tau} \propto  \frac{1}{\tau T} \propto \frac{1}{\tau^{2/3 \div 3/4} } \np
\stp
Thus during a 1D expansion of a conformal gas the system will move closer to
equilibrium.

Compare this scale invariant theory  to a gas with a very definite cross
section $\sigma_o$.
For a constant $\sigma_o$ the hydrodynamic
expansion parameter evolves as 
\st
\frac{\eta}{(e + \pr)}\frac{1}{\tau} \propto  \frac{1}{s\sigma \tau}  \propto  \left(\frac{1}{\tau} \right)^{0 \div 1/4}  \np
\stp
Thus,  with a  constant cross section, the gas  
will move  neither away nor toward equilibrium as 
a function of time. Non-equilibrium physics will make the matter evolve slowly 
toward equilibrium. 

Now we will compute the analogous effects for a three dimensional expansion. 
In the conformal case $\eta\propto T^3$ and $T \propto \frac{1}{\tau}$, so that
the final result is 
\st
\frac{\eta}{(e + \pr)}\frac{1}{\tau} \propto  \frac{1}{\tau T} \propto 
\left(\frac{1}{\tau} \right)^{0 \div 1/4}  \np
\stp
Thus a conformal gas expanding isentropically in three dimensions also moves neither
away nor towards equilibrium, though entropy production will cause it to 
slowly equilibrate.
Similarly for gas with a constant cross section the hydrodynamic 
parameter evolves as 
\st
\frac{\eta}{(e + \pr)}\frac{1}{\tau} \propto  \frac{1}{s\sigma \tau}  \propto  
\tau^{2 \div 5/4} \np
\stp
In estimating  this last line we have used \Table{temptab}. Thus 
we see that a gas with fixed cross-sections which expands
in three dimensions very rapidly breaks up.

The preceding results are summarized in \Table{Vistab}. 
Essentially the heavy ion collision proceeds along the following line of
reasoning. First, there is a one dimensional expansion where the temperature
is the  dominant scale in the problem. 
The parameter which controls the applicability of hydrodynamics
$\eta/[(e + \pr)\tau]$ decreases as a function of time; hydrodynamics gets
better and better, evolving according to the upper left corner of \Table{Vistab}.  As 
the system expands and cools toward the transition region additional scales 
enter the problem. Typically at this point $\tau \sim 4\,{\rm fm/c}$  the
expansion also becomes three dimensional.   The system then enters the 
lower right corner of  \Table{Vistab}  and very quickly the nucleus-nucleus collision starts to break up. We note that it is necessary to introduce some scale 
into the problem in order to see this freezeout process. For a conformal liquid 
with $\eta \propto T^3$ the system never freezes out even for a 3D expansion. 
This can be seen by looking at the upper-right corner of the table and 
noting that the hydrodynamic expansion parameter behaves as
\st
   \frac{\eta}{(e + \pr) \tau } \propto \left(\frac{1}{\tau} \right)^{0\div 1/4} \nc
\stp
and therefore approaches a constant  (or slowly equilibrates) at late times.
From this discussion we see that the temperature dependence of the 
shear viscosity is ultimately responsible for setting the  duration
of the hydrodynamic expansion.
\begin{table}
\begin{center}
\begin{tabular}{c|c|c}
Model &  1D Expansion & 3D  Expansion  \\  \hline
$\eta \propto T^3 $ & 
$ \frac{\eta}{(e + \pr)\tau} \propto \left(\frac{1}{\tau}\right)^{2/3\div 3/4}$ & 
$ \frac{\eta}{(e + \pr)\tau} \propto \left(\frac{1}{\tau} \right)^{0 \div 1/4}$   \\[5pt]  \hline
$\eta \propto \frac{T}{\sigma_o}$  
& $\frac{\eta}{(e + \pr)\tau } \propto \left(\frac{1}{\tau}\right)^{0 \div 1/4} $ & $\frac{\eta}{(e + \pr)\tau } \propto  \tau^{2 \div 5/4} $  \\[5pt]
\end{tabular}
\caption{
Dependence of the hydrodynamic expansion parameter $\eta/[(e + \pr)\tau]$ as
a function of time for two different functional forms for $\eta$ ($\eta \propto T$ and $\eta
\propto T^3$) and two expansion types (1D and 3D). A range of powers is given;  the
first power corresponds to ideal hydrodynamics and the second power 
corresponds to an estimate of non-equilibrium evolution.
\label{Vistab} }
\end{center}
\end{table}

\subsection{Second  Order Hydrodynamics}
\label{second}

In the previous sections we developed  the first order theory of
relativistic viscous hydrodynamics.
In the first-order theory there
are reported instabilities which are associated with the gradient
expansion\cite{Lindblom:1985}. 
Specifically, in the first-order theory the stress tensor is instantly specified 
by the constituent relation  and this
leads to acausal propagation\cite{Hiscock:1983zz} and ultimately 
the instability. Nevertheless, it was generally understood that
one could write down any  relaxation model  which conserved energy and momentum  and which  included some notion of entropy, and the results of such a model would be indistinguishable from the
Navier Stokes equations\cite{KadanoffMartin,Forster,Lindblom}. 
Many hydrodynamic models were written
down\cite{Israel:1979wp,LG,PJC,Ottinger,LG2} starting with 
a phenomenological model by Israel and Stewart\cite{IS,Israel:1979wp} and M\"uller\cite{Muller}. 
For example in the  authors own work the strategy was to write down a fluid model
(based on \Ref{Ottinger})
which relaxed  on some time scale to the Navier Stokes equations, solve these
model equations on the computer, and finally to verify that the results are
independent of the details of the model\cite{Dusling:2007gi}. Thus the goal was to solve the Navier Stokes equations and to estimate the effects of higher order terms. 

Recently an important work by R.~Baier,  P.~Romatschke,  D.~T.~Son,  A.~O.~Starinets and M.~A.~Stephanov (hereafter BRSSS) 
clarified and classified the nature of these higher
order terms\cite{BRSSS}.
An important impetus for this work came from the AdS/CFT correspondence\cite{BRSSS,Bhattacharyya:2008jc,Natsuume:2007ty}. 
Many of the fluid models discussed above were motivated by kinetic theory.
However, in the strongly coupled $\N=4$  plasma, 
kinetic theory is not applicable, and the 
precise meaning of these models was vague.
BRSSS determined  precisely in what sense these second order
viscous equations are theories and in what sense they are models. 
Simultaneously the  Tatta group  
completed the calculation of the second order 
transport coefficients in $\N=4$ SYM theory and clarified the 
hydrodynamic nature of black branes in the process\cite{Bhattacharyya:2008jc}.

The spirit of
the BRSSS analysis is the following: 
\begin{enumerate}
\item Write the stress tensor  as an expansion in all
possible second order gradients of conserved charges and external fields
which  are allowed by the symmetries. The transport coefficients 
are the coefficients of this gradient expansion.
\item In this expansion temporal derivatives can be rewritten as spatial derivatives using lower order equations of motion.
\item The conservation laws $\partial_\mu T^{\mu\nu} = 0$  and 
 the associated constituent relation dictates the dynamics of 
the conserved charges in the presence of the external field. By
adjusting the transport coefficients, this
dynamics will be able to reproduce  all  the retarded correlators 
of the  microscopic theory.
\end{enumerate}
In general, for a theory with  conserved baryon number there are many terms.
By focusing on a theory without baryon number and also assuming that the fluid 
is conformally invariant, the number of possible second order terms is 
relatively small. The  classification of gradients in terms of their 
conformal transformation properties 
was very useful, both theoretically and  phenomenologically.  At a theoretical
level there are a manageable number of terms to write down. At a phenomenological
level 
the gradient expansion converges more
rapidly  when only those second order terms which are allowed by conformal invariance are included (see \Sect{viscous_model}). Subsequently when additional conformal breaking terms 
are added, the conformal classification provides a useful
estimate for the size of these terms, {\it i.e.} quantities 
that scale as $T^{\mu}_{\mu} = e -3 p$ should be estimated differently
than those that scale as energy density itself. In retrospect,
this classification is an ``obvious" generalization of the first order
Navier-Stokes equations.

Proceeding more technically,  in analogy to the constituent relation 
of the Navier-Stokes theory \Eq{stress},
BRSSS determine that the possible forms of the gradient expansion  
in a conformal liquid are 
\bg
\label{constituent}
\pi^{\mu\nu} &=& -\eta \sigma^{\mu\nu} 
 + \eta \tau_\pi \left[ \llangle D\sigma^{\mu\nu}\rrangle + \frac{1}{d-1} \sigma^{\mu\nu} \partial \cdot u \right]  \nn
 & & +  \lambda_1 
\llangle \sigma^{\mu}_{\phantom{\mu} \lambda} \sigma^{\nu \lambda } \rrangle + 
\lambda_2  \llangle \sigma^\mu_{\phantom{\mu} \lambda} \Omega^{\nu \lambda} \rrangle + 
\lambda_3 \llangle \Omega^\mu_{\phantom{\mu} \lambda} \Omega^{\nu \lambda} \rrangle  \nc
\nd
where  the vorticity tensor is defined as
\st
 \Omega^{\mu\nu} = \frac{1}{2} \Delta^{\mu\alpha} \Delta^{\nu\beta} \left( \partial_\alpha u_\beta - \partial_\beta u_\alpha \right)  \nc
\stp
and $d=4$ is the number of space-time dimensions.
Conformal invariance  forces a particular combination of
second derivatives to have a single coefficient
\st
   \tau_\pi  \left[ \llangle D\sigma^{\mu\nu}\rrangle + 
\frac{1}{d-1} \sigma^{\mu\nu} \partial \cdot u \right]  \np
\stp
The time derivative  $D\sigma^{\mu\nu}$ may be expanded out using
lower order equations of motion  if desired.
The constituent relation (\Eq{constituent}) and the conservation laws
form 
the second order equations of motion of a conformal fluid. They are 
precisely analogous to the first order theory.  As in 
the first order case, these equations are also acausal.

To circumvent this issue,
BRSSS (following the spirit of  earlier work by Israel and Stewart \cite{IS,Israel:1979wp} and M\"uller \cite{Muller})
promote the constituent relation to a dynamical equation for the  viscous
components of the stress tensor $\pi^{\mu\nu}$. 
Using the
lower order relation $\pi^{\mu\nu} = -\eta \sigma^{\mu\nu}$,  
the (conformal) dependence of $\eta$ on temperature  $\eta \propto T^{d-1}$, 
and 
the ideal equation of motion \Eq{idealeom}, the following equation
arises for $\pi^{\mu\nu}$
\bg
\label{constituent2nd}
\pi^{\mu\nu} &=& -\eta\sigma^{\mu\nu} - \tau_\pi \left[ \llangle D\pi^{\mu\nu} \rrangle + \frac{d}{d-1} \pi^{\mu\nu} \nabla \cdot u \right] \nn
& &
+ \frac{\lambda_1}{\eta^2} \llangle \pi^{\mu}_{\phantom{\mu}\lambda} \pi^{\nu\lambda} \rrangle 
- \frac{\lambda_2}{\eta} \llangle \pi^{\mu}_{\phantom{\mu}\lambda} \Omega^{\nu \lambda}   \rrangle 
+ \lambda_3 \llangle \Omega^{\mu}_{\phantom{\mu} \lambda} \Omega^{\nu\lambda} \rrangle \np
\nd
From a numerical perspective the resulting equation of 
motion is now first order in time derivatives,
hyperbolic and causal.  The modes in this (and similar) models
have been studied in \Refs{Lindblom,Hiscock:1983zz,BRSSS} 

Nevertheless it should be emphasized  that the domain of validity of the resulting equations is still the same as \Eq{constituent}, {\it i.e.}   the hydrodynamic regime. 
Thus for instance
the second order equations should  be used  in a regime where 
\[
 \left|\pi^{\mu\nu} + \eta \sigma^{\mu\nu} \right| \ll  \left|\eta  \sigma^{\mu\nu} \right| \np
\]
Outside of this regime there is 
no guarantee that  entropy production 
predicted by this model will be positive during the course of the evolution \cite{BRSSS}.  
It should also be emphasized that this is 
not a unique way to construct a hydrodynamic model 
which reduces to \Eq{constituent} in the long wavelength limit  -- see 
\Ref{Schaefer:2009dj} for an example discussed in these terms.  What is guaranteed is 
that any conformal model or dynamics (such as conformal kinetic theory \cite{York:2008rr,Betz:2008me} or the dynamics predicted by AdS/CFT \cite{BRSSS,Bhattacharyya:2008jc}) 
will be expressible in the long wavelength limit in terms of the 
gradient expansion given above. 

There is an important distinction between the first and second order 
theories \cite{KadanoffMartin,Schaefer:2009dj,Forster}. In the first order theory, 
the ideal motion is damped, and there are corrections 
to the ideal motion of order the inverse Reynolds number
\st
    {\rm Re}^{-1} \equiv \frac{\eta}{(e+ \pr)  L^2} \Delta t  \sim \frac{\ell_{\rm mfp}}{L} \, \frac{ v_{\rm th} \Delta t}{L}\nc
\stp
where $L$ is the characteristic spatial dimension of the system,  
$\Delta t$ is the time of observation, and $v_{\rm th}$ is a typical
quasi-particle velocity. Thus for sufficiently
long times the viscous corrections become large and must be resummed
by solving the Navier-Stokes equations to capture the 
damping of the fluid motion.  Once this is done however, the 
remaining higher order terms  (which are captured by the second 
order theory) are {\it uniformly} small and modify the Navier-Stokes
solution by an amount  of order 
\[ 
\ell_{\rm mfp}^2/L^2  \np
\]
Often this makes these higher order terms difficult to measure
in normal laboratory liquids \cite{Schaefer:2009dj}.

For completeness we record the model equations which have been 
discussed in the heavy ion literature \cite{Romatschke:2007mq,Song:2008si,Song:2007ux,Huovinen:2008te}.  
\begin{enumerate}
\item 
The first of these is 
the simplified Israel-Stuart  equation, 
\st
  \pi^{\mu\nu} = - \eta\sigma^{\mu\nu}  - \tau_{\pi} \llangle D\pi^{\mu\nu} \rrangle \np 
\stp 
Since the derivatives do not appear as the combination
\st
\label{dpisimplified}
   \llangle D\pi^{\mu\nu} \rrangle + \frac{d}{d-1} \pi^{\mu\nu} \nabla \cdot u  \nc
\stp
but  rather involve $\llangle D\pi^{\mu\nu} \rrangle$ separately, this model
does not respect conformal invariance.
\item The second model is the full Israel-Stewart equation which has
the following form\cite{MurongaTerms} 
\bg
\label{fullisrael}
\pi^{\mu\nu} &=& 
-\eta \sigma^{\mu\nu}  - 
\tau_{\pi} \llangle D\pi^{\mu\nu} \rrangle +  
\frac{1}{2} 
\pi^{\mu\nu} 
 \frac{\eta T}{\tau_\pi}  \,
\partial_\rho 
\left( \frac{\tau_\pi}{\eta T} u^{\rho} \right)  + 2\tau_\pi \,  \pi^{\alpha (\mu} \Omega^{\nu) }_{\phantom{\nu)}\alpha }  \nc \\
             &\rightarrow& 
-\eta \sigma^{\mu\nu}  - 
\tau_{\pi} \left[ \llangle D\pi^{\mu\nu} \rrangle + \frac{d}{d -1} \pi^{\mu\nu} \nabla\cdot u\right]  
 + 2\tau_\pi \,  \pi^{\alpha (\mu} \Omega^{\nu) }_{\phantom{\nu)}\alpha }  \np
\nd
In the  last line we have used the conformal relation,  $\eta T/\tau_\pi \propto T^{d+1}$ and  equation of motion, $D(\ln T) =-1/(d-1) \nabla\cdot u$\,.
The model is equivalent to taking
$\lambda_1 = \lambda_3=0$ and $\lambda_2 = -2 \eta \tau_\pi$\,.  
\end{enumerate}

There has been some effort to compute the coefficients of
the gradient expansion both 
at strong and weak coupling.  The gradient expansion in \Eq{constituent} 
implies that the relative size of the coefficients  is $(\ell_{\rm mfp}/L)^2$, and 
this is of order $[\eta/(e+\pr)]^2$ in a relativistic theory.
The strong coupling results\cite{BRSSS,Bhattacharyya:2008jc} are listed in \Table{second}.
At weak coupling,  the results of  
kinetic calculations are also listed in the table\cite{York:2008rr} (see also \Ref{Betz:2008me}).
In kinetic theory the  physics of these higher order terms stems 
from the streaming terms and the collision integrals\cite{York:2008rr,BRSSS}. 
To first order in the gradients,
the distribution is modified from its equilibrium form 
\st
         n \rightarrow n + \delta f \nc
\stp
where $\delta f \propto p^{i}p^{j} \sigma_{ij}$ -- see \Sect{kinetics}.
Substituting this correction back into the 
Boltzmann equation 
\st
   P^{\mu} \partial_{\mu} f = -C[f] \nc
\stp
leads to several terms which are responsible for  the 
second order corrections to hydrodynamics.  We enumerate these contributions:
\begin{enumerate}
\item 
The $\tau_\pi$ and $\lambda_2$ terms are the result of  streaming of the first viscous correction $P^{\mu} \partial_{\mu} \delta f$ and do not involve the 
collision integral.  The common origin of these terms  ultimately 
explains the relation between them,  $\lambda_2 = -2\eta\tau_\pi$.
\item  The contribution to  $\lambda_1$ (the visco-elastic $\pi\pi$ term) reflects the streaming $P^{\mu}\partial_\mu \delta f $  and  the non-linearities 
of the collision integral, $C[\delta f]$. 
\item
Finally the 
vorticity-vorticity term does not appear on the LHS of the Boltzmann equation 
and therefore this term vanishes in kinetic theory \cite{BRSSS,Betz:2008me,York:2008rr}. In the strong coupling limit the absence of a vorticity-vorticity coupling is not understood.
\end{enumerate}

\begin{table}
\begin{center}
\begin{tabular}{c|c|c|c}
     Quality      &  $\N=4$ SYM &  QCD Kinetic Theory & Relaxation Time \\ [1ex]  \hline  
\rule{0pt}{3ex}  $\eta\tau_\pi$
& $4-2\,\ln(2)\simeq 2.61$ & 5.9 to 5.0  \quad (due to $g$)  &  6 \\[1ex]
\rule{0pt}{3ex}  $\lambda_1$    & 2  & 5.2 to 4.1 \quad (due to $g$) &  6 \quad ($\equiv \eta\tau_\pi$)  \\[1ex]
\rule{0pt}{3ex} $\lambda_2$ &  $-4\,\ln(2) \simeq -2.77$  
 & -11.8 to -10 \quad ($\equiv -2\eta\tau_\pi$) & -12 \quad ($\equiv -2 \eta\tau_\pi$) \\[1ex]
\rule{0pt}{3ex} $\lambda_3$ & 0  
 & 0 & 0 \\[1ex]
\end{tabular}
\end{center}
\caption{Compilation of values of (rescaled) second order transport quantities ($\eta\tau_\pi, \lambda_1,\lambda_2, \lambda_3$). 
{\it All} numbers in this 
table should be {\it multiplied} by $\eta^2/(e+\pr)$.
The complete strong coupling results are from an 
amalgamation of \protect \Ref{BRSSS} and \protect \Ref{Bhattacharyya:2008jc}. The weak
coupling results are from \protect \Ref{York:2008rr} and the relaxation
time approximation was studied in \protect \Ref{BRSSS} and clarified 
in \protect \Ref{York:2008rr}.
 Hydrodynamic simulations of the heavy ion event
are not sensitive to these values.  In a theory where $\lambda_1 =\eta \tau_\pi$ the second order corrections to a viscous $0+1$ dimensional Bjorken 
evolution vanish.
\label{secondhydro} }
\end{table}

In the relaxation time approximation discussed in \Sect{kinetics} (with
$\tau_R \propto E_\p$)  
the coefficient $\tau_\pi$ is readily calculated 
with linearized  kinetic theory for a massless gas\cite{BRSSS,York:2008rr}
\bg
\eta \tau_\pi &=&  6  \frac{\eta^2}{e +\pr} \nc
\nd
The kinetic theory relations  $\lambda_2=-2 \eta\tau_\pi $ and $\lambda_3 =0$
are respected for the same reasons as the full theory.
Also in the  relaxation time approximation one finds,
$\lambda_1 = \eta \tau_\pi$. 
In the full kinetic 
theory the {\it difference} $\lambda_1 - \eta \tau_\pi$ 
reflects the deviation from the quadratic ansatz discussed 
\Sect{kinetics}, and  to a much lesser extent the  non-linearities 
of the collision integral.
Nevertheless the relation  $\lambda_1 =\eta \eta_\pi$ 
almost holds indicating the dominance of the streaming term.
Overall the 
relaxation time approximation provides a good first estimate of these coefficients in kinetic theory.  This is important because the second order corrections
to a $0+1$ Bjorken expansion  vanish if $\lambda_1 = \eta \tau_\pi$ -- see below.

From a practical perspective a majority of simulations have used the 
full Israel-Stewart equations\cite{Romatschke:2007mq,Song:2008si,Molnar:2008xj} 
and treated $\tau_\pi$ 
as a free parameter, varying  $\left[\eta/(e+\pr)\tau_\pi \right]$ down  from the relaxation time value  by a factor of two.  
While 
it is gratifying that higher order transport coefficients can be computed 
and  classified,
the final phenomenological results (see \Sect{viscous_model}) 
are insensitive to the precise value 
of all second order terms for $\eta/s \lsim 0.3$ \cite{Luzum:2008cw,Song:2008si,Dusling:2007gi}.   Thus, the full hydrodynamic simulations corroborate 
the estimate given in \Sect{applicable} for the range of validity of hydrodynamics.

\subsection{Summary}

We have discussed various orders in the gradient expansion of 
hydrodynamics. Here we 
would like to summarize these results for a $0+1$ dimensional Bjorken
evolution. The equation of motion  for the Bjorken expansion 
is 
\st
   \frac{de}{d\tau} = - \frac{e + T^{zz} }{\tau} \nc
\stp
where $T^{zz}\equiv \tau^2 T^{\eta\eta}$ is the stress tensor at mid space-time rapidity, $\eta_s=0$. The stress tensor  through second order is\cite{Danielewicz:1984ww,BRSSS,Luzum:2008cw}
\st
\label{Bj2nd}
 T^{zz} = \pr \, - \, \frac{4}{3} \frac{\eta}{\tau} \,  +  \,  
\left(\lambda_1 - \eta \tau_\pi \right) \frac{8}{9\tau^2}   \np
\stp
Each additional term reflects one higher order in the hydrodynamic expansion 
parameter $[\eta/(e + p)\tau]$ discussed  in \Sect{applicable}.  We have 
made use of the intermediate results
\st
  \sigma^{\mu\nu} =  {\rm diag}\left(\sigma^{\tau\tau}, \sigma^{xx},\sigma^{yy},\tau^2 \sigma^{\eta\eta} \right) = 
\left(0,\, \frac{2}{3\tau},\, \frac{2}{3\tau},\, -\frac{4}{3\tau} \right) \nc
\stp
and
\st
 \llangle \sigma^{\mu \lambda }\sigma_{\lambda}^{\phantom{\lambda}\nu} \rrangle   = 
{\rm diag}\left( 0, -\frac{4}{9\tau^2} ,-\frac{4}{9\tau^2}, \frac{8}{9\tau^2} \right) \np
\stp
Notice in \Eq{Bj2nd}
that there is a cancellation between the relaxation terms 
$\sim \eta\tau_\pi D\sigma$ and 
the visco-elastic response\cite{Luzum:2008cw}, $\lambda_1 \sigma\sigma$. 
In  kinetic theory 
the difference $\eta\tau_\pi - \lambda_1$ 
is determined primarily from the deviation of $\delta f$ from the 
quadratic ansatz (see \Sect{kinetics}).
Thus $\eta\tau_\pi$ is expected to be approximately equal to $\lambda_1$
due to the overall kinematics of the streaming term\cite{York:2008rr}, 
$P^{\mu} \partial_\mu \delta f$. 
Examining \Table{second} we see that for a relaxation time approximation 
$\eta\tau_\pi = \lambda_1$ and the second order corrections to  
a $0+1$ Bjorken expansion of a conformally invariant fluid vanish!
This cancellation 
is partially present for the full kinetic theory and for the strongly interacting theory.

\begin{figure}
\begin{center}
\includegraphics[height=2.6in]{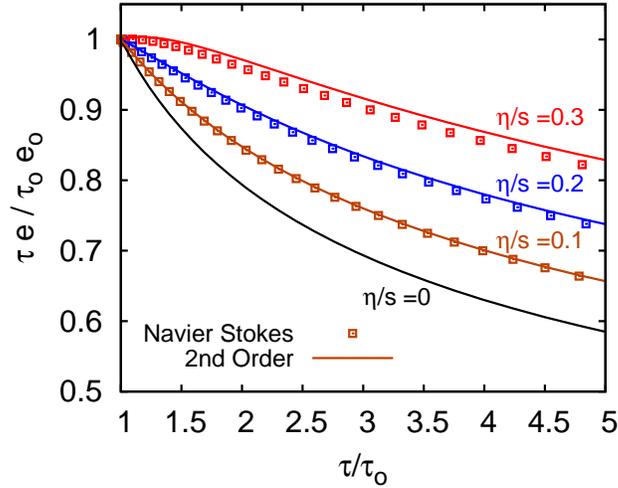}
\end{center}
\caption{The energy density ($\times \tau $) relative to the 
initial energy density ($\times \tau_o$) for a $0+1$ dimensional
Bjorken expansion.  The temperature is $T_o \simeq 300\,{\rm MeV}$ 
and $\tau_o\simeq 1\,{\rm fm}$, so that $1/\tau_oT_o \simeq 0.66$. 
The second order correction is smaller than expected due 
to a cancellation between the relaxation term $\sim \eta \tau_\pi D\sigma$ 
and the viscoelastic term\protect\cite{Luzum:2008cw},  $\sim \lambda_1 \sigma\sigma$. In 
a relaxation time approximation the second order correction vanishes (see text).
\label{bjfig}
}
\end{figure}

\Fig{bjfig} shows the how the different orders in \Eq{Bj2nd} influence 
the evolution of the energy density 
for a conformally invariant equation of state $(\pr = e/3 = sT/4)$  
and various values of $\eta/s$.
For definiteness we have used the $\N=4$  ratios
for the second
order transport  coefficients but this makes little difference since only the
combination $\lambda - \eta\tau_\pi$ matters -- see \Table{second}.
Finally we have used the estimates of 
\Sect{applicable} for the initial temperature $T_o$ and time $\tau_o$
\[
     \frac{1}{\tau_o T_o} =  0.66  \left(\frac{1\, {\rm fm} }{\tau_o} \right) \left( \frac{300\,\mbox{MeV}}{T_o} \right) \np
\]
Generally the effect of second order terms is small (due to the 
cancellation) and
the value of the first order terms drive the correction to 
the ideal evolution.


\section{Kinetic Theory Description}
\label{kinetics}

In \Sect{hydro_basic} 
we discussed various aspects of viscous hydrodynamics as
applied to heavy ion collisions. 
Since ultimately the experiments measure particles,  there is a need 
to convert the hydrodynamic information into particle spectra. 
This section will provide an introduction
to the matching between the  kinetic and hydrodynamic descriptions. 
This will be important when comparing the hydrodynamic models 
to data in \Sect{viscous_model}.
In addition, since \Sect{transport} 
discussed various calculations of  the shear 
viscosity in  QCD, this section we will sketch briefly
how  these kinetic calculations are performed. Good summaries 
of this set of steps are provided by \Refs{AMYLLog,Arnold:2006fz,degroot}.

In kinetic theory
the spectrum of particles in a volume $\Sigma$ is given by the Cooper-Frye formula\cite{CF} 
\st
  E \frac{\dd^3N}{\dd^3\p} =  \frac{1}{(2\pi)^3} \int_{\Sigma} \dd\Sigma_\mu P^{\mu} \, f(-P\cdot u) \np
\label{CooperFrye1}
\stp
Note that when  $\Sigma$ 
is a three volume at fixed time,
$\dd\Sigma_\mu = \left( \dd V , 0,0,0 \right)$, and this formula reduces to the 
traditional  result.    
Four vectors are denoted with capitol letters $P^{\mu} = (E_\p, \p)$,
and the  equilibrium distribution function is denoted with 
\st
  n(-P\cdot u) =  \frac{1}{\exp(-P\cdot u/T) \pm 1}  \np
\stp
We will also use a suffix notation, $n_\p=n(-P\cdot U)$ and $f_\p = f(-P\cdot u)$.
The distribution function obeys the Boltzmann equation 
\st
\label{boltz_basic}
 \partial_t f_\p +  v_\p \cdot \partial_x f_\p  = -\int_{234} 
\Gamma_{12\rightarrow 34} \left(f_1 f_2  - f_{3} f_{4} \right) \nc
\stp
where $v_\p=\partial E_\p/\partial \p = \p/E_\p$,
and  we have assumed $2\rightarrow2$ scattering 
with  classical statistics for simplicity,
 $f_\p=\exp(-E_\p/T)$.
The momenta are labeled as  $f_2=f_{\p_2}$ and $f_1=f_{\p_1}$ with $\p_1 \equiv \p$. The integral over the phase space is  abbreviated
\st
 \int_{234} = \int \frac{\dd^3\p_2}{(2\pi)^3} \frac{\dd^3\p_3}{(2\pi)^3} \frac{\dd^3\p_4}{(2\pi)^3}  \nc
\stp
and the transition rate $\Gamma_{12\rightarrow 34}$ 
for $2\rightarrow 2 $ scattering 
is related to the usual Lorentz invariant matrix element $\left|\mathcal M\right|^2$ by
\st
 \Gamma_{12\rightarrow34}  = 
\frac{\left| \mathcal M \right|^2 }{(2 E_1) (2 E_2) (2 E_3) (2E_4) }  (2\pi)^4
\delta^4(P_{1} + P_{2} - P_{3} - P_{4}) \np
\stp
The generalization of what follows  to a multi-component gas with quantum statistics
is left to the references\cite{AMYLLog}.

During a viscous evolution  the spectrum will be modified from
its ideal form 
\st
\label{viscous_corrections}
  f =  n_\p + \delta f_\p \nc
\stp
and this has important phenomenological 
consequences\cite{Teaney:2003kp}.
The modification of the distribution function depends on the details 
of the microscopic interactions. In a linear approximation the deviation
is proportional to the strains  and can be calculated in kinetic theory.
When the most important strain is shear,  the deviation $\delta f$ is proportional  $\sigma_{ij}$ .
Traditionally we parameterize the viscous 
correction to the distribution in the  rest frame of the medium by\footnote
{ When quantum statistics  are taken into account this 
should be written
\[
 \delta f_\p = -n_\p (1 \pm n_\p)  \, \chi(|\p|) \hat \p^{i} \hat \p^{j} \sigma_{ij} \nc
\] 
where the overall minus is introduced because in the Navier-Stokes theory $\pi^{\mu\nu}=-\eta \sigma^{\mu\nu}$
} $\chi(|\p|)$ 
\bg
\label{chiequ}
 \delta f_\p = -n_\p \, \chi(|\p|) \, \hat \p^{i} \hat \p^{j} \sigma_{ij} \np
\nd
Then the stress tensor in the  local rest frame  is
\st
  T^{ij} = p \delta^{ij} - \eta \sigma^{ij} = \int \frac{d^3\p}{(2\pi)^3} \frac{p^{i} p^{j} }{E_\p}  \left[n_\p + \delta f_\p\right] \np
\stp
Substituting \Eq{chiequ} for $\delta f_\p$  and 
using  rotational symmetry we have
\st
\label{shearvisc}
  \eta = \frac{2}{15}\int \frac{d^3\p}{(2\pi)^3} \frac{p^2}{E_\p} \chi(|\p|) n_\p \np
\stp
Thus we see that the form of the viscous correction to the 
distribution function determines the shear viscosity.

To calculate the transport coefficients the Boltzmann 
equation is analyzed in the rest frame of a particular location $\x_o$. In 
a neighborhood of this point the temperature and flow fields are
\st
  u^{\mu}(\x,t) \simeq  (1 , u^{i}(\x,t))\nc   \qquad T(\x,t) \simeq  T_o + \delta T(\x,t) \nc
\stp
where $u^{i}(\x_o,t) = \delta T(\x_o,t) =0$.
The equilibrium distribution function in this neighborhood  is
\st
\label{equil}
 n(-P\cdot u) \simeq  
n_\p^o  + n_\p^o \left( \frac{E_\p}{T_o^2} \,\delta T(\x,t)  + \frac{p^i u^{i}(\x,t)}{T_o} \right) \nc
\stp
where we have used the short hand notation, $n_\p^o = \exp(-E_\p/T_o)$.
We can now substitute the distribution function into the 
Boltzmann equation and  find an equation the  $\delta f$. 
The left hand side  of the Boltzmann equation involves gradients, and
therefore only the equilibrium distribution needs to be considered. 
Substituting \Eq{equil} into the l.h.s. of \Eq{boltz_basic},
using the ideal  equations of motion
\bg
   \partial_t u^{i}  &=&   -\frac{\partial^{i} \pr}{(e + \pr) } \nc \\
   \partial_t e  &=&  - (e + \pr) \partial_i u^{i}\nc
\nd
and several thermodynamic relationships
\bg
     c_v &=& \frac{de}{dT} \nc \\
     \frac{n}{e + \pr} d(\mu/T)   &=&  \frac{1}{T(e +\pr)} d\pr   + d\left( \frac{1}{T} \right) = 0  \nc
\nd
we find that\footnote{We have tacitly assumed that the dispersion curve $E(\p)$ does 
not depend on the temperature. This is 
fine as long as we are not considering the bulk viscosity \cite{Arnold:2006fz}.}
\st
\label{lhsboltz}
 \partial_t f_e  + v_\p  \cdot \partial_\x f_e = 
\frac{n_\p}{E_\p}\left[ \left(  \frac{|\p|^2 }{3 T }   - \frac{E_\p^2}{T} \frac{(e + \pr) }{T c_v}    \right) \partial_i u^{i}   
  +  \frac{ p^{i} p^{j}} {2 T } \sigma_{ij} \right] \np
\stp
The result is proportional  to two strains $\partial_i u^i$ and $\sigma_{ij}$
which are ultimately responsible for the bulk and shear viscosities 
respectively.
For a massless conformal gas we have $|\p|^2 = E_\p^2$, $Tc_v = 4 e$ and 
$e+p = (4/3) e$. The result is that the term proportional
to $\partial_i u^i$ vanishes and consequently the  
bulk viscosity  is  zero in this limit. Subsequently, we will consider only 
 the modifications due to the shear viscosity and 
refer to \Sect{transport} for a more complete discussion of bulk viscosity.

In \Eq{lhsboltz}, the l.h.s. of the Boltzmann equation is evaluated at the point $\x_o$.
We also evaluate
the r.h.s. of the
Boltzmann equation at the point $\x_o$  to linear order in $\delta f$  
\begin{align}
\label{boltz}
n_\p^{o} \,
      \frac{ p^{i} p^{j}} {2 T E_\p } \sigma_{ij} 
  &= -\int_{234} \Gamma_{12\rightarrow34} \, n_\p^o n_2^o \left[ \frac{\delta f(\p)}{n_\p^o} + \frac{\delta f_2}{n_2^o } - \frac{\delta f_3}{n_3^o} - \frac{\delta f_4}{n_4^o} \right] \np
\end{align}
In writing this equation we have made use of the detailed balance relation
\st
  n_1 n_2 \Gamma_{12\rightarrow 34} = n_3 n_4 \Gamma_{12\rightarrow 34} \np
\stp
\Eq{boltz} should be regarded as a matrix equation for the distribution
function $\delta f(\p)$. Although $\delta f(\p)$ or $\chi(p)$ can be determined numerically
by straight forward discretization and matrix inversion,
a variational
method is preferred in practice\cite{Baymetal,AMYLLog}.  
After determining $\delta f(\p)$ or $\chi(|\p|)$ the shear 
viscosity can be determined from \Eq{shearvisc}.     

Inverting the integral equation in \Eq{boltz} requires a detailed knowledge of the  microscopic
interactions. Lacking such detailed knowledge, one  can resort to a relaxation
time approximation, writing the Boltzmann equation as  
\st
   \partial_t f + v_\p \cdot \partial_\x f =  -\frac{1}{\tau_R(-P\cdot U)} \frac{(-P\cdot U)}{E_\p} \,\delta f \nc
\stp
where $\tau_{R} (-P\cdot U)$ is a momentum dependent relaxation time. 
In the local rest frame  this reduces to 
\st
   \partial_t f + v_\p \cdot \partial_\x f =  -\frac{1}{\tau_R(E_\p)} \,\delta f \np
\stp
By fiat the correction to the distribution function is simple
\st
 \delta f  =  - n_\p \frac{\tau_R(E_\p) }{2 T E_\p}  p^{i} p^{j} \sigma_{ij} \np
\stp

First we  consider
the case where the relaxation time  grows with energy
\st
\label{etaur}
  \tau_R(E_\p) = \mbox{Const} \times \frac{E_\p}{T}
   \qquad \delta f =  - \mbox{Const} \times  n_\p  \frac{p^{i} p^{j} }{2T^2}\sigma_{ij}
\stp
The form of this correction is known as the quadratic ansatz 
and was used by all hydrodynamic simulations so far\cite{Romatschke:2007mq,Dusling:2007gi,Luzum:2008cw,Song:2008si}. 
Substituting
this form into \Eq{shearvisc}  one determines that for a classical gas of arbitrary
mass the  constant is\cite{Teaney:2003kp}
\st
\label{quad_ansatz}
  \mbox{Const} = \frac{\eta}{e + \pr} \nc \qquad \mbox{with} \qquad   \tau_R(E_\p) \propto E_\p \np
\stp
For a Bose or Fermi gas we have the replacement   
$n_\p \rightarrow n_p(1 \pm n_p)$ and \Eq{quad_ansatz} is  approximate holding 
at the  few percent level.  For a mixture  of different classical particles
with  one  {\it common} relaxation time \Eq{quad_ansatz} also holds. In practical
simulations this correction is written covariantly and 
the phenomenological field  $\pi^{\mu\nu}=-\eta \sigma_{\mu\nu}$ is used, leading
to the final result
\st
 \delta f = \frac{1}{2(e + \pr) T^2} P^{\mu} P^{\nu} \pi_{\mu\nu} \np 
\stp

The quadratic ansatz may seem arbitrary, but it is 
often a good model of collisional energy 
loss  and weak scattering.
For instance an analysis
of the leading-log Boltzmann equation (along the lines of \Eq{boltz}) 
shows that the quadratic ansatz  describes the full results 
to  10-15\% accuracy\cite{AMYLLog}.
However, this agreement is in part an artifact of
the leading-log, or soft scattering,
approximation. For example, in the leading-log plasma  the energy loss 
of a ``high" $p_T$ quark from the  bath is given by\cite{Braaten:1991we}
\st
\label{dpdtbrattenthoma}
\frac{\dd p}{\dd t} = \left(N_c + \frac{N_f}{2} \right) \frac{C_F g^4 T^2}{24\pi} 
\log\left(T/m_D\right) \np
\stp
From this formula we see that the energy  loss is constant 
at high momentum and therefore the relaxation time scales  as
$\tau_R \propto p$ in agreement with \Eq{etaur}.
In reality \Eq{dpdtbrattenthoma} is decidedly wrong at large momentum 
where radiative energy loss becomes increasingly
significant and can shorten the relaxation time. 
Indeed when collinear radiation is included in the Boltzmann 
equation the quadratic ansatz becomes increasingly poor \cite{GuyPrivate}.  
In \Sect{viscous_model}
we will consider 
a relaxation time which is independent of energy 
as an extreme alternative
\st
   \tau_{R} \propto \mbox{Const}  \nc   
\stp 
and explore the phenomenological consequences of this ansatz.

As discussed in \Sect{viscous_model},  the differential elliptic flow $v_2(p_T)$  is sensitive  to the  form of these corrections, while the integrated 
$v_2$
is constrained by the underlying hydrodynamic variables,
and is largely independent of these details. 
This last remark should
be regarded with caution as it has not been fully quantified.  



\section{Viscous Hydrodynamic Models of Heavy Ion Collisions}
\label{viscous_model}
\begin{figure}
\begin{center}
\includegraphics[width=3.5in]{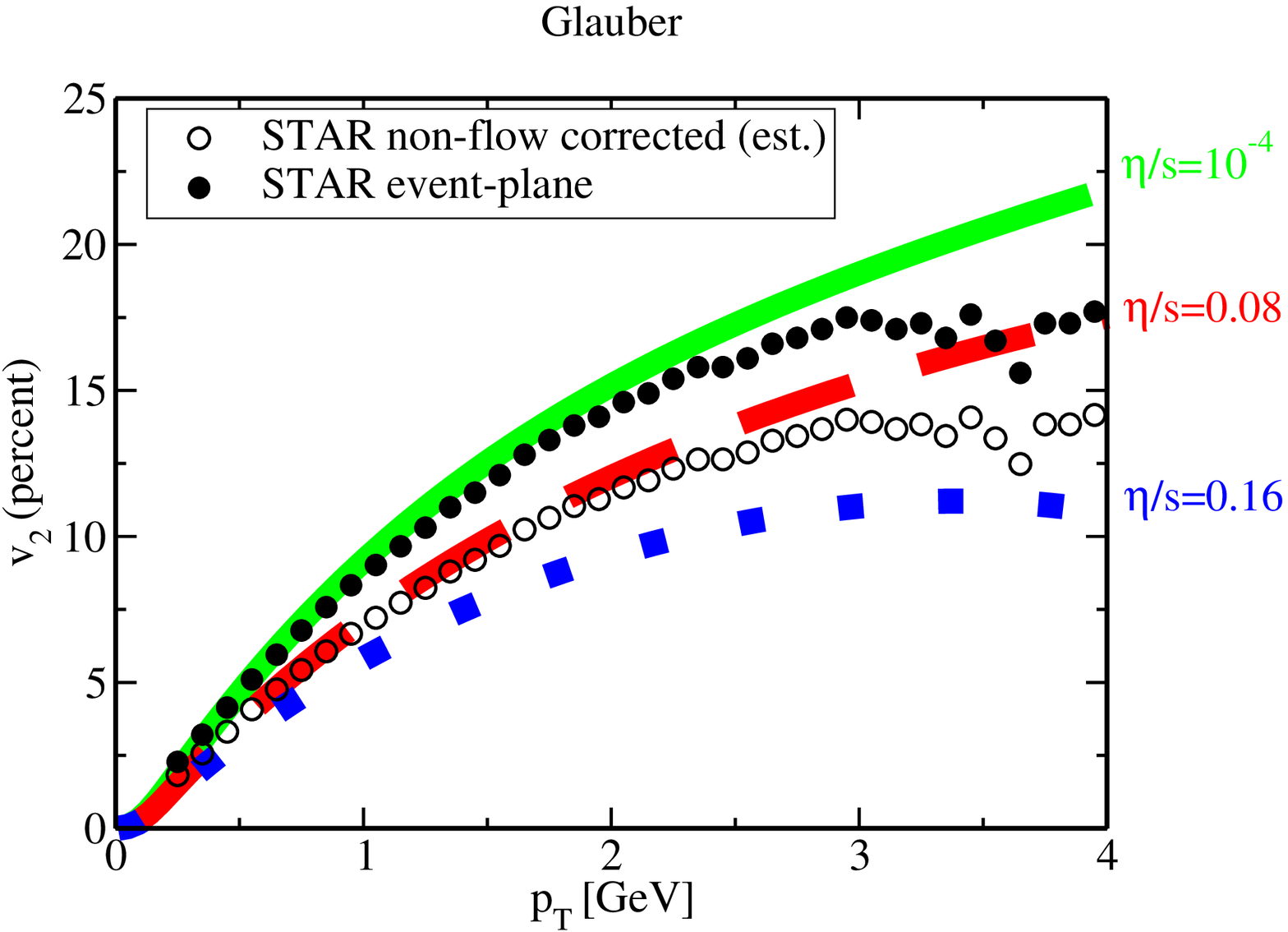}
\includegraphics[width=3.5in]{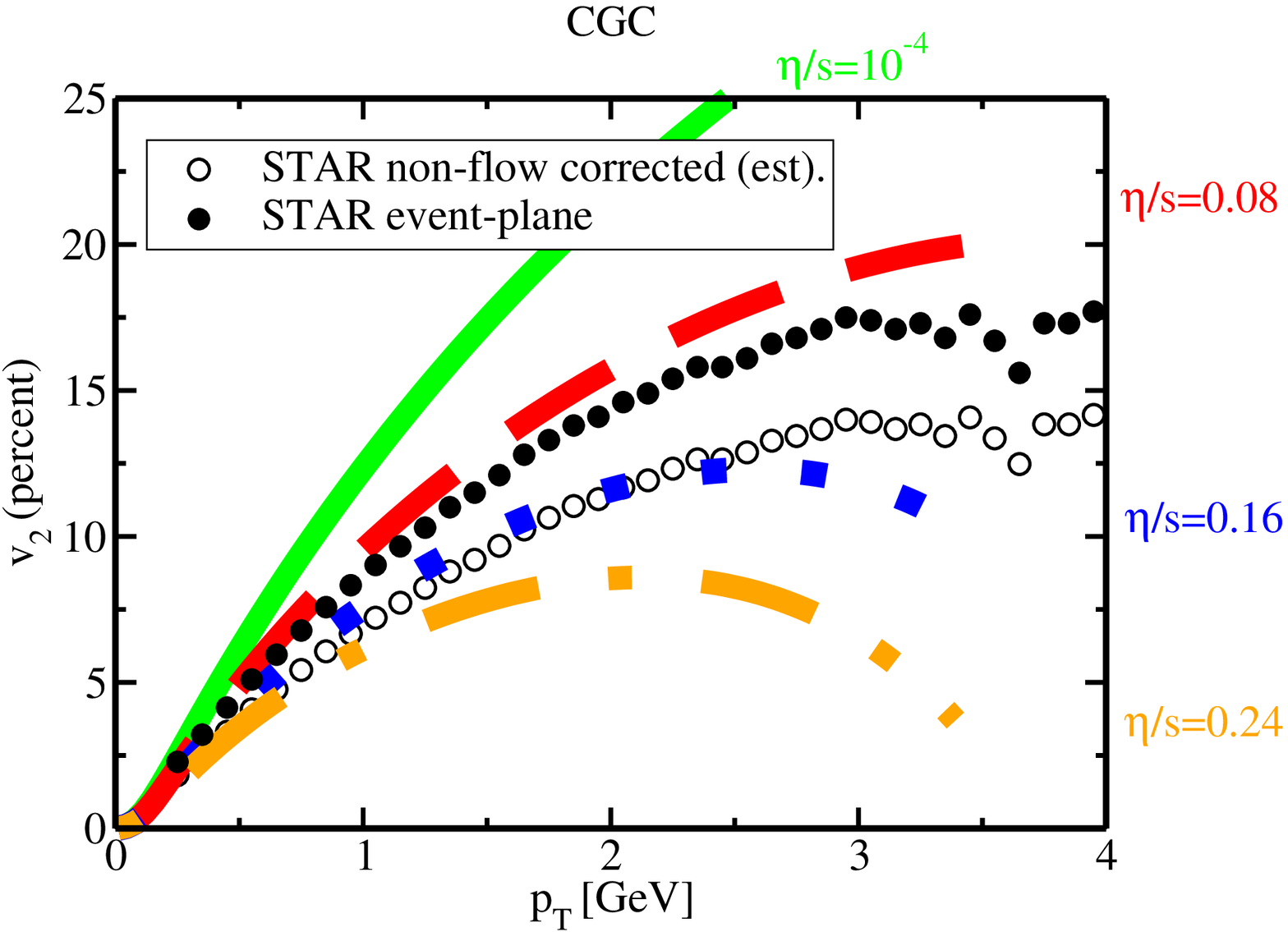}
\end{center}
\caption{Figure from \protect \Ref{Luzum:2008cw} which shows how  elliptic 
flow depends on shear viscosity. The theory curves  are most 
dependable for $p_T \lsim 1.5\,{\rm GeV}$ and should be compared
to the ``non-flow corrected" data. The Glauber and CGC initial 
conditions have different eccentricities as described in the text.
\label{LR_money_plot} 
}
\end{figure}

At this point we are in a position to discuss several viscous hydrodynamic models
which have been used to confront the elliptic flow  data.  
To initiate discussion, we show
simulation results for $v_2(p_T)$ from 
Luzum and Romatschke in Fig.~\ref{LR_money_plot}. 
Comparing the simulation to the ``non-flow corrected" data for $p_T \lsim 1.5\,{\rm GeV}$, 
we can estimate an allowed range for the shear viscosity, $\eta/s \approx 0.08 \leftrightarrow 0.16$.
Below we will place this estimate in context by
culling figures from related works.

A generic implementation of viscous hydrodynamics consists of several parts:  
\begin{enumerate}
\item At an initial time $\tau_o$ the energy density and flow velocities 
are specified.
For
Glauber initial conditions,  one takes for example 
\st
  e (\tau_o,\x_\perp)  \propto \frac{\dd N_{\rm coll}}{\dd x\,\dd y} \nc
\label{ebc}
\stp
where the overall constant is adjusted to reproduce the multiplicity in
the event.
The simulations assume Bjorken boost invariance  with the ansatz
\st
 e(\tau,\x_\perp,\eta) \equiv e(\tau,\x_\perp) \nc \\
\stp
\st
u^{\mu}(\tau,\x_\perp,\eta) = \left(u^{\tau}, u^{x},u^y,u^\eta\right)  = 
\left(u^{\tau}(\tau,\x_\perp),\, u^x(\tau,\x_\perp),\, u^y(\tau,\x_\perp), \, 0\right)  \np
\stp
In cartesian coordinates $u^{z}=u^{\tau} \sinh(\eta_s)$ and 
$u^{t}=u^{\tau}\cosh(\eta_s)$. The calculations typically assume 
zero transverse flow velocity at the initial time $\tau_o$
\st
   u^x(\tau_o,\x_\perp) = u^{y}(\tau_o,\x_\perp) = 0\nc \qquad u^{\tau}(\tau_o,\x_\perp) = 1\np
\stp
The strains are taken  from the Navier stokes 
theory for example 
\st
 \pi^{\mu\nu}(\tau_o,\x_\perp) = {\rm diag} \left(\pi^{\tau\tau},
\pi^{xx},\pi^{yy}, \tau^2 \pi^{\eta\eta} \right) =  \left(0, \; \frac{2}{3} \frac{\eta}{\tau}, \;
\frac{2}{3} \frac{\eta}{\tau},\; - \frac{4}{3} \frac{\eta}{\tau} \right)  \nc
\stp
and reflect the traceless character of shear stress.
\item The equations of motion are then solved.
  Viscosity 
modifies  the hydrodynamic variables,
$T$ and $u^{\mu}$, and also modifies 
off diagonal components of the 
stress tensor  through the viscous corrections $\pi^{\mu\nu}$.
\item A ``freezeout" condition is specified either by specifying 
a freezeout temperature or a kinetic condition.  During 
the time evolution a freezeout surface  is constructed. 
For instance the freezeout surface in \Fig{LR_money_plot} is the space-time three 
volume $\Sigma$ where $T_{\rm fo}\simeq 150\,{\rm MeV}$. 
\item Finally, in order to compare to the data, particle
spectra are computed by matching the hydrodynamic theory
onto kinetic theory. Specifically, on the 
freezeout surface final particle spectra are computed
using \Eq{CooperFrye}. Roughly speaking this  ``freezeout" procedure is equivalent to running the hydro 
up to a particular proper time $\tau_f$ or temperature $T_f$ 
and declaring that the thermal spectrum of 
particles at that moment is the measured particle spectrum. 
\end{enumerate}
There are many issues associated with each of these items.  The next subsections
will discuss them one by one.

\subsection{Initial Conditions}
\label{initcond}

First we note that the   hydrodynamic  fields are initialized at a time $\tau_0 \simeq 1\,{\rm fm/c}$, which is arbitrary to a certain extent. 
Fortunately, both in kinetic theory and hydrodynamics 
the final results are not particularly sensitive this 
value  \cite{DM_private,Luzum:2008cw}.
Also, all of the current simulations have assumed Bjorken boost invariance.
While this assumption should be relaxed, 
past experience with ideal hydrodynamics shows 
that the mid-rapidity elliptic flow 
is not substantially modified\cite{Hirano:2004en}. 
Above we have discussed one possible initialization of the hydro 
which makes the  energy density proportional to the number 
of binary collisions, {\it e.g.} the Glauber curves of \Fig{LR_money_plot}. 
Another reasonable option 
is to make the entropy proportional to the 
number of participants \cite{Dusling:2007gi}
\st
  s(\tau_0,\x_\perp) \propto \frac{\dd N_p}{\dd x \, \dd y}  \np
\label{spart}
\stp 
As a limit one can take the CGC model discussed in 
\Sect{elliptic}.   
Finally it is generally assumed that the
initial transverse flow is  zero
\st
 u_{x}(\tau_0,\x_\perp)=u_y(\tau_0,\x_\perp) = 0 \np
\stp 
This assumption should probably lifted in future calculations and  a more
reasonable  (but still small) estimate is given in \Ref{Vredevoogd:2008id}. 

Examining \Fig{LR_money_plot} and \Fig{raju_eccentricity} 
we see see that there is a significant
and predictable linear dependence on the eccentricity. When extracting
the shear viscosity from the data, this uncertainty in the eccentricity 
leads to a factor of two uncertainty in the final results for  the
shear viscosity. 
As emphasized in \Sect{elliptic}, the CGC model should 
be thought of as an upper limit  to the anisotropy that 
can be produced in the initial state. Therefore, the 
uncertainty in $\eta/s$ is probably not as large as dispersion
in the curves would 
indicate.  In ideal 
hydrodynamics,  the  spread in $v_2(p_T)$ resulting 
from the different initializations
specified by \Eq{spart} and \Eq{ebc}  was  studied\cite{Kolb:2001qz}, 
and is  small compared to the difference between 
the Glauber and CGC curves  in \Fig{LR_money_plot}.
 
\begin{figure}
\begin{center}
\includegraphics[height=2.5in]{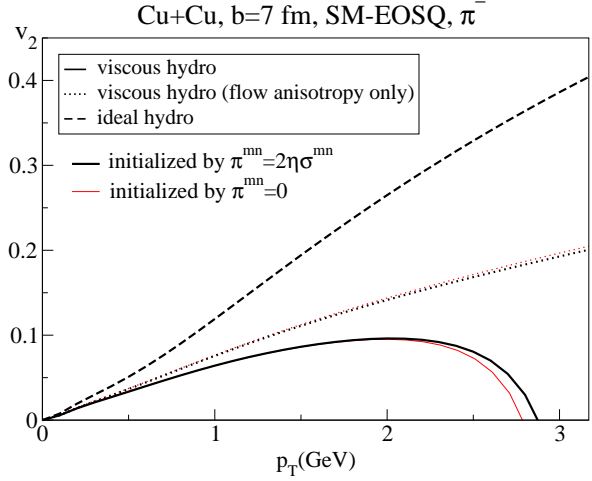}
\end{center}
\caption{
Figure from  \protect \Ref{Song:2007ux} studying the independence 
of the final results on the initialization of $\pi^{\mu\nu}$. 
Note that due to different metric and symmetrzation conventions 
the Navier Stokes  limit is $2\eta\sigma^{mn}$ rather than $-\eta\sigma^{\mu\nu}$ adopted here. 
\label{huichao_pi} 
} 
\end{figure}

Once the initial conditions for the temperature 
and the flow velocities are specified, 
the off diagonal components of the stress tensor
$\pi^{\mu\nu}$  are determined by the spatial gradients  in $T$ and $u^{\mu}$. 
To second order this is given by \Eq{constituent} and  there is no ambiguity
in this result.  Time derivatives
may be replaced with spatial derivatives using lower
order equations of motion to second order  accuracy.
In the phenomenological theory $\pi^{\mu\nu}$ 
is promoted to a dynamical variable. Clearly the appropriate 
initial condition for this variable is something which deviates from $-\eta \sigma^{\mu\nu}$ by second order terms.
However,  the extreme choice $\pi^{\mu\nu}=0$ was studied to 
estimate how initial non-equilibrium effects  could alter the final results.  This 
is just an estimate since the relaxation of these fields far from equilibrium is not 
well captured by hydrodynamics.  On the other hand, comparisons with full
kinetic theory simulations show that the 
Israel-Stewart model does surprisingly well at reproducing the  relaxation
dynamics of the full simulation\cite{Huovinen:2008te}.  
From a practical perspective, even with  this extreme
choice $\pi^{\mu\nu}=0$, the  stress tensor relaxes to the expected form 
$\pi^{\mu\nu}=-\eta \sigma^{\mu\nu}$  relatively quickly.  The result  is
that $v_2(p_T)$ is insensitive to the different initializations of $\pi^{\mu\nu}$. 
This can be seen clearly from \Fig{huichao_pi}.  

\subsection{Corrections to the Hydrodynamic Flow}
\label{flow_correct}

\begin{figure}
\begin{center}
\includegraphics[width=0.48\textwidth]{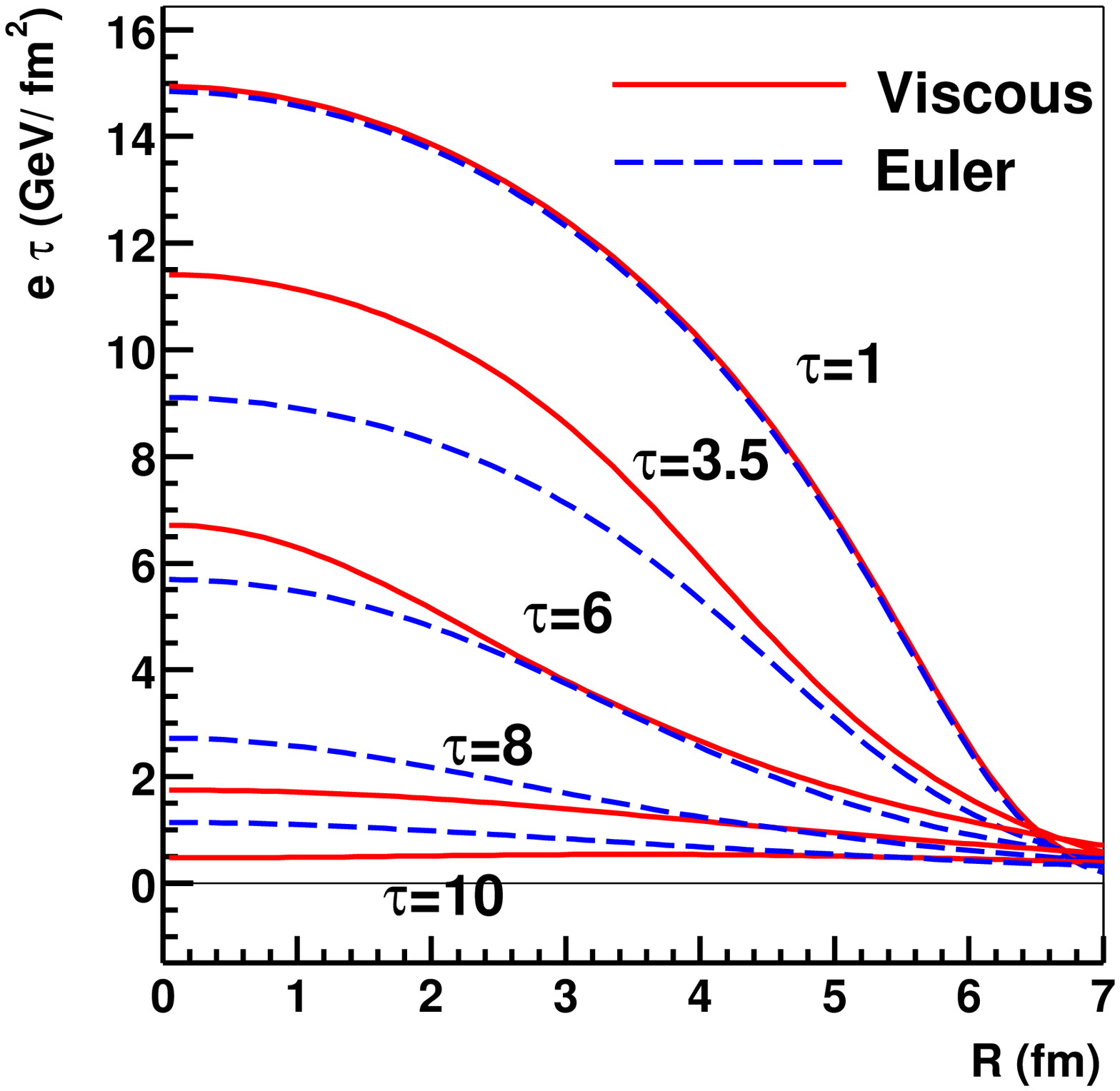}
\hfill
\includegraphics[width=0.48\textwidth]{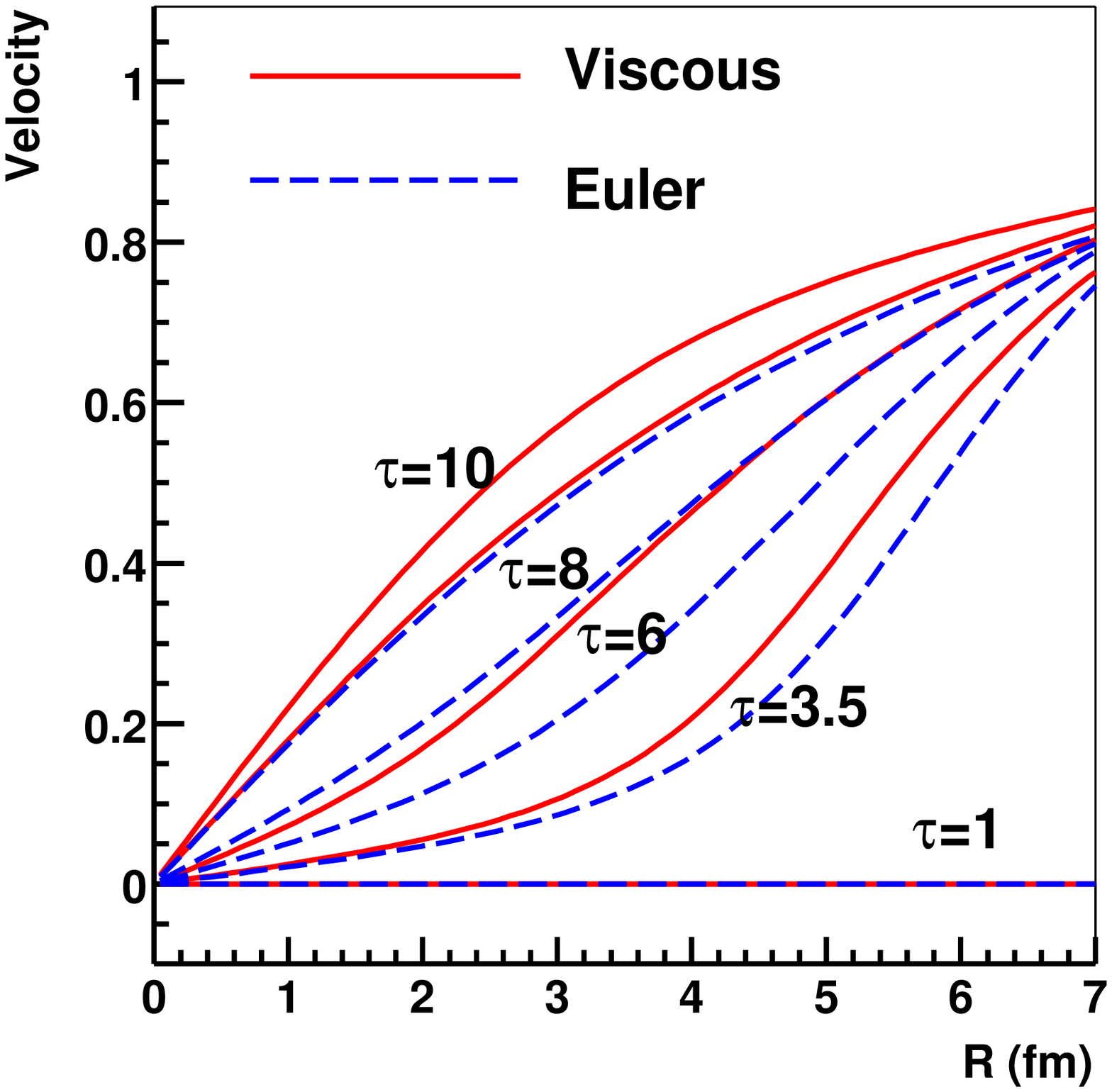}
\end{center}
\caption{A central AuAu simulation 
with an ideal gas equation of state $p=e/3$ and $\eta/s=0.2$ which compares the
viscous and Euler evolution\protect\cite{DT1}.  The left figure shows the
energy density ($\times \tau$) for different times. The right figure shows the
velocity for different times.  \label{etau1d} }
\end{figure}

Once the initial conditions are specified, the equations of motion 
can be solved. 
First we address the size of the viscous corrections 
to the  temperature and flow velocities.
The magnitude of
the viscous corrections depends  on 
the size of the system and the shear viscosity.   
\Fig{etau1d}  
shows a typical result  for the AuAu system.  
As seen from the figure the 
effect of viscosity on the temperature and flow velocities is 
relatively small, of order $\sim 15\%$ for $\eta/s \simeq 0.2$.

An explanation for this result is the following\cite{DT1}: In the 
first moments of the collision the system is expanding  in the longitudinal
direction and the pressure in the longitudinal directions is reduced
\st
  \tau^2 T^{\eta\eta} = \pr - \frac{4}{3} \frac{\eta}{\tau} \np
\stp
At first sight, this means that the system cools more slowly and
indeed this is initially true.
However, since the shear tensor is traceless 
there is an increase in the transverse pressure which 
is uniform in all directions
\st
   T^{xx} = T^{yy} = \pr + \frac{2}{3} \frac{\eta}{\tau} \nc
\stp
and which ultimately increases the radial flow. Since the radial flow is 
larger in the viscous case, the system ultimately cools faster.

\begin{figure}
\begin{center}
\includegraphics[width=0.49\textwidth]{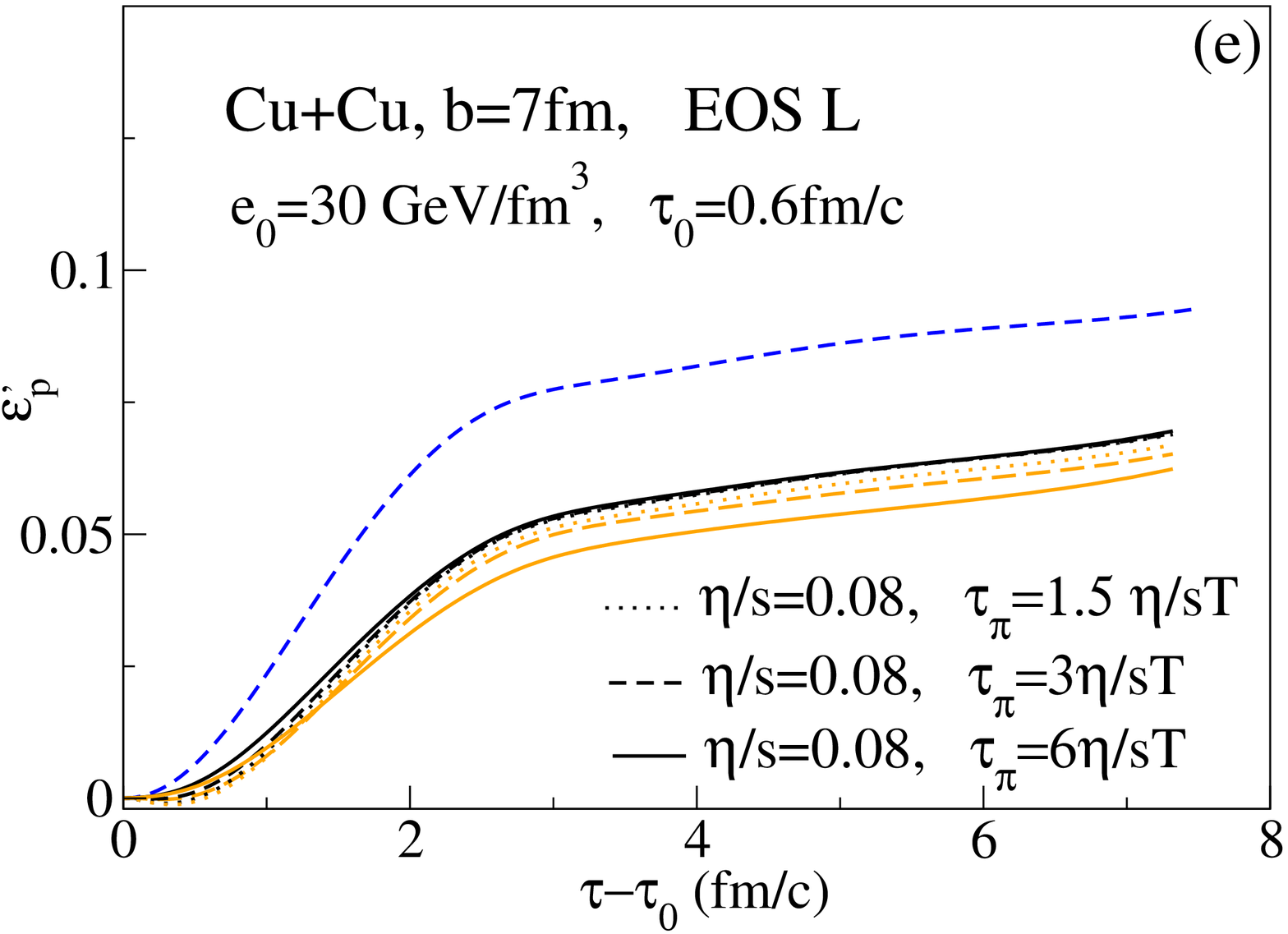}
\includegraphics[width=0.49\textwidth]{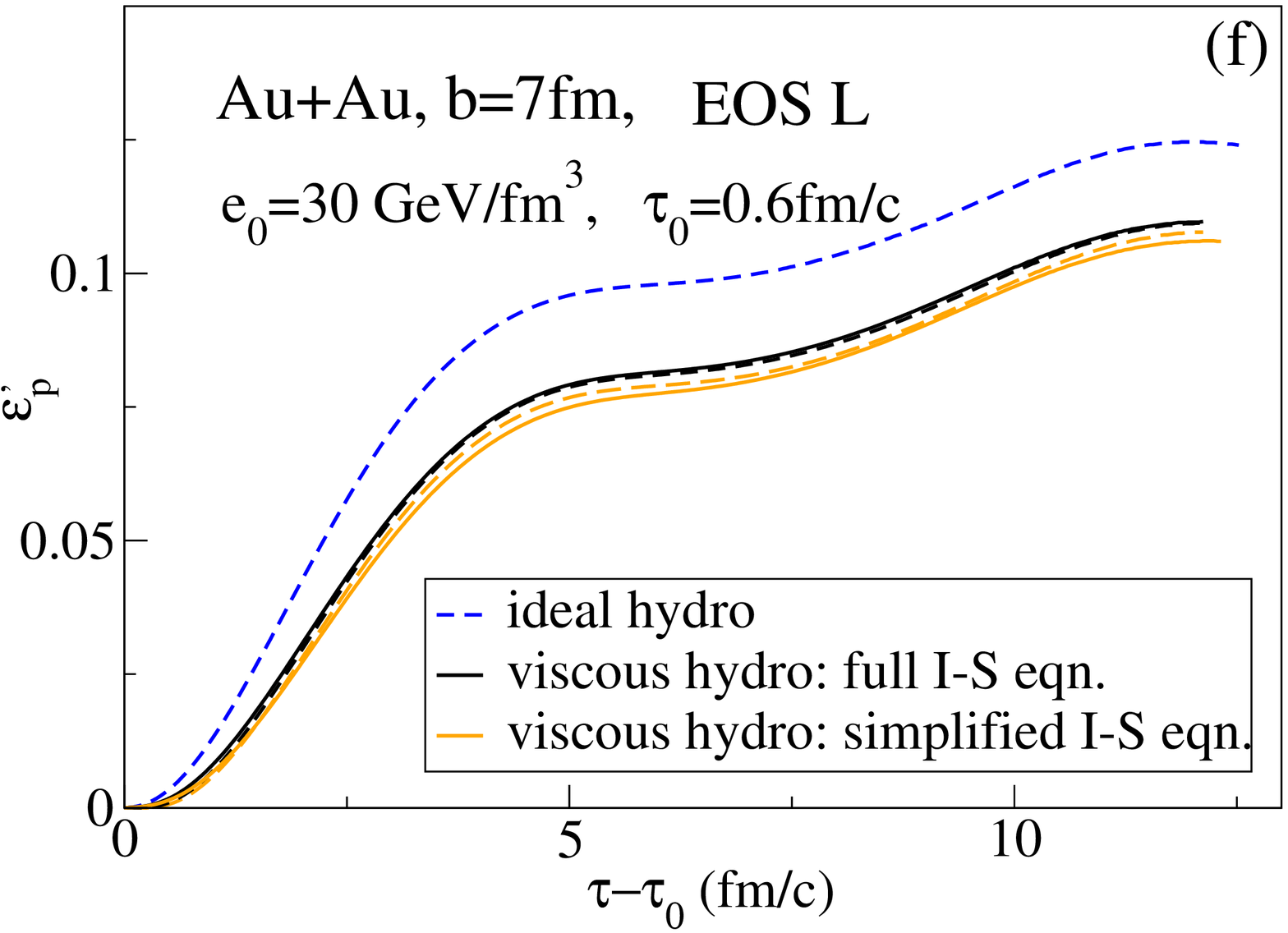}
\end{center}
\caption{Figure from \protect \Ref{Song:2008si}
comparing the development of the flow anisotropy $e_p'$ (\Eq{epp}) in viscous hydrodynamics  relative to the ideal hydrodynamics.
The  lower band of curves 
are all representative of 
viscous hydro and
 differ only in how the second order corrections
are implemented.
The anisotropy differs from ideal hydro 
because the anisotropy involves the viscous difference, $\Pi^{xx} - \Pi^{yy}$. 
\label{huichao_ep2} }
\end{figure}

Having discussed the dependence of $T$ and $u^{\mu}$, we 
turn to a quantity which largely dictates  the final elliptic
flow -- the momentum anisotropy \cite{KSH}.   The momentum 
anisotropy is defined as\footnote{
Note there is a misprint in the original definition of $e_p'$ in \Ref{KSH}. Eq.~(3.2) of that work used a double bracket notation indicating
an energy density weight which should only be a single bracket as above. 
This double bracket definition (rather than \Eq{epp}) 
was used subsequently 
in \Ref{Dusling:2007gi} which is why the associated curve
differs from other recent works \cite{Luzum:2008cw,Song:2008si}. 
Further the definition should perhaps include a factor of $u^{0}$ 
so that the integrals would be boost invariant, $i.e.$  
$\int \dd\Sigma_{\mu} u^{\mu}\, (T^{xx} - T^{yy} ) =\tau \Delta \eta_s \int d^2\x \, u^{0} (T^{xx} - T^{yy})$. However, typically the integral is dominated near the 
center where $u^{0} \simeq 1$. An alternative definition $S^{\mu\nu\rho}_{\rm hydro}$ is suggested in \Sect{particles}. }
\st
\label{epp}
  e_p' =\frac{ \llangle T^{xx} - T^{yy} \rrangle }{ \llangle T^{xx}  + 
T^{yy} \rrangle }  = \frac{\int d^2\x_\perp  \left( T^{xx} - T^{yy} \right)}{\int d^2\x_\perp  \left( T^{xx}  + T^{yy} \right) } \np
\stp
\Fig{huichao_ep2} illustrates how this momentum anisotropy increases as 
a function of time in the CuCu and AuAu systems.
Although the flow fields $T(\x_\perp,\tau)$ and $u^{\mu}(\x_\perp,\tau)$ are 
quite similar between 
the ideal and viscous cases, the ideal anisotropy is significantly 
reduced by viscous effects. 
The reason for this reduction is  because the viscous stress tensor anisotropy,  $T^{xx} - T^{yy}$,
involves the difference
\[
\Pi^{xx} - \Pi^{yy} \nc
\]
in addition to the temperature and flow velocities.
This additional term  
is ultimately responsible for the deviation of  $e_p'$ between
the ideal and viscous hydrodynamic calculations. At later times
there is some modification of $e_p'$, due to the flow itself, 
but this is dependent on freezeout.
The deviation $\Delta \Pi=\Pi^{xx} - \Pi^{yy}$  will have important
phenomenological consequences in determining the viscous correction 
to the elliptic flow spectrum.


\subsection{Convergence of the Gradient Expansion}
\label{convergence}

\Fig{huichao_ep2} also compares the simplified Israel-Stewart equation \Eq{dpisimplified}
to the full Israel-Stewart equation \Eq{fullisrael} as a function of the relaxation time 
parameter $\tau_\pi$. The result supports much of the discussion given 
in \Sect{second}. In the roughest approximation neither the simplified Israel-Stewart
equation nor the full Israel-Stewart equation depend on the relaxation time 
parameter $\tau_\pi$. 
When an ideal gas equation of state is used the dependence on $\tau_\pi$ is stronger especially for the simplified Israel-Stewart equation\cite{Song:2008si}.  Note also that  the dependence on $\tau_\pi$ is 
stronger in the smaller CuCu system than in AuAu.
However, in the  conformally invariant full Israel-Stewart 
equation the dependence on $\tau_\pi$ is negligible, indicating that the 
result is quite close to the first order Navier Stokes theory.  The more rapid convergence of the gradient expansion in conformally invariant fluids 
is a result of  the fact that the derivatives in the conformal case 
come together as
\[
     \tau_\pi \left[ \llangle D \pi^{\mu\nu} \rrangle  +  \frac{4}{3} \pi^{\mu\nu} \nabla \cdot u \right] \np
\]
We have selected one figure out of many\cite{Song:2008si,Dusling:2007gi,Luzum:2008cw}. The 
result of the analysis is that the  flow fields  and $v_2(p_T)$ are 
largely independent of the details of the second order terms at least for $\eta/s \lsim 0.3$. For this range of parameters, hydrodynamics 
at RHIC is an internally consistent theory.

\subsection{Kinetic Theory and Hydrodynamic Simulations} 
\label{freezeout}

Clearly viscous hydrodynamics is an approximation which is not 
valid at early times and near the edge of the nucleus. This failure
afflicts the current viscous  
calculations at a practical level  right at the moment of initialization.  For
instance, the longitudinal pressure 
\st
  T^{zz} \equiv \tau^2 T^{\eta\eta}  = \pr  -\frac{4}{3} \frac{\eta}{\tau_0} \nc
\stp
eventually becomes negative near the edge of the nucleus indicating
the need to transition to a kinetic description\cite{Martinez:2009mf}. 
(Note that $\pr\propto T^4$ while $\eta \propto T^3$ , so at 
sufficiently low temperatures the viscous term is always dominant regardless
of the magnitude of  $\eta/s$.)  
The current calculations simply limit 
the size of this  correction through the phenomenological Israel-Stewart 
model. For example, one approach would be to take 
\st
\tau^2 T^{\eta\eta} = \pr + \tau^2 \Pi^{\eta\eta}  \nc
\stp
with
\st
  \tau^2 \Pi^{\eta\eta} =
\begin{cases}
-\frac{4}{3} \frac{\eta}{\tau_o}  &  \mbox{while $4/3\, \eta/\tau_0 < 0.9\, \pr$  }  \\
 -0.9 \,\pr                           &  \mbox{otherwise  }
\end{cases} \np
\stp
This ad-hoc fix  is clearly not nice and points to the larger problem  of freezeout which is difficult to address with hydrodynamics itself.

\begin{figure}
\begin{center}
\includegraphics[height=3.0in]{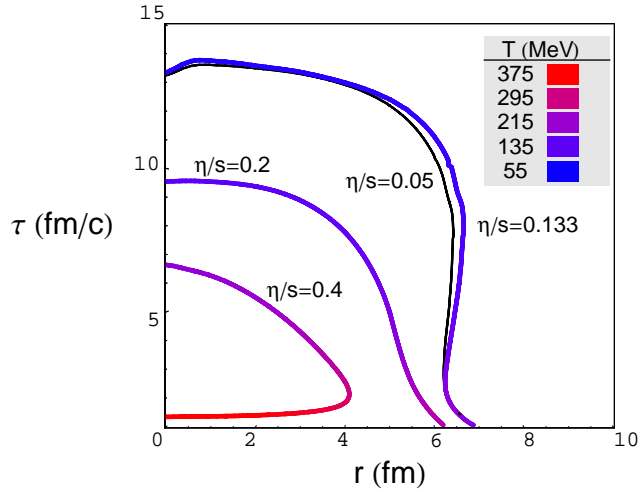}
\end{center}
\caption{This 
figure estimates for a conformal gas with equation of state $p=e/3$ and
constant $\eta/s$ the space time region described by viscous
hydrodynamics \protect \cite{Dusling:2007gi}. The contours are where $(\eta/\pr)
\,\partial_\mu u^{\mu} = 0.6$ 
for different values of $\eta/s$. For the smallest value  $\eta/s=0.05$ the
system freezes out at a time of order $\sim 40\,{\rm fm}$. This unrealistically 
long time reflects the conformal nature of the gas as discussed in \Sect{time}.  For comparison 
we have shown the $(\eta/\pr)\, \partial_\mu u^{\mu} = 0.225$ contour for 
$\eta/s=0.05$. 
\label{kevin_freezeout} }
\end{figure}

Freezeout is the colloquial term for the transition from a hydrodynamic
to a kinetic regime and is impossible to separate cleanly from the viscosity
itself  in a realistic nucleus-nucleus collision. 
Clearly as the shear viscosity is made smaller and smaller, a larger and 
larger space time volume is described by hydrodynamics.  To estimate the
size of the relevant space time region we remark that  hydrodynamics 
is valid when  
the relaxation time $\tau_R$ is much 
smaller than the inverse expansion rate, 
$\tau_R \partial_{\mu} u^{\mu}\ll 1$.
Therefore, in the simulations one can estimate the region of validity 
by monitoring the expansion rate relative to the relaxation time\cite{Heinz:1987ca,Hung:1997du}.
Specifically, freezeout is signaled when
\st
\label{freezecond}
     \frac{\eta}{\pr} \partial_{\mu} u^{\mu} \sim \frac{1}{2} \np
\stp
This combination  of parameters can be motivated from the kinetic 
theory estimates \cite{Reif}.  The pressure is $\pr \sim e
\llangle v_{\rm th}^2\rrangle  $,  with $\llangle v_{\rm th}^2\rrangle $ the 
typical quasi-particle velocity.
The viscosity is of order $\eta \sim e \llangle v_{\rm th}^2 \rrangle \tau_{R}$ 
with $\tau_R$ the relaxation time.  Thus 
hydrodynamics breaks down when
\st
    \frac{\eta}{\pr} \partial_{\mu} u^{\mu} \sim \tau_R \partial_{\mu} u^{\mu} \sim \frac{1}{2}  \np
\stp
\Fig{kevin_freezeout} estimates the space-time region described
by viscous hydrodynamics.  Examining this figure we see that 
the time duration of the hydrodynamic regime is a relatively strong function of $\eta/s$ at least for a conformal gas. In reality the behavior of the shear viscosity 
near the transition region will control when the hydrodynamics will end. 

\begin{figure}
\begin{center}
\includegraphics[height=3.0in]{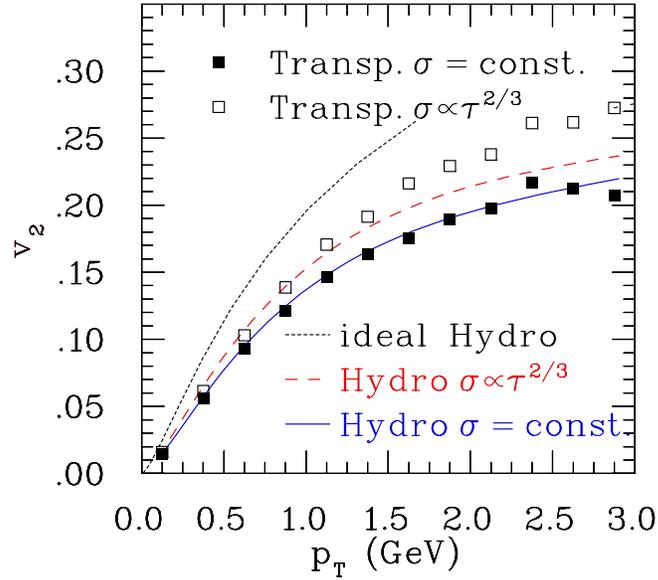}
\end{center}
\caption{Elliptic flow from \protect \Ref{Molnar:2008xj} for a massless classical gas with a 
constant cross section in kinetic theory (Transp. $\sigma=\mbox{const}$) and viscous hydrodynamics (Hydro $\sigma=\mbox{const}$). 
The $\sigma \propto \tau^{2/3}$ case is not discussed in 
this review but is an effort to simulate a gas with $\eta\propto T^3$.
  }  
\label{pasi_v2pt} 
\end{figure}

Clearly the surface to volume ratio in \Fig{kevin_freezeout} is
not very small. Hydrodynamics is a terrible approximation  
near the edge.  There is a need for a model which smoothly transitions
from the hydrodynamic regime in the center to a kinetic or free streaming 
regime at the edge.   Near the phase transition kinetic theory 
may not be a good model for  QCD, but it has the virtue that it gracefully implements this hydro to kinetic transition.  
Although the interactions and the quasi-particle picture of kinetic
theory may be decidedly incorrect, this is unimportant in the 
hydrodynamic regime. In the hydrodynamic regime  the only properties 
that determine the evolution of the system are the equation of state, $\pr(e)$, 
and the shear viscosity and bulk viscosities, $\eta(e)$ and
$\zeta(e)$.  In the sense that kinetic theory provides a reasonable guess as to how
the surface to volume ratio influences the forward evolution, these models
can be used to estimate  the shear viscosity and the estimate  may be  more
reliable than the hydrodynamic models.  A priori one should demand that the
kinetic models have the same equation
of state and the same shear viscosity as expected  from QCD,  $\eta \propto T^3$. 
For instance  in a kinetic 
model of massless particles with a constant cross section $\sigma_o$
(such as studied in \Ref{Drescher:2007cd,Gombeaud:2007ub,Huovinen:2008te}) 
the shear viscosity scales
as $\eta \propto T/\sigma_o$. This difference with QCD should be 
kept in mind when extracting conclusions about the heavy ion reaction. 
Further, many transport models conserve particle number, which is an additional 
conservation law not inherent to QCD; this also changes the dynamics. 
Keeping these reservations in mind we examine  \Fig{pasi_v2pt} from \Ref{Molnar:2008xj}.
This figure shows a  promising comparison 
between kinetic theory with a constant cross section ($\sigma_o = {\rm Const}$) and 
a viscous hydrodynamic calculation with $\eta \propto T/\sigma_o$.
The case with $\sigma \propto \tau^{2/3}$ will not be discussed in 
this review, but is an attempt  to mimic a fluid which has 
$\eta \propto T^3$. 

What is exciting about this figure is the fact that the hydrodynamic 
conclusions are largely supported by  the results of a similar kinetic 
theory. This gives considerable confidence that surface to volume 
effects are small enough that the hydrodynamic conclusions 
presented in \Fig{LR_money_plot} are largely unchanged by particles 
escaping from the  central region. 
More formally, the opacity is 
large enough to support hydrodynamics.

\begin{figure}
\begin{center}
\includegraphics[height=2.5in]{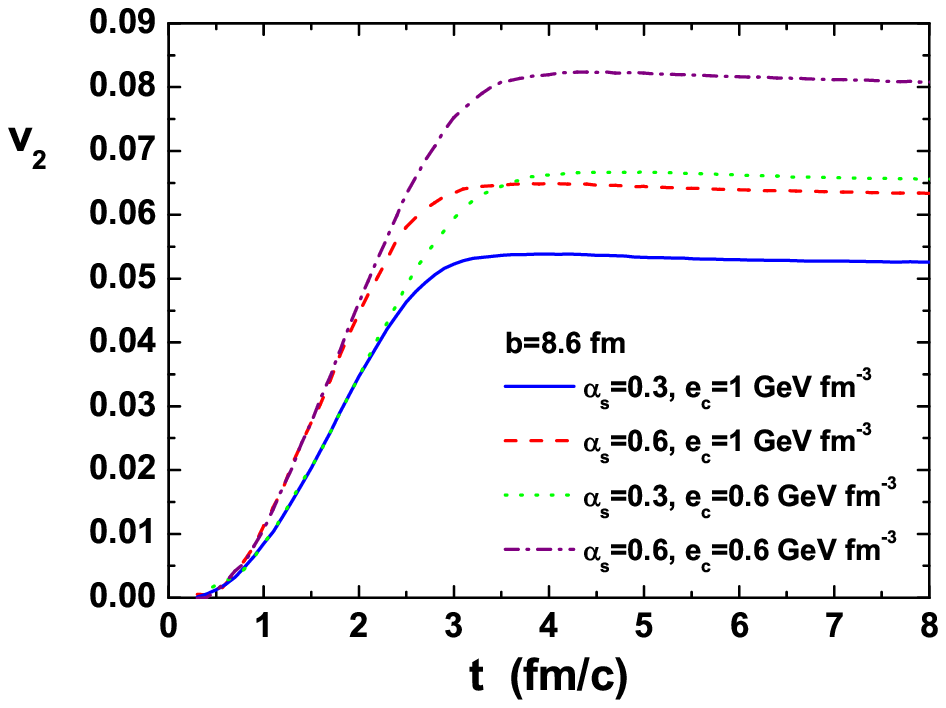}
\includegraphics[height=2.5in]{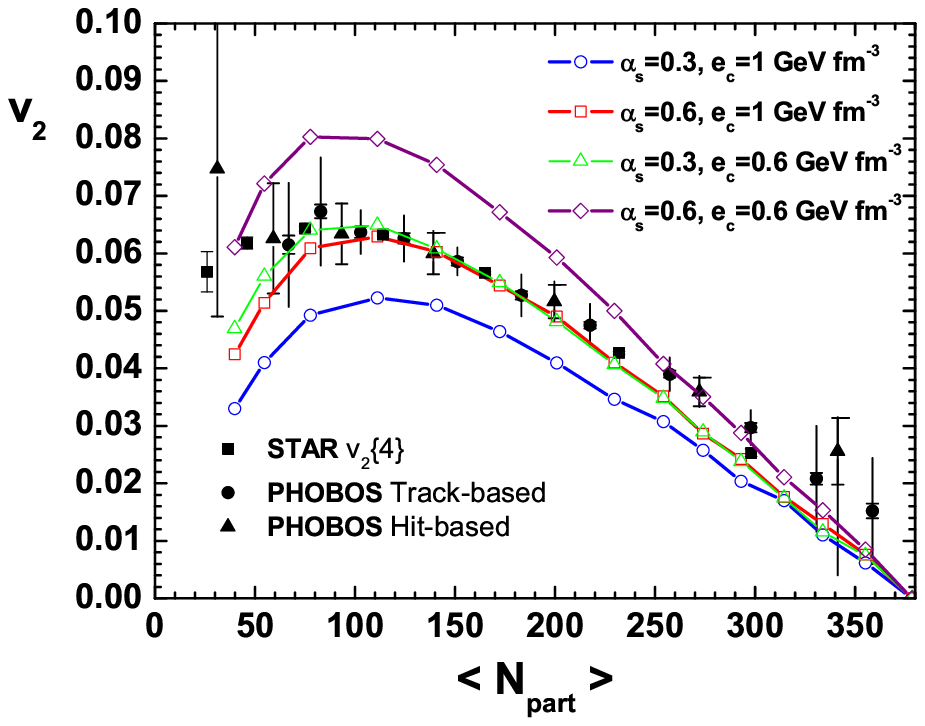}
\end{center}
\caption{ 
(Top) The development  of elliptic flow $v_2$ as a function 
of time in the BAMPS model\protect \cite{Xu:2008av}. 
(Bottom) The final elliptic flow as a function of centrality. 
The shear viscosity 
to entropy ratio $\eta/s$ corresponding  to the 
model parameter $\alpha_s=0.3 \leftrightarrow 0.6$ was estimated in \protect \Ref{Xu:2007ns,El:2008yy} and is  $\eta/s=0.16 \leftrightarrow 0.08$. 
The evolution is stopped when the energy density reaches a critical value of $e_c$.
\label{carsten_fig} }
\end{figure}

There have been other kinetic calculations which are working towards 
extracting  $\eta/s$ from the heavy ion data \cite{Ferini:2008he,Greco:2008fs,Xu:2004mz}. 
In particular  
\Refs{Xu:2008av,Xu:2007jv,El:2008yy,Xu:2007ns} used a kinetic theory implementation of $2\rightarrow2$ and 
$2\rightarrow3$ interactions motivated by weak coupling QCD \cite{Xu:2004mz}. The simulation  also calculates the Debye 
scale self consistently, {\it i.e.} in equilibrium one sets $m_D^2 \propto g^2 T^2$ 
and  $T$ changes with time. Out of equilibrium this mass scale  $m_D$ is 
determined from the momentum distribution of particles. 
Consequently this model respects the symmetry properties of high
temperature QCD, {\it i.e.} the model  has $\eta \propto T^3$ and does not conserve 
particle number.  For the model parameter $\alpha_s=0.3\leftrightarrow0.5$ 
(which is only schematically related to the running coupling) the shear to 
entropy ratio is $\eta/s=0.16\leftrightarrow0.08$ \cite{Xu:2007ns,El:2008yy}.
The model (known as BAMPS) is conformal and never freezes out as discussed in \Sect{time}. The current implementation of BAMPS simply stops the kinetic 
evolution when the energy density reaches a critical value, $e_c\simeq 0.6
\leftrightarrow 1.0\,{\rm GeV}/{\rm fm}^3$. This 
is an abrupt way to introduce a needed scale into the problem  and schematically approximates
the rapid variation of the shear viscosity in this energy density range. 
\Fig{carsten_fig} shows the time development of elliptic flow in this model
 which can reproduce the observed flow only for $\eta/s=0.16 \leftrightarrow
 0.08$.  The time development 
of $v_2$ seen in \Fig{carsten_fig} shows that it is very difficult to
separate precisely the shear viscosity in the initial stage from the freezeout
process controlled by $e_c$.

Clearly the transition from a hydrodynamic regime to a kinetic regime is
important to clarify in the future. In the meantime most hydrodynamic groups
have invoked an  ad-hoc freezeout prescription. In  
\Refs{Romatschke:2007mq,Luzum:2008cw,Song:2008si,Song:2007ux} the hydrodynamic codes were run until a typical 
freezeout temperature, $T_{\rm fo} \simeq150\,{\rm MeV}$. Technically 
the freezeout surface is constructed by 
marching forward in time and  triangulating the space-time  surface with
constant temperature.
In \Ref{Molnar:2008xj} a surface of constant particle density 
was chosen $n\simeq 0.365/{\rm fm}^3$, 
and in \Ref{Dusling:2007gi} the chosen surface was motivated by  the kinetic 
kinetic condition in \Eq{freezecond}.
Ideally  this could be improved by dynamically coupling the
hydrodynamic evolution to a kinetic description or by simulating  the
entire event with a kinetic model which closely realizes the equation 
of state and transport coefficients used in the hydrodynamic simulations. 

\subsection{Particle Spectra} 
\label{particles}

Finally we turn to the particle spectra in viscous hydrodynamics.  
Ideally the system would evolve through the approximate phase transition
down to sufficiently low temperatures where 
the dynamics could be described either with viscous  
hydrodynamics or with  the kinetic theory of a Hadron Resonance Gas (HRG).
In reality this does not seem particularly likely since the system 
is already expanding three dimensionally and the scales are approximately
fixed (see \Sect{time}).  The estimates of the shear viscosity to entropy
ratio in a hadronic gas are reliable for $T \lsim 130\,{\rm MeV}$ and
do not support this optimistic picture (see \Sect{transport}).  
In seems quite unlikely that there is equilibrium evolution 
in the HRG below a temperature of $T\simeq150\,{\rm MeV}$.  
Clearly the dynamics is extremely complex 
during the quark-hadron transition which takes  place 
for an energy density of $e\simeq 0.5 \leftrightarrow 1.2\,{\rm GeV}/{\rm fm^3}$.
In
this range, the temperature 
changes by only $\Delta T \simeq 20\,{\rm MeV}$. However,
the hydrodynamic simulations evolve this complicated 
region  for a significant period of time, $\tau \simeq 4\,{\rm fm} \leftrightarrow 7\,{\rm fm}$.  This transition region can 
be seen from the inflection in the AuAu plots in \Fig{huichao_ep2}. 

The pragmatic approach to this complexity  
is to compute  the 
quasi-particle spectrum of hadrons at a temperature of  $T\simeq150\,{\rm MeV}$.
Since the HRG describes the QCD thermodynamics well, this
pragmatism is fairly well motivated.
The approach conserves energy and momentum and when viscous corrections
are included also  matches the strains across the transition.
In ideal hydrodynamics simulations the subsequent evolution of the 
hadrons has been followed with  hadronic  cascade models \cite{Nonaka:2006yn,Teaney:2001av,Hirano:2004en}. The 
result of these hybrid models is that the hadronic 
rescattering is essentially unimportant for the $v_2(p_T)$ observables presented here.

Technically, the procedure is the following:
along the freezeout surface the spectrum of particles
is computed with the Cooper-Frye formula
\st
\label{CooperFrye}
 E \frac{\dd N^a}{\dd^3\p}  = \frac{d_a}{(2\pi)^3} \int_{\Sigma} \dd\Sigma_{\mu} P^{\mu} \, f^a(-P\cdot u/T) \nc
\stp
where $a$ labels the particle species,  the distribution 
function is, 
\st
  f^a(-P\cdot u) = n^a(-P\cdot u/T) + \delta f^a(-P\cdot u/T) \nc
\stp
and $d_a$ labels the spin-isospin degeneracy factor for each particle included (see \Sect{kinetics}).
In practice, the Boltzmann approximation is often sufficient. 
In \Ref{Luzum:2008cw} all particles were included up to mass of $m_{\rm res} < 2.0\,\rm{GeV}$ and then subsequently decayed. 
In other works  a simple single species gas was used to study various aspects 
of viscous hydrodynamics divorced from this 
complex reality\cite{Dusling:2007gi,Molnar:2008xj}. 

All of the viscous
models used the  quadratic ansatz discussed in \Sect{kinetics},  writing
the change to the distribution function of the $a$-th particle type as
\st
  f^a \rightarrow n^a + \delta f^a \nc
\stp
with $\delta f^a$ given by 
\st
\label{df}
 \delta f^a = \frac{1}{2(e+\pr) T^2} n^a (1 \pm n^a)  P^{\mu} P^{\nu} \pi_{\mu\nu} \np
\stp

Before continuing  we review the elements 
that go into a complete hydrodynamic calculation.  First 
initial conditions are specified (see \Sect{initcond}) ; 
then the equations are solved with the viscous term (see \Sect{flow_correct}) ;
after this a freezeout surface is specified (see \Sect{freezeout} for
the limitations of this); finally we compute spectra 
using \Eq{CooperFrye} and \Eq{df}. This particle spectra 
can ultimately be compared to the observed elliptic flow.
With this oversight we 
take a  more nuanced look at \Fig{LR_money_plot}.    

\begin{figure}
\begin{center}
\includegraphics[height=2.5in]{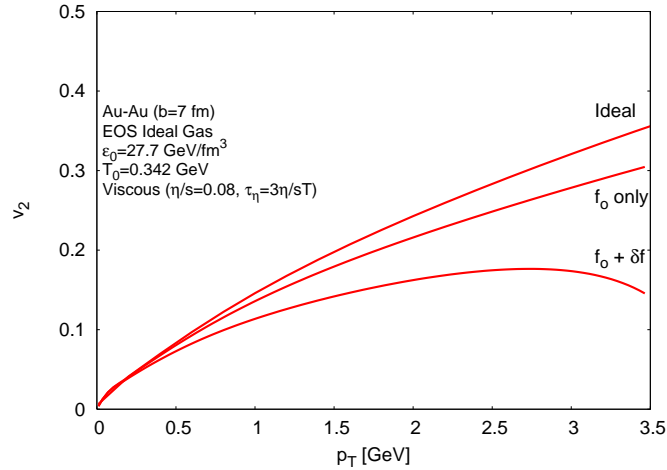}
\caption{$v_2(p_T)$ based on  \protect \Refs{Dusling:2007gi,DuslingPrivate}  
showing the $v_2(p_T)$ with ($f_o +\delta f$) and
and without ($f_o$ only) the viscous modification of the distribution function.  The result depends to a certain extent on freezeout and the freezeout 
temperature here is $T_{\rm fo}=130\,{\rm MeV}$.
\label{kevin_dffig} }
\end{center}
\end{figure}

To separate the viscous modifications of
the flow variables ($T$ and $u^{\mu}$) and the viscous modifications of the distribution function, we turn to \Fig{kevin_dffig}.  
Examining this figure we see that a significant part  of the
corrections due to the shear viscosity are from the distribution function rather than the flow.  
Although the magnitude of the flow modifications depends on the
details of freezeout,  this dependence on $\delta f$  is the typical  and somewhat distressing result.
We emphasize however that it is inconsistent to drop the modifications due to
$\delta f$.  The result of dropping the $\delta f$ means that the 
energy and momentum of the local fluid cell $T^{\mu\nu}u_{\nu}$
is  matched by the particle content, but the off diagonal strains
$\pi^{\mu\nu}$ are not 
reproduced.  The  assumption underlying 
the comparison of viscous hydrodynamics to data is that the form of these off
diagonal strains is largely unmodified from the Navier Stokes 
limit during the freezeout process. As the particles are ``freezing out"
this is a reasonable assumption.  
However, since these strains are only partially
constrained by conservation laws, this assumption needs to be tested
against kinetic codes as already emphasized above. This freezeout 
problem is clearly an obstacle to a reliable extraction of $\eta/s$ 
from the data.  

\begin{figure}
\begin{center}
\includegraphics[height=2.5in]{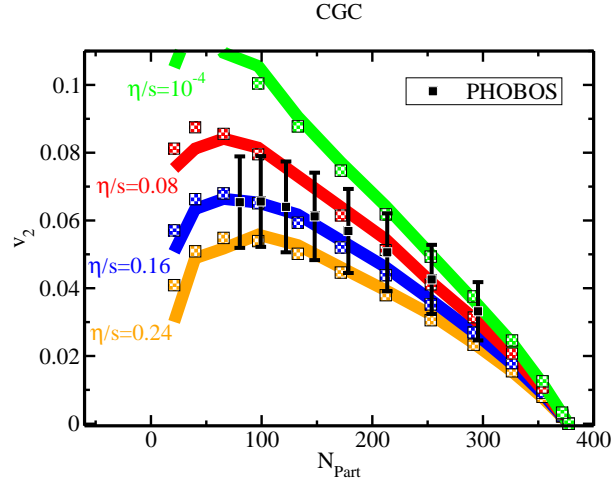} 
\end{center}
\caption{Dependence of  elliptic flow versus some centrality from \protect \Ref{Luzum:2008cw}.  The lines show the results of viscous hydrodynamics and 
the colored squares  show the anisotropy of the stress tensor $0.5\,e_p'$ (see \protect \Eq{epp}) for different values of $\eta/s$. The sensitivity 
to the quadratic ansatz is estimated in the text and corresponds to 
half the difference between the red ($\eta/s=0.08$) and blue ($\eta/s=0.16$) curves.  The heuristic 
rule $v_2 \simeq 0.5\,e_p'$ is motivated in the text.
\label{romatschke_luzum_v2b}
}
\end{figure}

Although the dependence on $\delta f$ in $v_2(p_T)$ is undesirable, 
the viscous modifications of the {\it integrated} elliptic flow  $v_2$
largely reflects the modifications to the stress tensor itself.  The observation
is that the stress 
anisotropy  $e_p'$ (see \Eq{epp}) determines the average flow 
according to a  simple  rule of thumb\cite{KSH,Luzum:2008cw}
\st
   v_2  \simeq \frac{1}{2} e_p' \np
\stp
\Fig{romatschke_luzum_v2b} shows $e_p'$ and $v_2$ as a 
function of centrality.
The figure  supports this rule and  suggests that  sufficiently
integrated predictions from hydrodynamics  do not depend
on the detailed form  of the viscous distribution.

To corroborate this conclusion 
we turn to an 
analysis originally presented by Ollitrault\cite{OllitraultSph} and subsequently generalized to
the viscous case\cite{Dusling:2007gi,DuslingInProg}.  
First we parameterize the 
single particle spectrum $dN/dp_T$  with an 
exponential  and $v_2(p_T)$ as linearly rising, {\it i.e.}
\st
 \frac{1}{p_T} \frac{dN}{dp_T} \simeq C e^{-p_T/T} \nc \qquad \mbox{and} \qquad  v_2 \propto p_T \np
\stp  
With this form one finds  quite generally 
that the $p_T^2$ weighted elliptic flow is twice the average $v_2$ 
\st
   2v_2   \simeq  A_2 \equiv \frac{\llangle p_x^2 - p_y^2 \rrangle }{\llangle p_x^2 + p_y^2 \rrangle} \np
\stp
The $p_T^2$ weighted elliptic flow has a much closer relationship 
to the underlying hydrodynamic variables.  Indeed
we will show how this simple rule of thumb arises and 
that it is largely independent of the details of $\delta f$.

To this end, we evaluate the sphericity tensor which will have
a simple relationship to $A_2$
\st
     S^{\mu\nu\rho} = S_{I}^{\mu\nu\rho} + S_{V}^{\mu\nu\rho} =  \int \frac{d^3\p}{(2\pi)^3 E_\p} 
  P^{\mu} P^{\nu} P^{\rho} \left[ n(-P\cdot u) + \delta f(-P\cdot u) \right]
\stp  
Then using the Cooper-Frye formula 
(with $\dd\Sigma_\mu  = (\tau \dd\eta_s\, \dd^2\x_\perp , 0,0,0)$) the asymmetry at any given moment  in 
proper time is 
\st
 A_2 =   \frac{ \int d^2\x_\perp \, (S^{0 xx} - S^{0 yy})} { \int d^2\x_\perp \; (S^{0 xx} + S^{0 yy}) }  \np
\stp

The sphericity tensor consists of an ideal piece and a viscous 
piece 
\st
    f\rightarrow n + \delta f \nc \qquad S^{\mu\nu\rho} \rightarrow  S_I^{\mu\nu\rho} + S_V^{\mu\nu\rho} \np
\stp 
First we consider the ideal piece and  work in
a classical massless gas approximation for ultimate simplicity.
The tensor is a third rank symmetric 
tensor  and  can be decomposed as
\st
 S^{\mu\nu\rho}_I = A(T) u^{\mu} u^{\nu} u^\rho + B(T) \left(u^{\mu} g^{\nu\rho} + \mbox{perms}  \right) \np
\stp
Here $A(T)$ and $B(T)$ are thermodynamic functions 
and 
are given   by
\st
 \frac{A}{6}  =  B = \int \frac{d^3\p}{(2\pi)^3} \frac{p^2}{3} n_\p =  (e + \pr) T \np
\stp
For Bose and Fermi gases this relation between $A(T)$, $B(T)$, and the enthalpy is approximate.
Thus the ideal 
piece of the sphericity tensor is largely constrained by 
thermodynamic functions. 
The viscous piece is largely constrained
by the  shear viscosity. 
As discussed in \Sect{kinetics} we parameterize  the $\delta f$ with $\chi(p)$
\st
  \delta f  = -n_\p \, \frac{\chi(p)}{(P\cdot U)^2}  P^{\mu}P^{\nu} \sigma_{\mu\nu}  \np
\stp
While the precise form of the viscous correction $\chi(p)$ depends on the
details of the microscopic interactions, it is constrained by the 
shear viscosity
\st
  \eta = \frac{2}{15} \int \frac{d^3\p}{(2\pi)^3} p \,\chi(p)  n_\p \np
\stp
Substituting the viscous parameterization into the definition of the sphericity
we find
\bg
S^{0xx} &=& -C(T) \left[u^{0} \sigma^{xx}  + 2 u^{x}\sigma^{x0}  \right] \nc \\ 
        &\simeq& -C(T)  u^{0} \sigma^{xx}  \nc
\nd
where $C(T)$ is 
\st
   C(T) = \frac{2}{15} \int \frac{d^3\p}{(2\pi)^3} p^2 \chi(p)  n_\p   \np
\stp
For simplicity we have assumed that the flow is somewhat non-relativistic so
that $u^{x} \pi^{x0}$  is  $O(v^2)$ compared to   $u^{0} \pi^{xx}$. 
To get a feeling for how sensitive the results are to the
quadratic  ansatz 
we will work with a definite functional form 
\st
\chi(p) = {\mbox{Const}} \times p^{2-\alpha} \np
\stp
In a relaxation time approximation discussed in \Sect{kinetics},  $\alpha=0$  
corresponds to a relaxation time which increases with linearly $p$ while $\alpha=1$ 
corresponds to a relaxation time independent of $p$.  Substituting this
ansatz  we find
\st
   C(T) \simeq  (6 - \alpha) T \eta\nc \quad\mbox{where} \quad 0 < \alpha < 1 \np
\stp

Having assembled the ingredients we can write down an approximate formula 
for $A_2$ 
\st
\label{A2}
 A_2  \simeq
\frac{ \int \dd^2\x_\perp \;   
  T u^{0}\left[  (e + \pr)  (u^x u^x  - u^y u^y)  +   (1 - \frac{\alpha}{6} ) \left(\pi^{xx} - \pi^{yy} \right) \right]   }
{ \int \dd^2\x_\perp \;
  T u^{0} \left[ (e + \pr)  (u^x u^x  + u^y u^y)  +  (e + \pr)/3 +  (1 - \frac{\alpha}{6} )  \left(\pi^{xx} + \pi^{yy} \right) \right]   } \nc 
\stp
with $0 < \alpha < 1$. 
This is the desired formula which expresses the observed elliptic flow 
in terms of the hydrodynamic variables.  To reiterate, the coefficient 
$\alpha$ changes the functional form
the viscous distribution function and  $0 < \alpha < 1 $ is a reasonable range  --- $\alpha=0$ is the usual quadratic ansatz.
It is useful to compare this formula 
to the definition of $e_p'$
\st
e_p'  = 
\frac{ \int \dd^2\x_\perp \;
   \left[  (e + \pr)  (u^x u^x  - u^y u^y)  +   \left(\pi^{xx} - \pi^{yy} \right) \right]   }
{ \int  \dd^2\x_\perp \;
   \left[ (e + \pr)  (u^x u^x  + u^y u^y)  +  2 \pr +  \left(\pi^{xx} + \pi^{yy} \right) \right]   } \np
\stp
Thus while $e_p'$ is not exactly equal to the $A_2$  of \Eq{A2},
it is close enough to explain the heuristic  rule, $2v_2 \simeq
e_p'$.

The overall symmetries and dimensions of the sphericity tensor  
suggests a definition for an analogous quantity in hydrodynamics
\st
  S^{\mu\nu\rho}_{\rm hydro} = T\left[ 
(e +\pr) u^{\mu} u^{\nu} u^{\rho} + \frac{1}{6} (e+\pr)(u^{\mu} g^{\nu\rho} + \mbox{perms})  + (u^{\mu} \pi^{\nu\rho} + \mbox{perms} ) 
\right] \np 
\stp
In fact in a classical massless gas 
approximation with the quadratic ansatz, the analysis sketched above shows
that 
\st
    S^{\mu\nu\rho}_{\rm hydro} = \frac{1}{6} S^{\mu\nu\rho}  \np
\stp
The important  point for this review is  that from \Eq{A2}
we see that the  {\it integrated} elliptic flow is relatively insensitive to
the quadratic form for the viscous distribution function.  
More specifically  the uncertainty is  $\sim 15\%$ of  $A_{2}^{\rm ideal} - A_2^{\rm viscous}$, {\it i.e.} 
about half the difference
between the blue ($\eta/s=0.16$) and red curves ($\eta/s=0.08$) in \Fig{romatschke_luzum_v2b}.  However, the estimate shows that changing the quadratic ansatz 
will move the curves systematically higher increasing the preferred 
value of $\eta/s$ to a certain extent.
The  quadratic ansatz  (which has been used by all hydrodynamic 
simulations) can be a poor 
approximation when collinear emission processes are included
in the Boltzmann description\cite{GuyPrivate}. 

Clearly addressing more completely the uncertainties 
associated with the particle 
content and microscopic interactions in \Fig{romatschke_luzum_v2b}  is a task for the future.
Nevertheless, it does seem that sufficiently integrated quantities will 
reflect rather directly the bulk properties of the hydrodynamic motion 
in a way that can be quantified.

\section{Summary and Outlook}
\label{summary}
The elliptic flow data presented in \Fig{raimond_v2pt} and \Fig{omega} provide strong 
evidence for hydrodynamic evolution in a deconfined phase of QCD.
It is difficult to think of other QCD based mechanisms
which could  explain the measured elliptic flow.  
This is because the
nuclear geometry  
has a size $R_{Au} \sim 4\,{\rm fm}$, which is large compared to the momentum 
scales in the problem $p \sim 0.6\,{\rm GeV}$. 
Given this large size, the  response of the 
nuclei to this initial geometry must develop over a relatively long time, $\tau\simeq 4\,{\rm fm}/c$.  Hydrodynamics is the appropriate framework 
to describe the collective motion over these long 
time scales.

We have discussed three advances
which  have  corroborated the hydrodynamic interpretation 
of the flow. 
The first advance is experimental and
a brief overview is provided in \Sect{elliptic}. These measurements now show quite clearly 
that the hydrodynamic response is from a deconfined phase (see \Fig{omega}).
Furthermore, the measurements also show 
that the flow decreases in smaller systems in a systematic way 
(see \Fig{raimond_v2pt}).  

The second advance is in viscous hydrodynamics.
There
has been important conceptual progress in understanding hydrodynamics
beyond the Navier-Stokes  order (see \Sect{second}).
The important result of this analysis (see \Fig{bjfig}) is that
for a Bjorken expansion 
there are generic kinematic cancellations between the second order terms
which reduce their relative importance. 
In fact in a $0+1D$ Bjorken expansion of 
a conformal gas in the relaxation time approximation, the second order
corrections vanish, while in other more general kinetic theories the 
second order corrections almost vanish. 
This understanding of second order hydrodynamics has spurred
additional progress in viscous simulations of the heavy ion event (see \Sect{viscous_model}). 
It is satisfying that viscous
hydrodynamics simulations are in better agreement with the data than ideal hydrodynamics and naturally explain several trends in the elliptic flow.
For example, viscosity explains the fall off of $v_2$    at high 
momentum (see \Fig{kevin_dffig} and \Fig{raimond_v2pt}) and the fall off of $v_2$  in more 
peripheral collisions (see \Fig{raimond_v2pt} and \Fig{romatschke_luzum_v2b}).
Many of these trends were previously reproduced by transport models\cite{Molnar:2001ux} 
and it is exciting to see that they are now reproduced with a macroscopic approach.

The final advance has been in kinetic theory 
simulations. 
Kinetic theory simulations smoothly interpolate between 
the hydrodynamic regime and free streaming. Since hydrodynamics
is universal ({\it i.e.} it only depends on $\pr(e)$ and $\eta(e)$), 
the kinetic theory results  can be used to estimate how 
still higher gradients would influence the hydrodynamic conclusions.
Ideally the results of these simulations would be largely independent 
of the details of the interactions.
The kinetic simulations have demonstrated the ability to reproduce the
viscous hydrodynamics (see \Fig{pasi_v2pt})), validating  the hydrodynamic
assumption that the dynamics near the edge does not significantly 
influence the time  development of elliptic flow.
In addition, various advances have allowed these
kinetic codes to simulate a fluid 
which does not conserve particle number and which has a shear 
viscosity,  $\eta\propto T^3$ (see \Fig{carsten_fig}). Thus, the estimates of $\eta/s$ from 
these simulations  are complementary to the macroscopic approach, and must be accepted  even if theoretical prejudice
rejects the microscopic details.
 
Examining the figures in this review, we see that there is 
consensus between hydrodynamic and kinetic models on 
several points\cite{Molnar:2001ux,Teaney:2003kp,Romatschke:2007mq,Xu:2007jv,Xu:2008av,Drescher:2007cd,Song:2007ux,Dusling:2007gi,Molnar:2008xj,Ferini:2008he}: 
\begin{itemize}
\item  
Shear viscosity is needed to reproduce trends in the data. 
This consensus 
has  yet to translate to an agreed upon lower bound on $\eta/s$ but
probably will in the not too distant future.
\item
It is impossible to reproduce the elliptic flow if the 
shear is too large:
\st
\eta/s \gsim 0.4 \np
\stp
This bound is quite safe and has been found by all groups which
have tried to reproduce the observed flow.
\item The preferred value of $\eta/s$ is (see \Fig{LR_money_plot} and \Fig{carsten_fig}) 
\[ 
\eta/s \simeq (1\leftrightarrow 3) \times \frac{1}{4\pi} \np 
\]
\end{itemize}
To reduce these constraints further several items need to studied
and quantified.
\begin{itemize}
\item At the end of the collision kinetic 
assumptions need to be made.  The uncertainties involved with 
the particle content and the quadratic ansatz to the distribution
need to be quantified to a much greater extent than has been 
done so far. The theoretical motivations and limitations 
of the quadratic ansatz have been discussed in \Sect{kinetics}.   
In \Sect{particles} we have taken the nascent steps 
to quantify these uncertainties 
with an independent analysis of  \Fig{romatschke_luzum_v2b} due to 
Luzum and Romatschke.
\item The 
nucleus-nucleus collision terminates as the system enters
the transition region and starts expanding three dimensionally.
Some estimates for this transition process have been 
given in \Sect{time}.  Clearly the transition region sets a very 
definite scale $e\simeq0.6\leftrightarrow 1.2\,{\rm GeV}/{\rm fm}^3$ which can not be ignored.
It is difficult to separate the rapid $\eta(e)$ dependence
in this region from the shear viscosity well into  the QGP phase (See \Fig{carsten_fig}).
This scale influences the  dependence of $v_2$ on 
centrality and hampers an extraction of $\eta/s$  from this 
size dependence. It will be important to categorize in a model independent
way ({\it e.g.} $\eta(e)$) how this scale influences the final flow and 
the final estimates of $\eta/s$.
\end{itemize}

Finally, while this review has focused squarely on elliptic flow, 
the short transport time scales estimated here have
implications for a large number of other observables -- energy loss\cite{Adams:2005dq,Adcox:2004mh},
the ridge and Mach cones\cite{Adler:2005ee,Alver:2008gk,Abelev:2008nd}, heavy quarks \cite{Abelev:2006db,Adare:2006nq}, and many more.
 Ultimately these observables will provide 
a more complete picture of strongly coupled dynamics near the QCD phase transition.

\section*{Acknowledgments}

I gratefully acknowledge help  from
  Raimond Snellings, Kevin Dusling, Paul
Romatschke, Peter Petreczky, Denes Molnar, Guy Moore, Pasi Huovinen, Tetsufumi Hirano, 
Peter Steinberg, Ulrich Heinz, Thomas Schaefer, and Carsten Greiner. Any misstatements and
errors
reflect the shortcomings of the author. I also am grateful to the 
organizers and participants of the Nearly Perfect Fluids Workshop 
which clarified the appropriate content of this review.
D.T. is supported by the U.S. Department of
Energy under an OJI grant DE-FG02-08ER41540 and as a Sloan Fellow. 


\end{document}